%% file: CMF-in-ALMA-IMF.tex
\documentclass[longauth]{aa} 
\usepackage{graphicx}
\usepackage{subfig}
\usepackage{lscape}
\usepackage{txfonts}
\usepackage[colorlinks=true,allcolors=blue]{hyperref}
\usepackage[colorinlistoftodos]{todonotes}
\usepackage{mathtools}
\usepackage{romannum}

\newcommand{\msun}{$M_\odot$}

\begin{document} 

   \title{ALMA-IMF}

   \subtitle{\Romannum{15} The core mass function in the high-mass star-formation regime} 

   \author{F. Louvet\inst{1,2}
\and 
P. Sanhueza\inst{3,4}
\and 
A. Stutz\inst{5}
\and 
A. Men'shchikov\inst{6}
\and 
F. Motte\inst{1}
\and 
R. Galv\'an-Madrid\inst{7}
\and 
S. Bontemps\inst{8}
\and 
Y. Pouteau\inst{1}
\and 
A. Ginsburg\inst{9}
\and
T. Csengeri\inst{8}
\and
J. Di Francesco\inst{10}
\and 
P. Dell'Ova\inst{11}
\and
M. Gonz\'alez\inst{6}
\and  
P. Didelon\inst{6}
\and 
J. Braine\inst{8}
\and 
N. Cunningham\inst{1}
\and
B. Thomasson\inst{1}
\and 
P. Lesaffre\inst{11}
\and 
P. Hennebelle\inst{6}
\and 
M. Bonfand\inst{25}
\and 
A. Gusdorf\inst{11,12}
\and 
R. H. \'Alverez-Guti\'errez\inst{5}
\and 
T. Nony\inst{7}
\and 
G. Busquet\inst{13,14,15}
\and 
F. Olguin\inst{16}
\and 
L. Bronfman\inst{2}
\and 
J. Salinas\inst{5}
\and 
M. Fernandez-Lopez\inst{17}
\and 
E. Moraux\inst{1}
\and 
H.L. Liu\inst{18}
\and
X. Lu\inst{19}
\and 
V. Huei-Ru\inst{16}
\and
A. Towner\inst{26} 
\and
M. Valeille-Manet\inst{8}
\and 
N. Brouillet\inst{8}
\and 
F. Herpin\inst{8}
\and 
B. Lefloch\inst{8}
\and 
T. Baug\inst{21}
\and 
L. Maud\inst{22}
\and 
A. Lopez-Sepulcre\inst{1,23}
\and 
B. Svodoba\inst{24}
          }

\institute{Univ. Grenoble Alpes, CNRS, IPAG, 38000 Grenoble, France. \email{fabien.louvet@univ-grenoble-alpes.fr}
\and
DAS, Universidad de Chile, 1515 camino el observatorio, Las Condes, Santiago, Chile
\and 
National Astronomical Observatory of Japan, National Institutes of Natural Sciences, 2-21-1 Osawa, Mitaka, Tokyo 181-8588, Japan
\and
Department of Astronomical Science, SOKENDAI (The Graduate University for Advanced Studies), 2-21-1 Osawa, Mitaka, Tokyo 181-8588, Japan
\and
Departamento de Astronom\'{i}a, Universidad de Concepci\'{o}n, Casilla 160-C, 4030000 Concepci\'{o}n, Chile
\and
Universit\'e Paris-Saclay, Universit\'e Paris Cit\'e, CEA, CNRS, AIM, 91191 Gif-sur-Yvette, France
\and
Instituto de Radioastronom\'ia y Astrof\'isica, Universidad Nacional Aut\'onoma de M\'exico, Morelia, Michoac\'an 58089, M\'exico
\and
Laboratoire d'astrophysique de Bordeaux, Univ. Bordeaux, CNRS, B18N, all\'ee Geoffroy Saint-Hilaire, 33615 Pessac, France
\and
Department of Astronomy, University of Florida, PO Box 112055, USA
\and
Herzberg Astronomy and Astrophysics Research Centre, National Research Council of Canada, 5071 West Saanich Road, Victoria, BC CANADA V9E 2E7
\and
Laboratoire de Physique de l'\'Ecole Normale Sup\'erieure, ENS, Universit\'e PSL, CNRS, Sorbonne Universit\'e, Universit\'e de Paris, Paris, France
\and
Observatoire de Paris, PSL University, Sorbonne Universit\'e, LERMA, 75014, Paris, France
\and
Departament de F\'isica Qu\`{a}ntica i Astrof\'isica (FQA), Universitat de Barcelona (UB), c. Mart\'i i Franqu\`{e}s, 1, 08028 Barcelona, Spain
\and
Institut de Ci\`{e}ncies del Cosmos (ICCUB), Universitat de Barcelona (UB), c. Mart\'i i Franqu\`{e}s, 1, 08028 Barcelona, Spain
\and
Institut d'Estudis Espacials de Catalunya (IEEC), 08340, Barcelona, Catalonia, Spain
\and
Institute of Astronomy, National Tsing Hua University, Hsinchu 30013, Taiwan
\and
Instituto Argentino de Radioastronom\'\i a (CCT-La Plata, CONICET; CICPBA), C.C. No. 5, 1894, Villa Elisa, Buenos Aires, Argentina
\and
Department of Astronomy, Yunnan University, Kunming, 650091, PR China
\and
Shanghai Astronomical Observatory, Chinese Academy of Sciences, 80 Nandan Road, Shanghai 200030, People's Republic of China
\and
Department of Astronomy, The University of Tokyo, 7-3-1 Hongo, Bunkyo, Tokyo 113-0033, Japan
\and
S. N. Bose National Centre for Basic Sciences, Block JD, Sector III, Salt Lake, Kolkata 700106, India
\and
ESO Headquarters, Karl-Schwarzchild-Str 2 D-85748 Garching
\and
Institut de Radioastronomie Millim\'etrique (IRAM), 300 rue de la Piscine, 38406 Saint-Martin-D'H\`eres, France
\and
National Radio Astronomy Observatory, PO Box O, Socorro, NM 87801 USA
\and
Departments of Astronomy and Chemistry, University of Virginia, Charlottesville, VA 22904, USA
\and
Steward Observatory, University of Arizona, 933 North Cherry Avenue, Tucson, AZ 85721, USA
\and 
Universit\'e Paris Cit\'e, Universit\'e Paris-Saclay, CEA, CNRS, AIM, F-91190 Gif-sur-Yvette, France
             }

   \date{Received January 25, 2023; accepted July 25, 2024}

 

\abstract{
The stellar initial mass function (IMF) is critical to our understanding of star formation and the effects of young stars on their environment. On large scales, it enables us to use tracers such as UV or H$\alpha$ emission to estimate the star formation rate of a system and interpret unresolved star clusters across the universe. So far, there is little firm evidence of large-scale variations of the IMF, which is thus generally considered "universal". Stars form from cores and it is now possible to estimate core masses and compare the core mass function (CMF) with the IMF, which it presumably produces. The goal of the ALMA-IMF large program is to measure the core mass function at high linear resolution (2700\,au) in 15 typical Milky Way protoclusters spanning a mass range of 2.5$\times 10^3$ to 32.7$\times 10^3$\,\msun. In this work, we used two different core extraction algorithms to extract $\approx$680 gravitationally bound cores from these 15 protoclusters. We adopt per core temperature using the temperature estimate from the PPMAP Bayesian method. A power-law fit to the CMF of the sub-sample of cores above the $1.64$\,\msun~completeness limit — 330 cores — through the maximum likelihood estimate technique yields a slope of $1.97\pm0.06$, significantly flatter than the 2.35 Salpeter slope. Assuming a self-similar mapping between the CMF and the IMF, this result implies that these 15 high-mass protocluster will generate atypical IMFs. This sample is the largest to date produced and analysed self-consistently, derived at matched physical resolution, with per-core temperature estimates and cores as massive as 150\,\msun. We provide the raw source extraction catalogues and the source derived size, temperature, mass, and spectral indices in the 15 protoclusters.
}

   \maketitle
%


\section{Introduction}
\label{s:intro}

In 1955, Edwin Salpeter reported that the high-mass tail of the initial mass function of stars, the IMF, could be represented by a power law in the form $\frac{{\rm d} N}{{\rm d}M}\propto M^{-\alpha}$ with $\alpha=2.35$. Since then, numerous studies reported its universality \citep[see the review by][]{bastian10}, and studied its origin \citep[e.g.][]{hennebelle-chabrier08}. To investigate the origin of the IMF, astronomers studied the fragmentation of molecular clouds, in the form of small substructures ($\lesssim$0.03\,pc) — the cores — and the link between the core mass function (CMF) and the IMF \citep[e.g.][]{motte98,fiorellino21}. Because of observational limitations, until 2018 these studies focused mostly on nearby clouds ($\sim$140-400\,pc), which mainly form low- and intermediate-mass stars; they all reported CMF slopes at the high-mass end compatible with the slope of the IMF, suggesting that the IMF directly inherits its shape from the CMF \citep[e.g.][]{alves07,konyves10}. 

In 2018, taking advantage of the angular resolution and sensitivity of the Atacama Large Millimetre/submillimetre Array (ALMA), \cite{mottenat} studied the high-mass protocluster W43-MM1, at a distance of $\simeq$5.5\,kpc. They reported a CMF with a high-mass tail flatter than the canonical IMF, with $\alpha$=1.96$\pm$0.13. Since then, other teams reported similar results in other high-mass protoclusters but often using single pointing observations, thus narrow fields not imaging pc-scale clouds \citep[e.g.][]{liu18,cheng18,kong19,oneill21}. These results cast doubt on the universality of the IMF or on the direct link between the CMF and the IMF, or even on both. However, the comparison of core samples between these studies — which are captured at different evolutionary stages — becomes complex or impossible when the observations are carried out at disparate physical resolutions and/or sensitivities. Thus, it became imperative to obtain a sample of high-mass protoclusters representative of the Milky Way observed homogeneously. 

To directly address this, we present in this article the results of the ALMA-IMF Large Program (PIs: F. Motte, A. Ginsburg, F. Louvet, P. Sanhueza). ALMA-IMF observed 15 high-mass protoclusters at comparable sensitivity and physical resolution. The main driver of the protocluster selection is a mass criteria, first choosing protoclusters with inner parts of at least 500 \msun, as described in detail in the Sect. 2.1 of \cite{motte22}. A second criteria is the distance of the protoclusters. Not too close by ($>2$\,kpc) to permit imaging a large area with mosaic observations and not too far ($<6$\,kpc) to prevent for excessive integration time. Our final sample consists of 15 protoclusters that were observed with continuum sensitivity below 60 mK at 1.3\,mm, and at similar physical resolution ($\sim\,$2\,kau).

In this article, we determine and analyse the combined CMF of all 15 protosclusters from ALMA-IMF. We show the 1.3 and 3\,mm continuum images in Sect.~\ref{s:obs}, first presented in \cite{ginsburg22}. We describe the source extraction in Sect.~\ref{s:extraction}. We describe the source selection and core mass derivation in Sect.~\ref{s:res}, paying special attention to freefree emission and temperature of the cores. We present the global CMF for all 15 fields. We discuss our results regarding the current knowledge about CMF in Sect.~\ref{s:disc}, and we summarize our conclusions in Sect.~\ref{s:concl}.


\section{Observations}
\label{s:obs}

The data were taken from	 the ALMA-IMF large program (Project ID: 2017.1.01355.L), entitled \textit{ALMA transforms our view of the origin of stellar masses}. \cite{motte22}, hereafter referred to as Paper~\Romannum{1}, described the project, its choice of targets, datasets, objectives and first results. \cite{ginsburg22}, hereafter referred to as Paper~\Romannum{2}, described the reduction pipeline for the continuum emission maps at 1.3\,mm and 3\,mm in the ALMA-IMF sample. Paper~\Romannum{2} presented two versions of the continuum maps: the \textsc{bsens} and the \textsc{cleanest}. The \textsc{bsens} maps were constructed using the full range of the continuum spectral windows — which may include significant molecular emission, especially toward hot cores. The \textsc{cleanest} maps were constructed flagging all channels contaminated by molecular emission (see Fig.~3 in Paper~\Romannum{2}). Since we wish to assemble the most reliable sample of cores, we use the \textsc{cleanest} continuum maps, publicly available on the ALMA-IMF website\footnote{\url{https://www.almaimf.com/}}. Table~\ref{t:alma-imf-fields} lists the fifteen regions investigated by ALMA-IMF together with observational characteristics, taken from Papers~\Romannum{1} and~\Romannum{2}, relevant to the present paper. We refer to Paper~\Romannum{1} for an in depth description of the 15 regions.

\input{alma-imf-fields.tex}

\section{Source extraction}
\label{s:extraction}

\subsection{Extraction tools}
\label{ss:extraction}

To extract sources in the continuum emission maps, we used two extraction tools: \textsl{getsf} \citep[v210414, ][]{sasha21}
and \textsl{GExt2D} (v210603, Bontemps et al., in prep.). This approach permits us to assess the biases inherent to our source extraction algorithms.

The multi-scale, multi-wavelength source and filament extraction method \textsl{getsf} supersedes the previous algorithms 
\textsl{getsources}, \textsl{getfilaments}, and \textsl{getimages} \citep[][respectively]{sasha12,sasha13,sasha17}. The \textsl{getsf} method separates the structural components of sources, filaments, and background in the observations into independent images of the different components. Source and filament extraction are done with their respective images, where the contributions of the other components are largely removed. For more details on \textsl{getsf}, illustrations of its applications, quantitative extraction performance, and comparisons with \textsl{getsources}, readers are referred to \cite{sasha21b,sasha21}.

\textsl{GExt2D} is based on the same idea as the \textsl{cutex}  tool \citep{molinari11}. It uses as a first step the second derivative of the maps to identify the locations of maximum curvature as indicative of compact sources. It starts from the strongest curvatures down to fainter and shallower fluctuations close to the ones expected from measured local noise. The second step involves source boundaries identification and removal of background emission as the interpolated values in the map inside the boundaries. The third step corresponds to automated Gaussian fitting at each locations of compact sources after subtracting the derived background. 

\subsection{Source extraction at native angular resolutions}
\label{ss:extractnative}

With both \textsl{getsf} and \textsl{GExt2D}, the source extraction strategy has the same two steps. Firstly, we detected source candidates in the continuum images at 1.3\,mm (Band 6) without primary beam correction. These images have a homogeneous noise level across the entire field. This choice permits the detection of sources with a constant signal-to-noise ratio threshold. Secondly, we measured the sources' fluxes and sizes in the continuum maps at 1.3\,mm and 3\,mm (Band 3) after primary beam correction. This way, the fluxes we obtained account for any attenuation resulting from being farther from the phase centre of the fields. This strategy replicates that applied in \cite{pouteau22}, hereafter referred to as Paper~\Romannum{3}, dedicated to the W43-MM2 and W43-MM3 protoclusters. Considering all 15 protoclusters together, we extracted 820 sources with \textsl{getsf}, and 930 with \textsl{GExt2D}. \textsl{GExt2D} detects the sources in the observed image, while \textsl{getsf} decomposes the image first and requires detecting a source in successive scales to validate its detection. Hence \textsl{getsf} is more conservative and extracts fewer sources than \textsl{GExt2D}. 

\begin{figure*} 
	\hspace*{-1cm}
    \centering
\subfloat{\includegraphics[trim=0cm 0cm 0cm 0cm, width=0.55\linewidth]{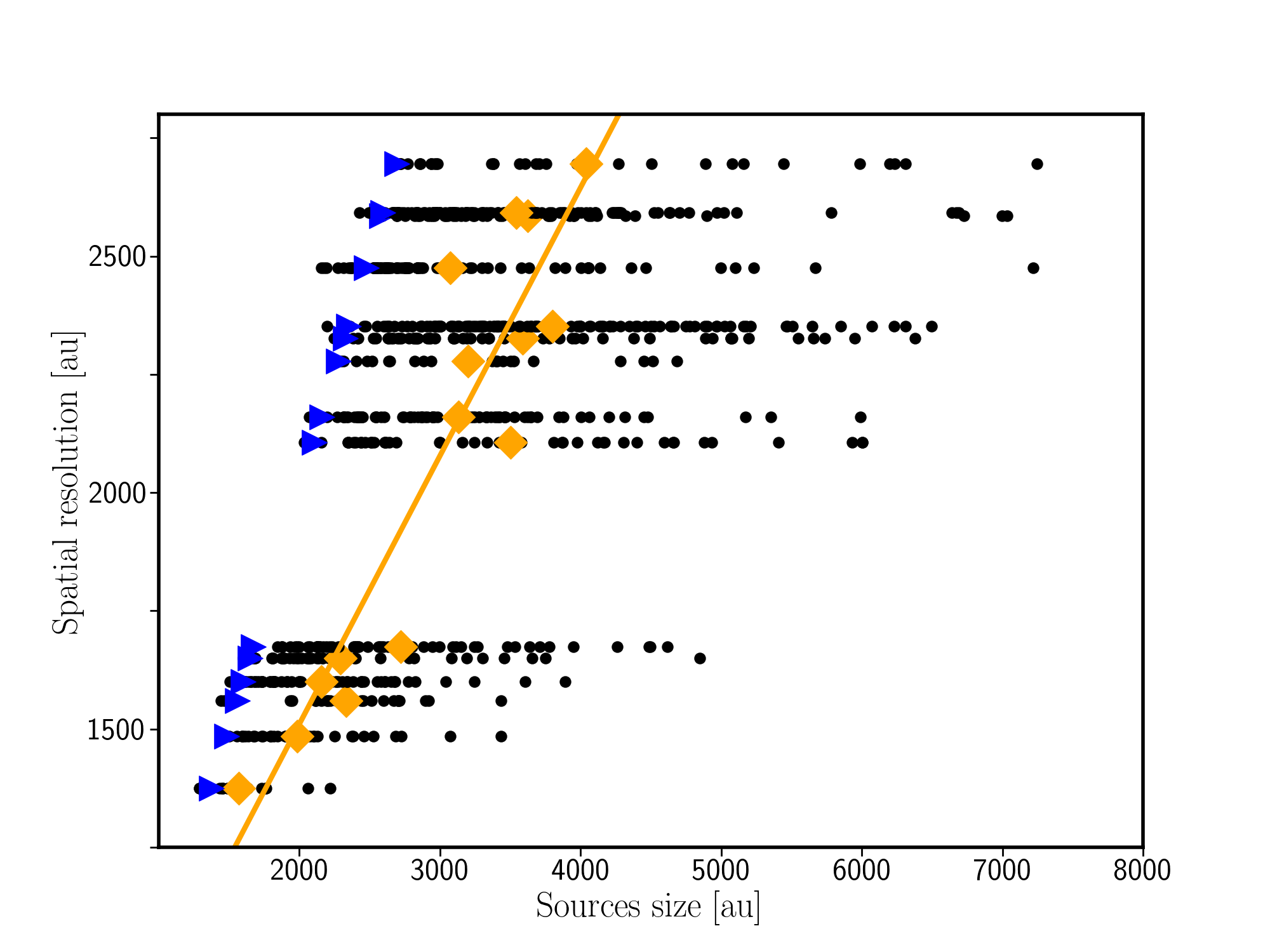}}
\subfloat{\includegraphics[trim=0cm 0cm 0cm 0cm, width=0.55\linewidth]{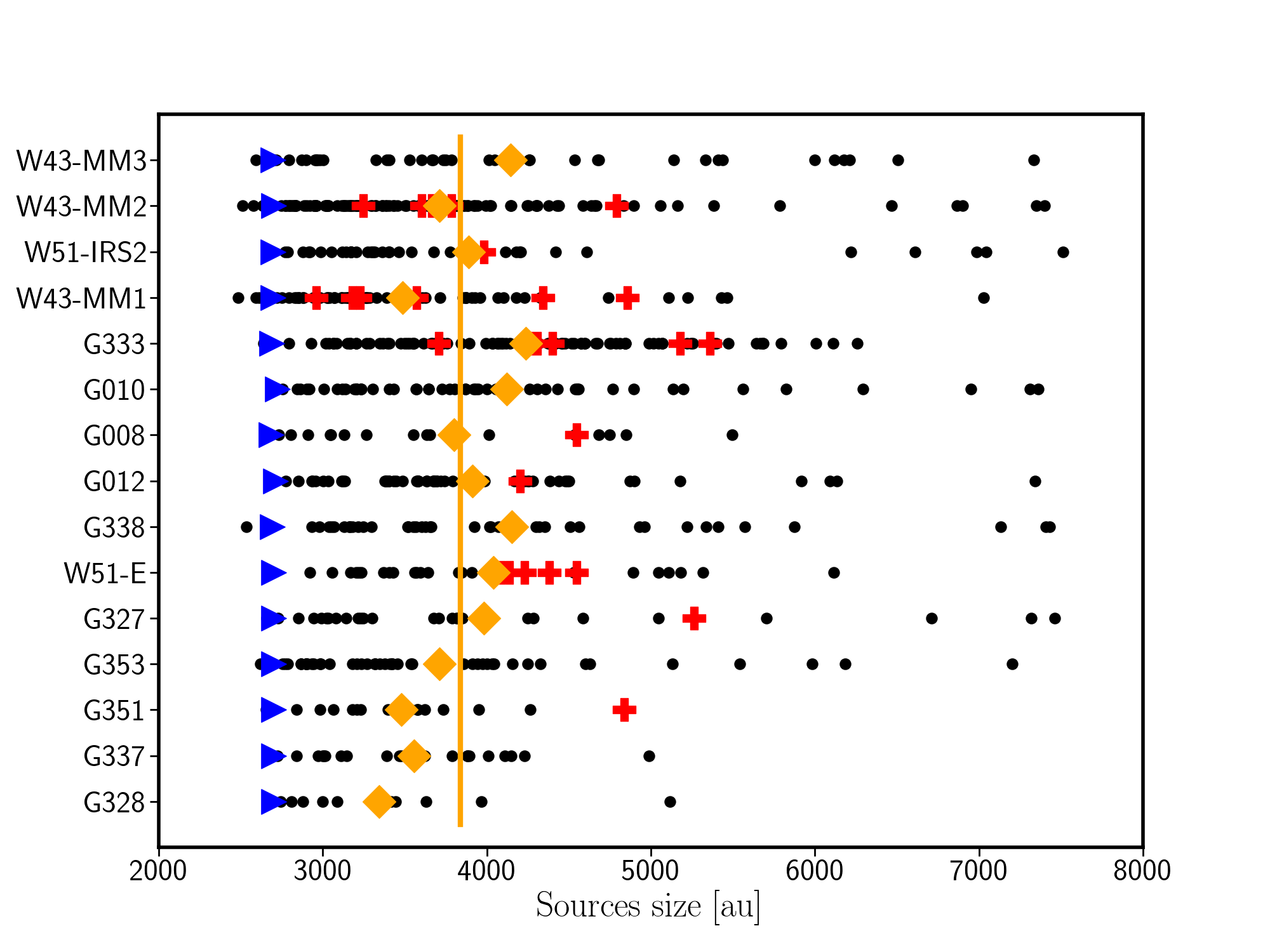}}
\caption
{ 
Scatter plots of the source size (black dots), extracted with \textsl{getsf}, before (left) and after (right) smoothing the maps to a homogeneous spatial resolution of 2700\,au. The blue triangles display the beam sizes. The orange diamonds display the mean sizes of sources. On the left panel, the \textsl{y-axis} is the linear scale (beam $\times$ distance) of the maps in Band 6, and the orange line shows the fit to the mean source sizes distribution for each field. On the right panel, the orange line shows the mean core size of all 15 protoclusters and the red crosses illustrate the cores exceeding 25\,\msun.
} 
\label{f:coresize}
\end{figure*}

The distances of the protoclusters range from 2.5 to 5.5\,kpc (see Table~1 in Paper~\Romannum{1}). The ALMA-IMF large program aimed to image the 15 protoclusters with the same spacial resolution of 2100\,au, requesting the corresponding angular resolution for each field. Despite this effort, the available antenna configurations at the observatory led to variations in the spacial resolutions by up to a factor of two (see Table~\ref{t:alma-imf-fields} and Paper~\Romannum{1}). These variations in physical scale impact the size and mass of the sources one may extract. Indeed, \cite{fkl21} performed convolution tests on projected density cubes obtained from hydro-dynamical simulations, plus column density maps created from \emph{Herschel} observations, and showed that the sizes and integrated fluxes of extracted sources closely follow the size of the linear scale. The left panel of Fig.~\ref{f:coresize} shows a linear dependency between the mean source size and the linear resolution. To obtain a homogeneous sample of sources the data must be smoothed to a matched spatial resolution.

\subsection{Source extraction at matched physical resolution}
\label{ss:extractsmoo}

\input{alma-imf-smoothing.tex}

\begin{figure*} 
    \centering
\includegraphics[trim=0cm 0cm 0cm 0cm, width=1.0\linewidth]{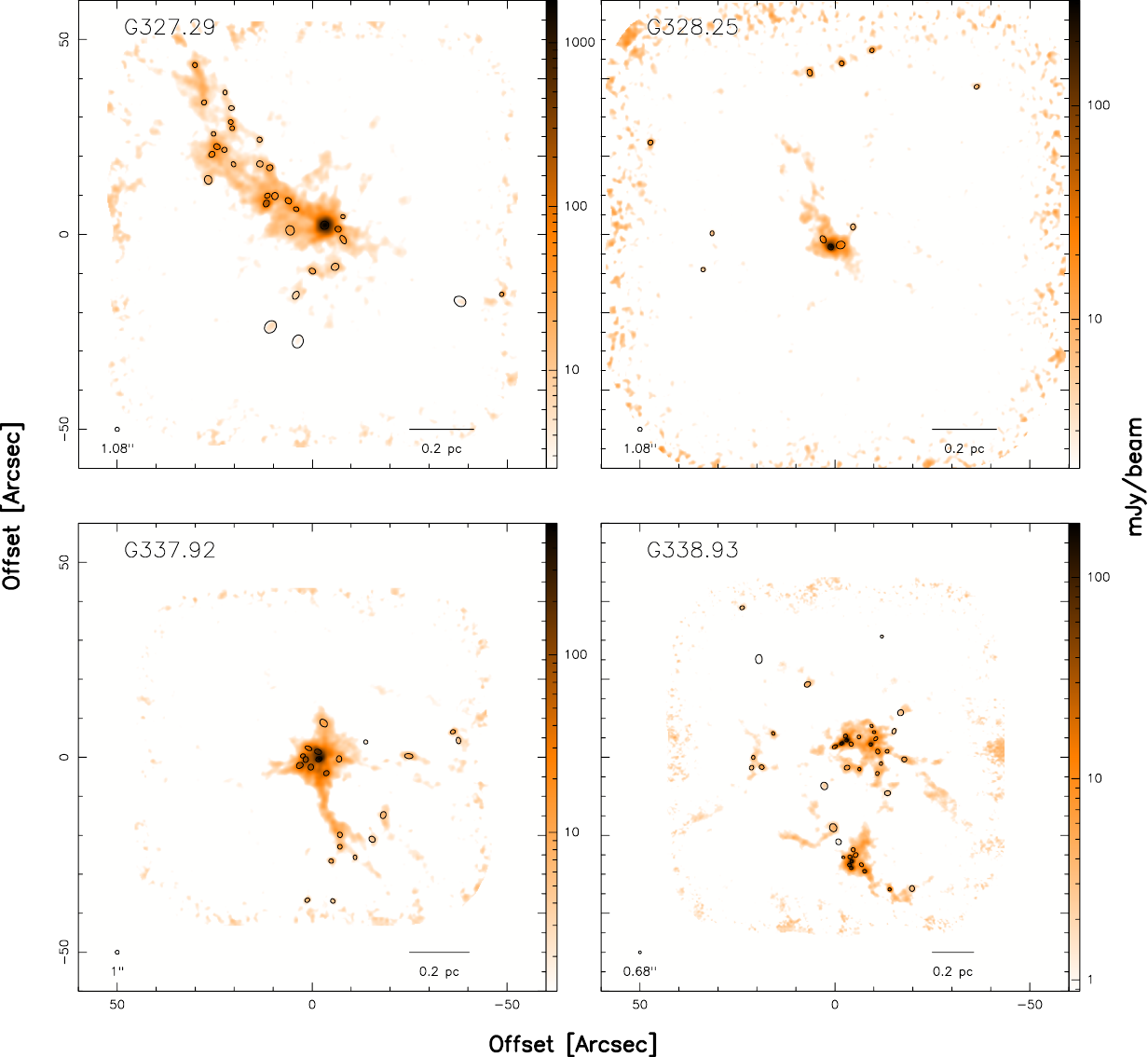}
\caption
{ 
The 15 ALMA-IMF protoclusters as traced by their continuum emission at 1.3\,mm at a matched spatial resolution of 2700\,au. The ellipses locate all the cores found by \textsl{getsf}. The name of the field is indicated in the top left corner of each panel and the beam size is shown in the bottom left corner. For each protocluster, the center coordinates are those specified in Table~\ref{t:alma-imf-fields}.
} 
\label{f:contmaps1}
\end{figure*}

To extract a spatially homogeneous core sample, we smoothed the 1.3 and 3\,mm maps to the lowest native resolution of $\simeq$2700\,au. We convolved each map with a Gaussian kernel whose width, $\theta_{\rm conv}$, is calculated as:
\begin{equation}
\theta_{\rm conv}=\sqrt{\left(\frac{2700\,{\rm au} }{D}\right)^2-\theta^2_{\rm native}}
\end{equation}
where $D$ is the distance to the target expressed in parsecs, and $\theta_{\rm native}$ is the native angular resolution of the map. Table~\ref{t:field-smoothing} lists the angular resolution of each field after smoothing. We present the smoothed continuum emission maps at 1.3\,mm in Fig.~\ref{f:contmaps1}. 

\addtocounter{figure}{-1}
\begin{figure*} 
    \centering
\includegraphics[trim=0cm 0cm 0cm 0cm, width=1.0\linewidth]{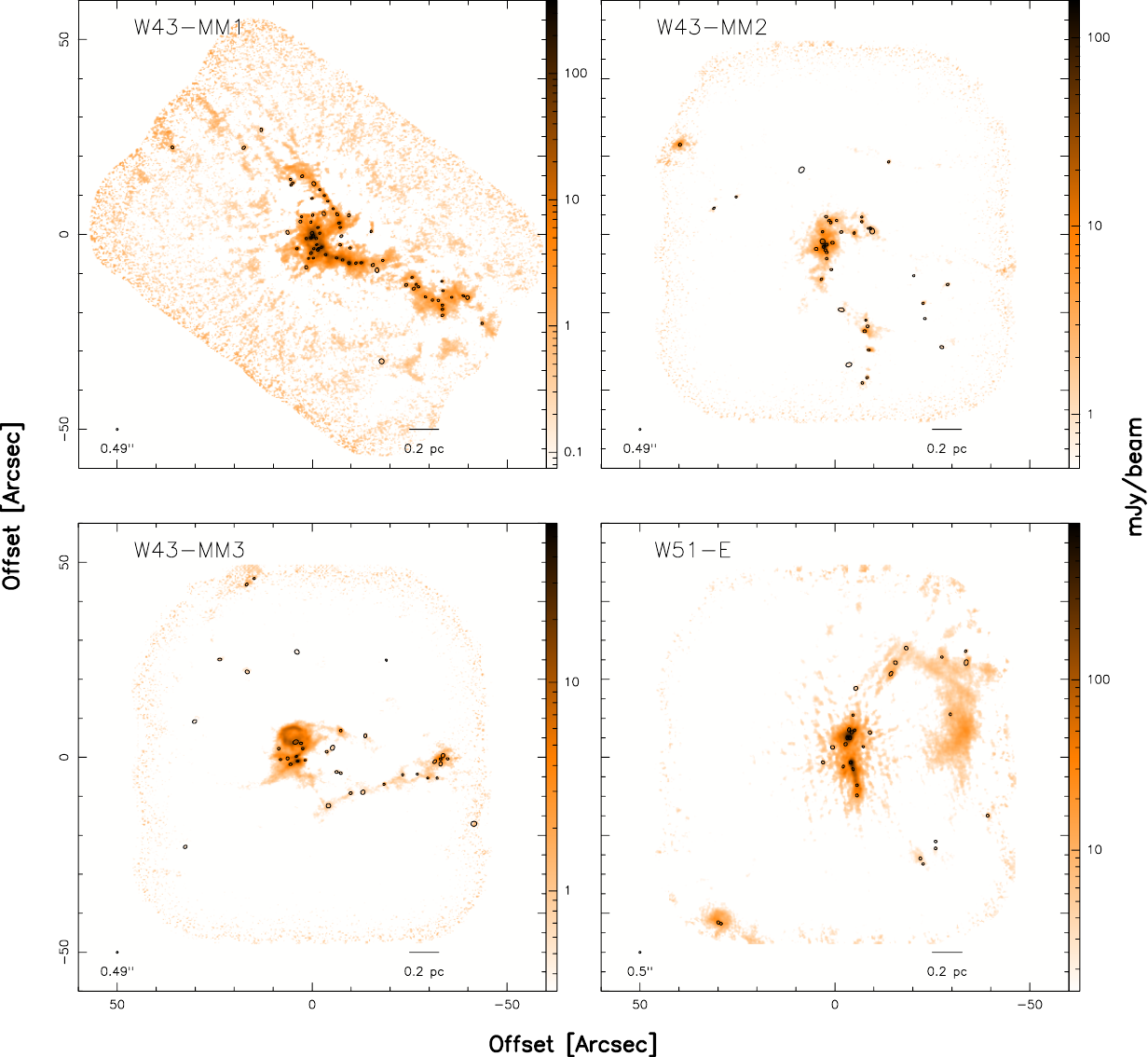}
\caption
{ 
Continuum emission maps (continued).
} 
\end{figure*}

\addtocounter{figure}{-1}
\begin{figure*} 
    \centering
\includegraphics[trim=0cm 0cm 0cm 0cm, width=1.0\linewidth]{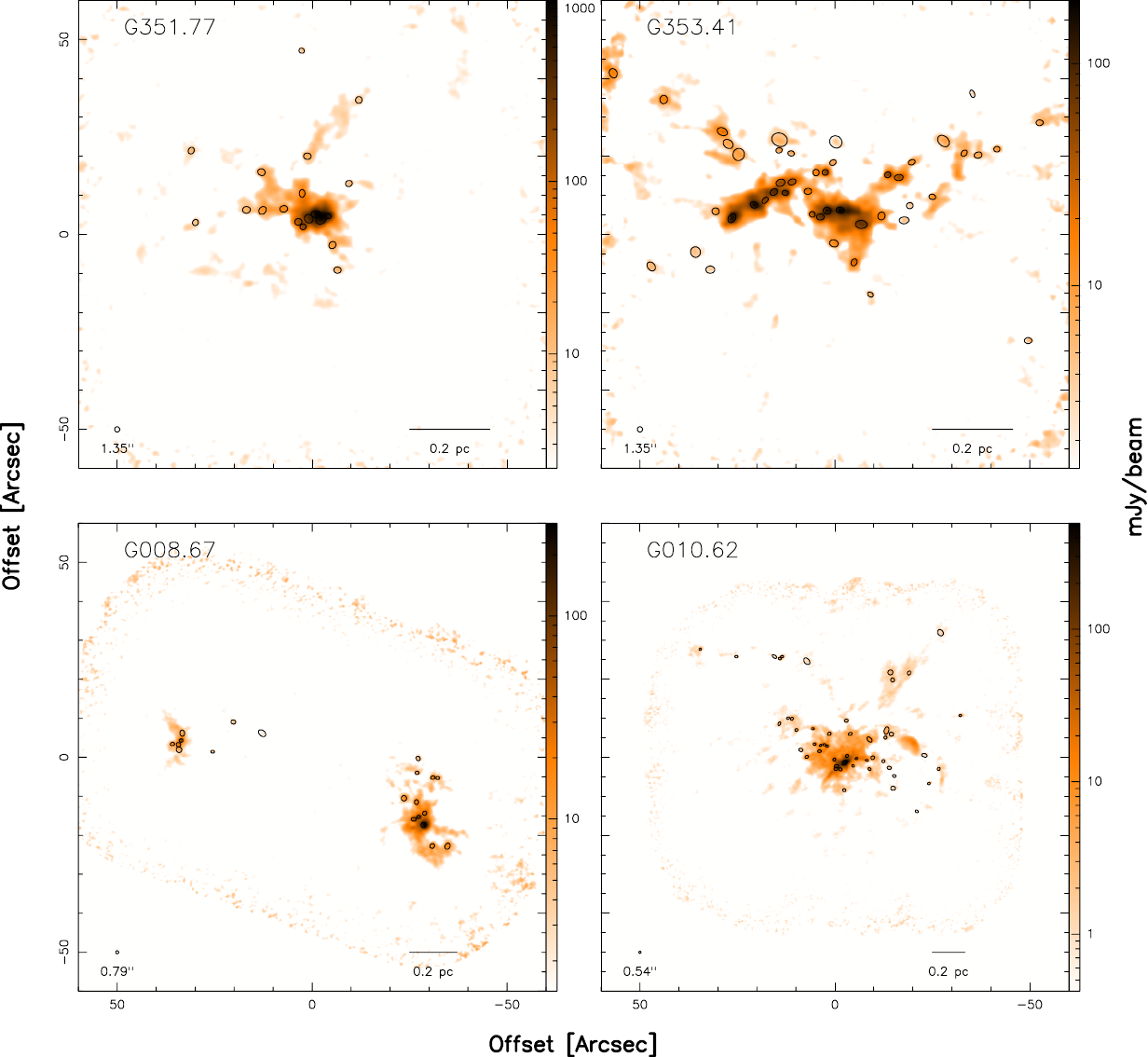}
\caption
{ 
Continuum emission maps (continued).
} 
\end{figure*}

\addtocounter{figure}{-1}
\begin{figure*} 
    \centering
\includegraphics[trim=0cm 0cm 0cm 0cm, width=1.0\linewidth]{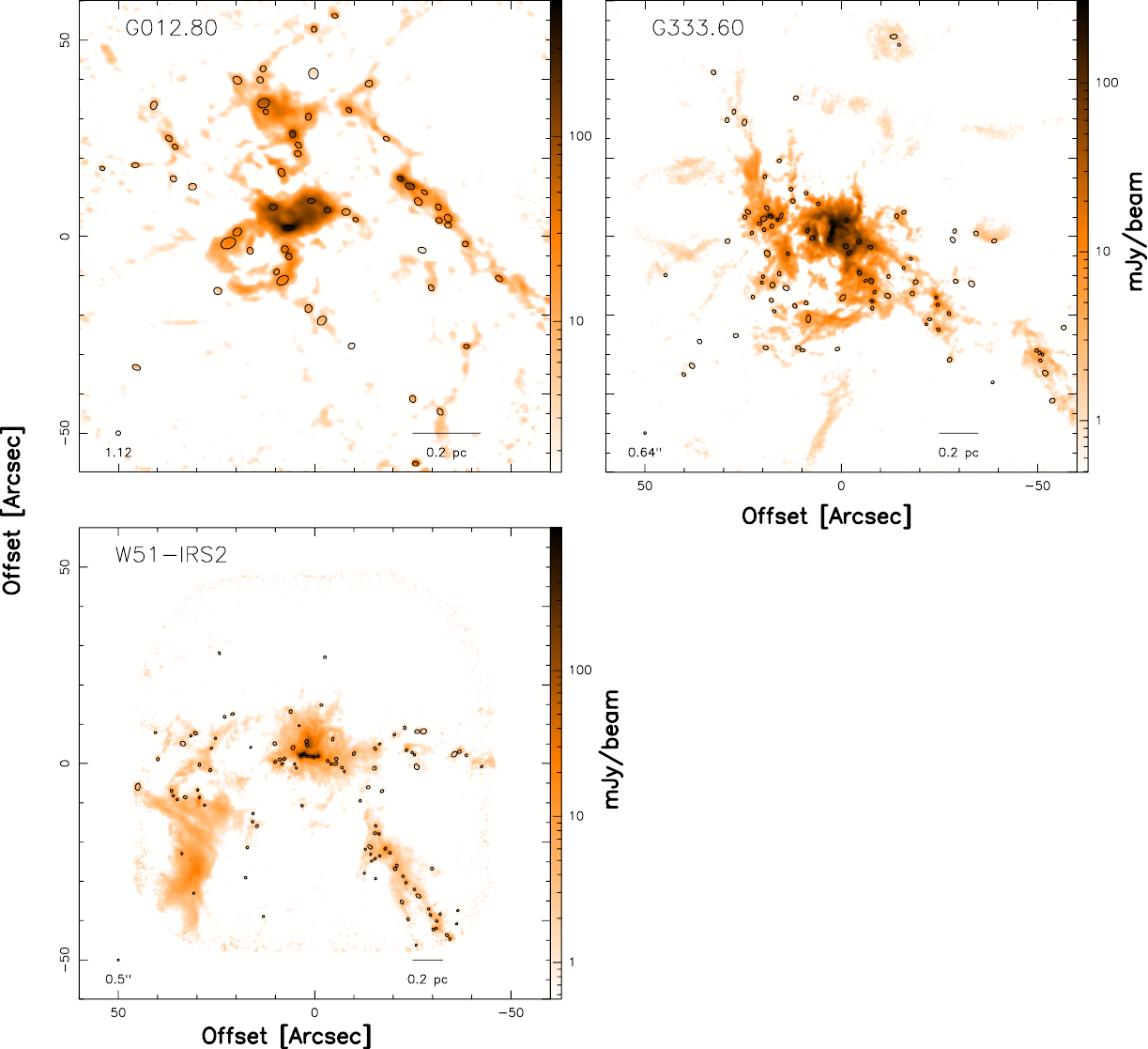}
\caption
{ 
Continuum emission maps (continued).
} 
\label{f:contmaps4}
\end{figure*}

We adopted the same strategy as explained in Sect.~\ref{ss:extractnative} to extract sources in the smoothed maps. In total, we retrieve about 680 sources with \textsl{getsf}, and 1020 with \textsl{GExt2D}\footnote{One can download the extraction catalogues from both \textsl{getsf} and \textsl{GExt2D}, at the native and smoothed angular resolutions on the website of the ALMA-IMF Large Program (\url{https://www.almaimf.com/}), and on \href{http://cdsportal.u-strasbg.fr/}{CDS} in a machine readable format.}. Table~\ref{t:field-smoothing} lists the number of sources extracted per field with both algorithms, Fig.~\ref{f:contmaps1} shows the sources extracted with \textsl{getsf} and Fig.~\ref{f:contmapsgag1} shows the sources extracted with both algorithms. The right panel of Fig.~\ref{f:coresize} shows the sizes of the sources we extracted with \textsl{getsf} for each field. The mean size is similar for each field with $L=3840\pm270$\,au.\footnote{A similar figure for \textsl{GExt2D} extractions is shown in Appendix~\ref{f:coresize-getext2d}.}


\section{Results}
\label{s:res}

\subsection{Getsf versus GExt2D extraction methods}
\label{ss:getsfvegext2d}

\begin{figure*} 
    \centering
\subfloat{\includegraphics[trim=0cm 0cm 0cm 0cm, width=0.5\linewidth]{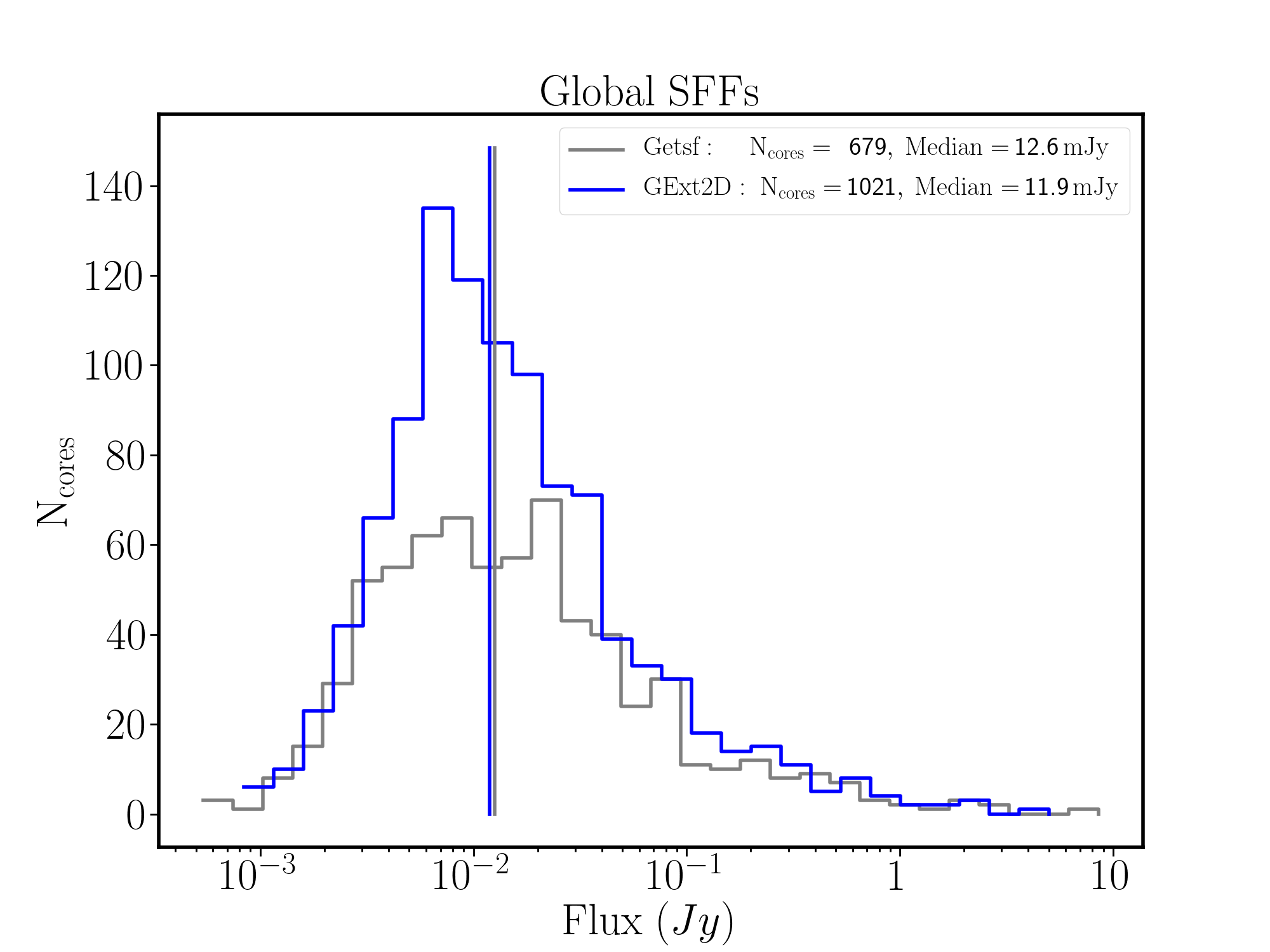}}
\subfloat{\includegraphics[trim=0cm 0cm 0cm 0cm, width=0.5\linewidth]{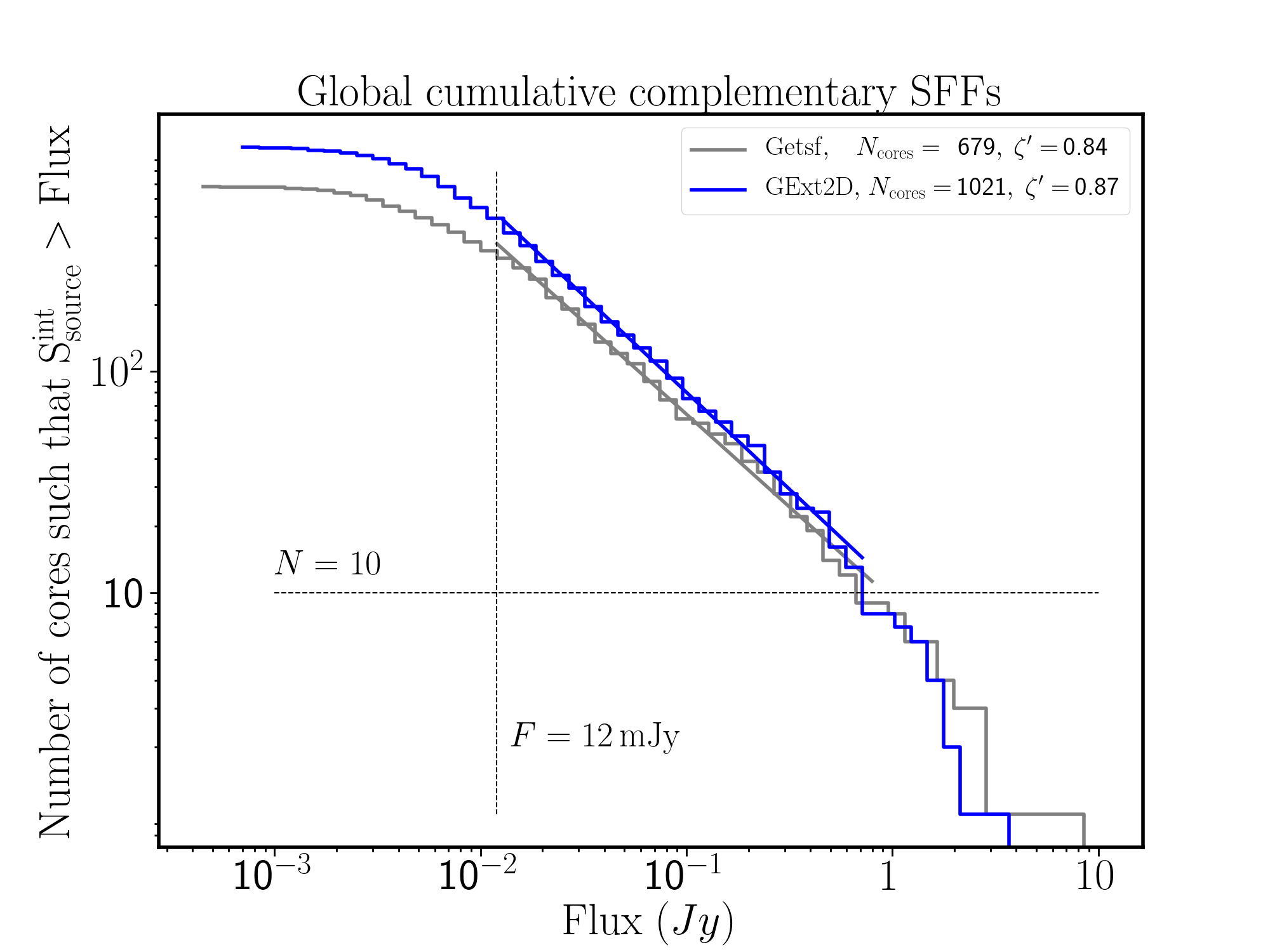}}
\caption
{ 
The left panel displays the source flux functions (SFFs) of the sources extracted by \textsl{getsf}, in grey, and by \textsl{GExt2D} in blue. The vertical blue and grey lines indicate the median values for each source catalogues, which differ by $\simeq$5\,\%. The right panel displays the complementary cumulative SFFs from the \textsl{getsf} source extraction (in grey) and from the \textsl{GExt2D} source extraction (in blue). The grey and blue lines display the best fits from linear regressions on the complementary cumulative SFFs from the median value at 12\,mJy up to the flux at which the number of cores is less than 10. The corresponding $\zeta$ power-law indexes are indicated in the top right corner.
} 
\label{f:cmfallnofree-free}
\end{figure*}

In total, we retrieve 677 sources with \textsl{getsf}, and 1020 with \textsl{GExt2D}. We see a good agreement between these extractions as $\sim$80\,\% of the sources found by \textsl{getsf} are also found by \textsl{GExt2D} (Table~\ref{t:field-smoothing}). To compare the source flux measurements, we constructed the source flux functions (SFFs) and complementary cumulative SFFs for both algorithms (Fig.~\ref{f:cmfallnofree-free}). The first function shows that the source fluxes range between 10\,mJy and 10\,Jy for both algorithms, with a similar median value at $\simeq$12\,mJy. It also shows that the additional sources found by \textsl{GExt2D} are low and intermediate flux sources from $\sim$30\,mJy to 200\,mJy. The complementary SFF permits to estimate the slope of the high-flux end of the SFF. We fit the cumulative SFF with a power law from the median value of the samples, that is 12\,mJy, up to the flux at which the number of cores is below ten\footnote{We illustrate the cumulative complementary SFFs with constant logarithm flux interval. This method introduces a constant weight for each bin, whatever the number of sources constituting the bin. Therefore it artificially creates a bias for relatively unpopulated bins. For this reason, we ignored the last 10 sources from the fit.}. The fitted high-flux slopes are similar for both catalogues, with $\zeta^{\prime}\simeq$\,0.85. These exponents, adjusted onto the complementary cumulative SFFs, correspond to power-laws with exponents $\zeta=\zeta^{\prime}+1\simeq 1.85$. This analysis proves that the overall statistics of source flux is independent of the two extraction algorithms we employed. In the following, we choose to focus on the extractions performed by \textsl{getsf} since it is more conservative than \textsl{GExt2D}, which is confirmed by the large fraction of \textsl{getsf} sources confirmed with \textsl{GExt2D}. Comparing the sources extracted at the native angular resolutions (see Sect.~\ref{ss:extractnative}) to the sources extracted in the smoothed maps (see Sect.~\ref{ss:extractsmoo}), there is better agreement between the \textsl{getsf} catalogues than between \textsl{GExt2D} catalogues. Indeed, $\simeq$95\% of the sources extracted in the smoothed maps were also extracted at the native angular resolutions with the \textsl{getsf} algorithm. This fraction drops to $\simeq$75\% in the case of the \textsl{GExt2D} extractions.

\subsection{From source flux to core mass}
\label{ss:fluxmass}

\subsubsection{Removing contaminated sources}

\begin{figure*} 
    \centering
\includegraphics[trim=0cm 0cm 0cm 0cm, width=1.0\linewidth]{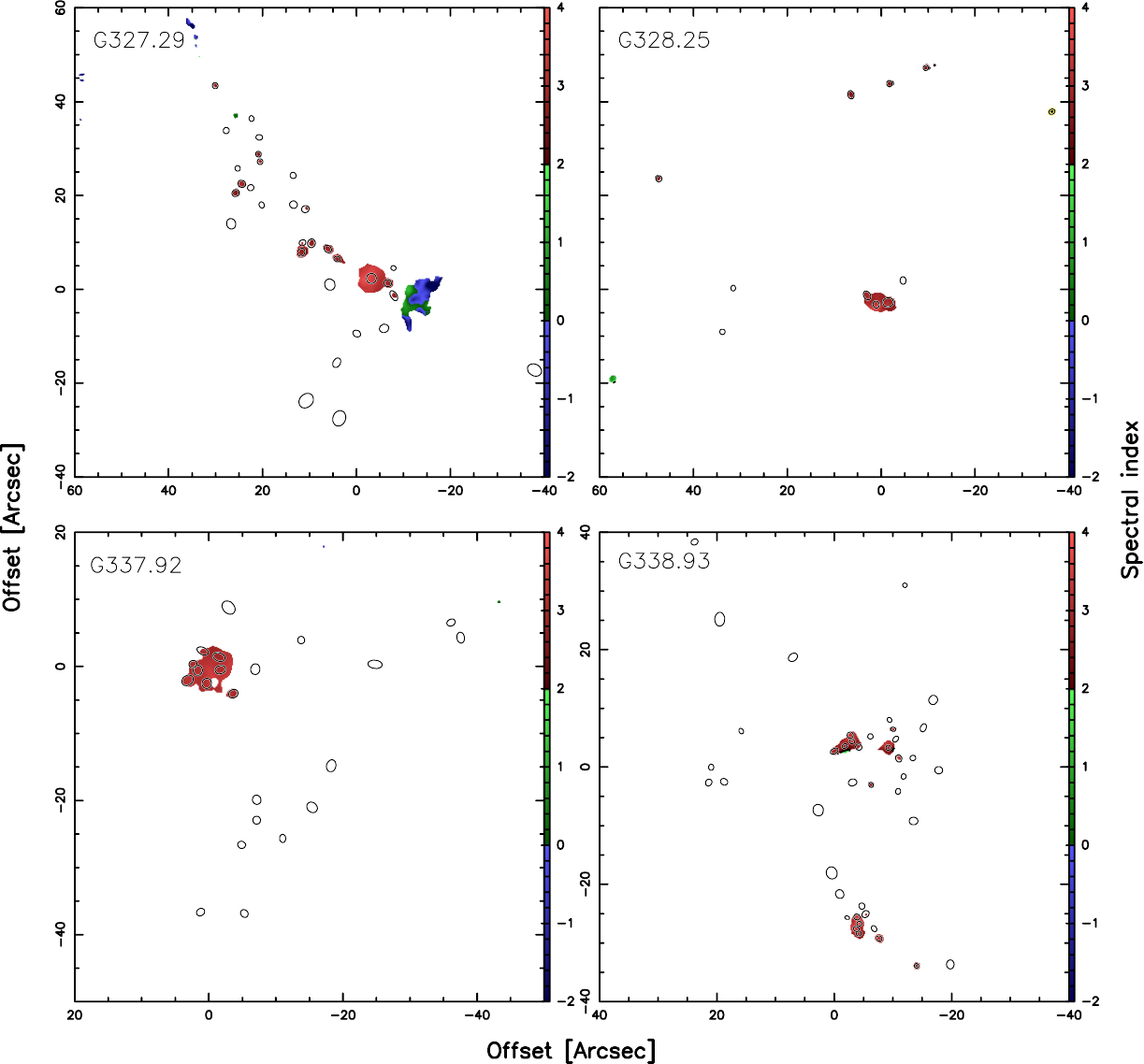}
\caption
{ 
The maps display the spectral index in the fifteen ALMA-IMF fields. We chose the colours such that red traces dust thermal emission, blue traces free-free emission and green likely traces a mixture of dust and free-free emissions. The maps display the spectral index value only where the 3\,mm intensity exceeds 3\,$\sigma$. The black ellipses display the sources extracted by \textsl{getsf}.
} 
\label{f:free-free1}
\end{figure*}

The integrated fluxes comprise thermal dust emission and potentially free-free emission for the more evolved sources. In the following, we filter the sources that are arguably contaminated by free-free emission. To do so, we used the integrated flux measurements at 1.3 and 3\,mm, when available, to compute their spectral index. To correct from the source size differences between 1.3 and 3\,mm extractions, we adjusted the integrated fluxes at 3\,mm linearly: $S^{\rm int*}_{\rm 3\,mm}=S^{\rm int}_{\rm 3\,mm}\times\left(\frac{\Theta_A^{\rm 1.3\,mm}\times \Theta_B^{\rm 1.3\,mm}}{\Theta_A^{\rm 3\,mm}\times \Theta_B^{\rm 3\,mm}}\right)$, where $\Theta_A$ and $\Theta_B$ are the major and minor axes of the sources, respectively, at 1.3 or 3\,mm. This operation corrects the 3\,mm fluxes by 17\,\% in average. This flux rescaling replicates the method we used in Paper~\Romannum{3}. This rescaling works both for an optically thick emission and for an optically thin emission of an isothermal protostellar envelope with a density profile $\rho(r)\propto r^{-2}$, where $r$ is the radius of the envelope. We then compute the spectral index as:
\begin{equation}
\label{e:specind}
\gamma=\frac{\log\left(S^{\rm int}_{\rm 1.3\, mm}/S^{\rm int*}_{\rm 3\, mm}\right)}{
\log\left(\nu_{\rm 1.3\, mm}/\nu_{\rm 3\, mm}\right)}
\end{equation}
where $\nu_{\rm 1.3\,mm}$ and $\nu_{\rm 3\,mm}$ are the central frequencies of the ALMA Band 6 and Band 3 respectively (see Table~\ref{t:alma-imf-fields}). We remove 68 sources with $\gamma<2$, which are presumably contaminated by free-free emission. These sources are represented by the pink ellipses in Fig.~\ref{f:free-free1}.

To investigate the $\simeq$500 sources that lack measurements at 3\,mm, we built the spectral index map for each field, using Eq.~\ref{e:specind} on a pixel-by-pixel basis by replacing $S^{\rm int}_{\rm 1.3\,mm}$ and $S^{\rm int}_{\rm 3\,mm}$ with the pixel intensities. We point out that the angular resolutions and pixel sizes of the images at 1.3\,mm equal that of the images at 3\,mm. We show the spectral index maps in Fig.~\ref{f:free-free1}. The red areas ($\gamma>2$) represent pixels with thermal dust emission, those in green ($0<\gamma<2$) pixels contaminated by free-free emission, and those in blue ($\gamma<0$) pixels dominated by free-free emission. We manually selected all the sources lying over free-free emission areas ($\gamma<2$, see yellow ellipses in Fig.~\ref{f:free-free1}). For the sources undetected at 3\,mm we estimate their flux, $S_{\rm 3\,mm}^{\rm upper}$, using the integrated flux in the aperture corresponding to the footprint of the source detected at 1.3\,mm. We emphasise that although \textsl{getsf} did not detect these 1.3\,mm sources as compact sources at 3\,mm they do have a background emission above 3$\sigma$ at 3\,mm, by definition of the spectral index maps. We use this upper limit to compute an upper limit on their spectral index as: 
\begin{equation}
\gamma=\frac{\log\left(S^{\rm int}_{\rm 1.3\, mm}/S^{\rm upper}_{\rm 3\, mm}\right)}{
\log\left(\nu_{\rm 1.3\, mm}/\nu_{\rm 3\, mm}\right)}
\end{equation}
\noindent and we rejected the 16 sources with spectral index below 1, those for which free-free emissions could substantially contribute to the emission.

These filtering rejected 84 of the 677 sources extracted by \textsl{getsf}. The degree of rejection depends on the evolutionary stages of the regions: $\simeq$\,0.5\% in the young regions, $\simeq$\,10\% in the intermediate regions, and $\simeq$\,25\% in the evolved regions. Table~\ref{t:core-extraction-getsf} lists, for each field, the number of sources rejected when applying these two levels of filtering.

\input{Stat-extractions-getsf.tex}

\addtocounter{figure}{-1}
\begin{figure*} 
    \centering
\includegraphics[trim=0cm 0cm 0cm 0cm, width=1.0\linewidth]{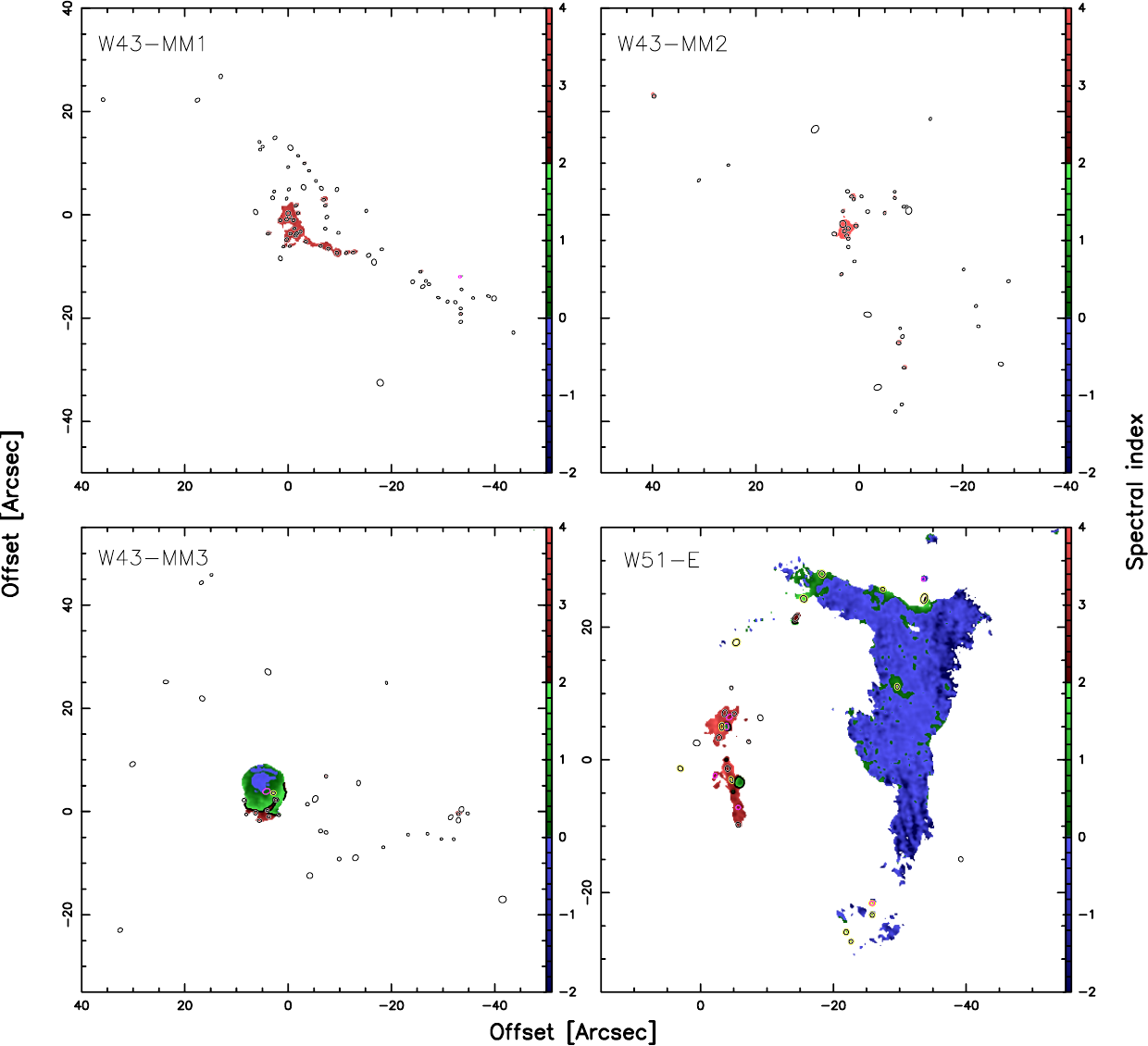}
\caption
{ 
Spectral index maps (continued). The ellipses in pink highlight sources automatically flagged due to their spectral index; those in yellow display the sources whose spectral index were further checked manually (see Eq.~\ref{e:specind}).
} 
\end{figure*}

\addtocounter{figure}{-1}
\begin{figure*} 
    \centering
\includegraphics[trim=0cm 0cm 0cm 0cm, width=1.0\linewidth]{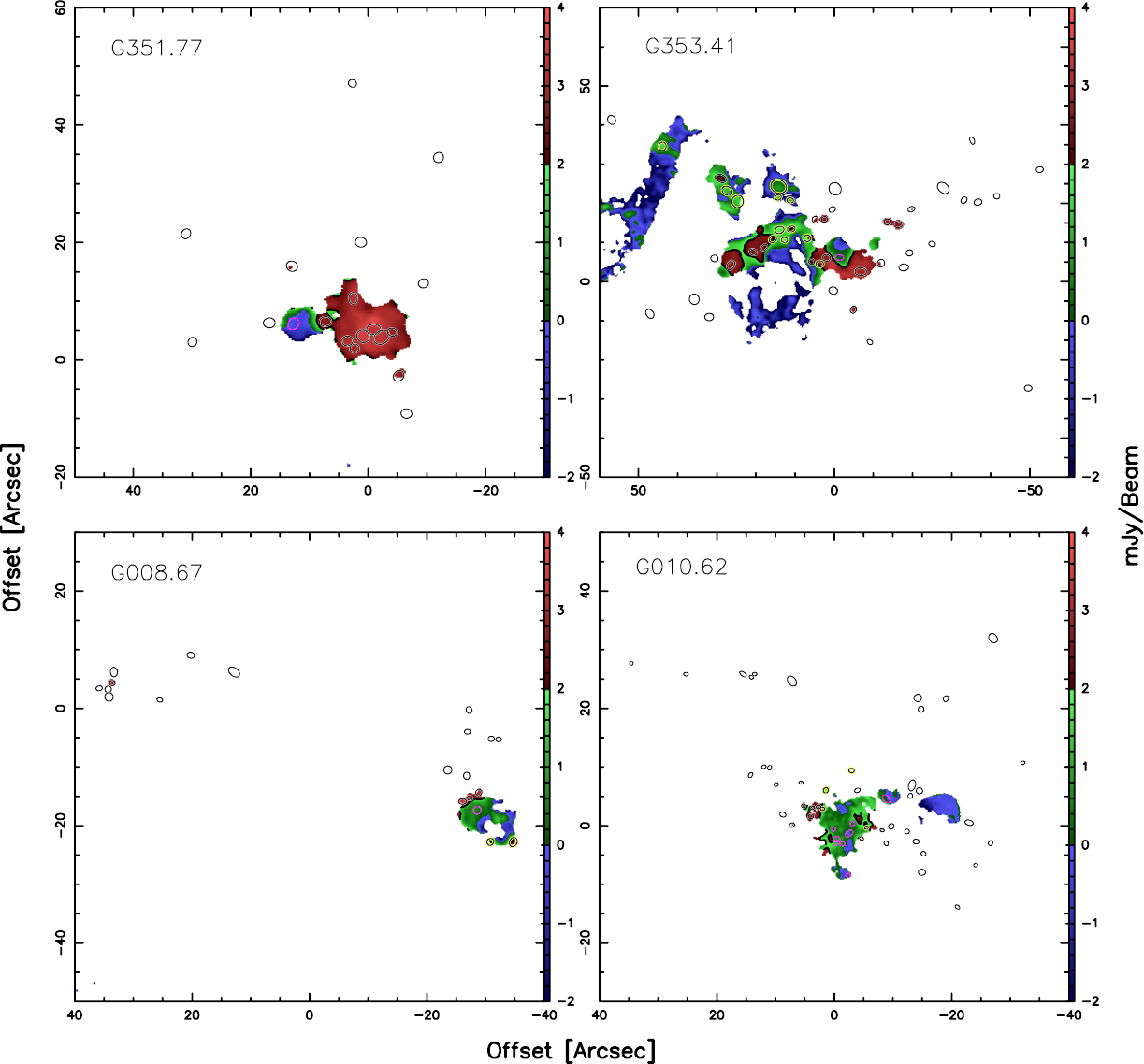}
\caption
{ 
Spectral index maps (continued).} 
\end{figure*}

\addtocounter{figure}{-1}
\begin{figure*} 
    \centering
\includegraphics[trim=0cm 0cm 0cm 0cm, width=1.0\linewidth]{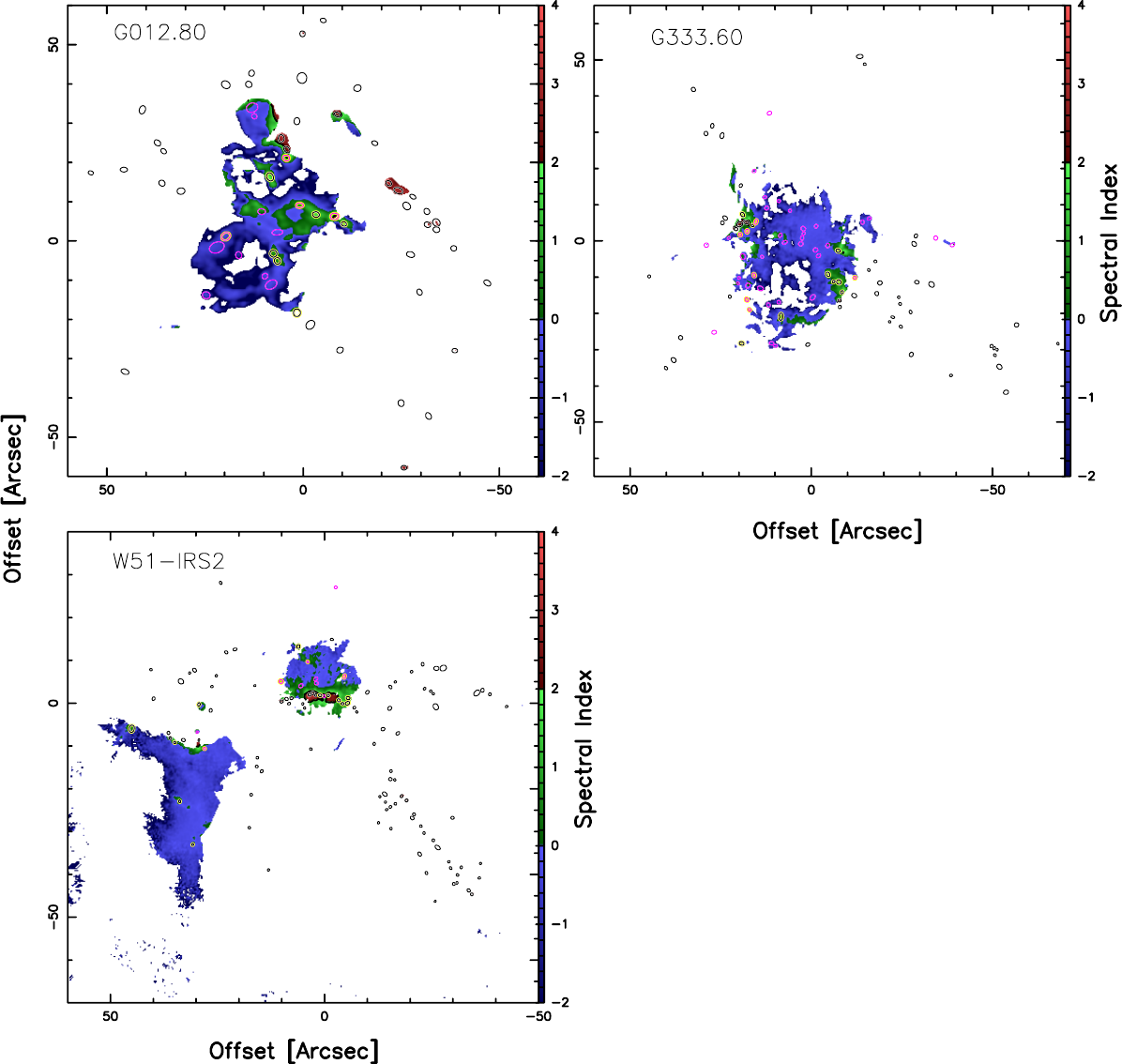}
\caption
{ 
Spectral index maps (continued).} 
\end{figure*}

\subsubsection{Core temperatures}
\label{sss:temperature}

Following the pilot study by \cite{mottenat}, we used the Bayesian procedure \textsl{PPMAP} \citep{marsh15,dellova24} to build the temperature maps of the ALMA-IMF protoclusters. The procedure uses several continuum emission maps to compute a cube of column densities as a function of dust temperature. Our \textsl{PPMAP} algorithm takes, along with the ALMA data in Band 6 decontaminated from free-free emission (Galv\'{a}n-Madrid et al. subm.), the following maps as input:
\begin{itemize}
\item SOFIA/HAWC+ data at 214 $\mu$m with an angular resolution of 19.0$^{\prime \prime}$ (only for G012.80, G351.77, W51-E and W51-IRS2, \citealt{vaillancourt16,pillai23}).
\item Herschel/PACS and Herschel/SPIRE data at 70, 160, 250, 350, 500 $\mu$m with angular resolutions of 5.6, 10.7, 17.6, 23.9 and 35.2$^{\prime \prime}$, respectively \citep{molinari10,motte10}.
\item APEX/SABOCA data at 350 $\mu$m with an angular resolution of 7.8$^{\prime \prime}$ \citep{lin19}.
\item APEX/LABOCA data at 870 $\mu$m with an angular resolution of 19.2$^{\prime \prime}$ \citep{schuller09}.
\end{itemize}
We refer to \cite{dellova24} for an in-depth description of the method. Finally, we obtain a unique temperature for each core at an angular resolution of 2.5$^{''}$. The derived temperatures vary from 19.4 to 62.8\,K (see Tables~\ref{t:core-extraction-smoothed-G008} to~\ref{t:core-extraction-smoothed-W51-IRS2}). The error associated with \textsl{PPMAP} derivation is $\approx$5\,K; to account for potential systematic contributions, we adopt a 25\% error on the \textsl{PPMAP} temperature estimates\footnote{We adopt an uncertainty of 5\,K for sources below 20\,K.}.

We adopted the PPMAP temperatures for all cores except for the hot core candidates for which we adopted the method proposed in \cite{bonfand24}: we cross-correlated the position of continuum sources (see Sect.~\ref{ss:extractsmoo}) with methyl formate (CH$_3$OCHO) emission maps that were observed by ALMA-IMF. Methyl formate forms at the surface of dust grains in lukewarm (30-40\,K) regions, and is then released in the gas phase when the temperature reaches $\sim$100\,K \citep[e.g.][]{garrod09}. Methyl formate can be used as a proxy to trace regions where heating is present and dust surface products have started to sublimate. Following the method discussed in \cite{bonfand24}, we set 100$\pm$50\,K to the 49 sources whose position corresponds to the peaks of extended methyl formate emission; these sources are classified as hot core candidates.

Additionally, we set 300$\pm$100\,K to sources n$^\circ$1, n$^\circ$2, and n$^\circ$1 in  W51-IRS2, W51-E and G327.29 respectively, following their detailed modelling by \cite{ginsburg17}, \cite{goddi20}, and the adopted temperature in \cite{bonfand24}, respectively. These three sources are associated with strong and extended emission structures in the methyl formate emission maps.\footnote{\cite{bonfand24} report 6 sources with temperatures estimated at 300$\pm$100\,K. The 3 additional sources with respect to ours are either classified as freefree contaminated (source W51-E-MF1) or not matching our continuum extractions (sources W51-E-MF2 and W51-IRS2-MF3).}

\subsubsection{Core mass and boundedness}
\label{ss:masses}

We converted the measured integrated flux at 1.3\,mm, $S^{\rm int}_{\rm 1.3\,mm}$, into a mass using the formula presented in Paper~\Romannum{3} that includes a first order correction for the optical thickness:
\begin{equation}
\label{e:mass}
M_{\rm core}=-\Omega^{\rm 1.3\, mm}_{\rm beam}\;\frac{D^2}{\kappa_{\rm 1.3\, mm}}\; \frac{S^{\rm int}_{\rm 1.3\, mm}}{S^{\rm peak}_{\rm 1.3\, mm}} \ln\left(1- \frac{S^{\rm peak}_{\rm 1.3\, mm}}{\Omega^{\rm 1.3\, mm}_{\rm beam}\,B(T,\nu)} \right)
\end{equation}

\noindent where $D$ is the distance to the target, $\kappa_{\rm 1.3\,mm}=0.1\,(\nu/1000\,{\rm GHz})^{\beta}\,{\rm cm^2\,g^{-1}}$ is the dust opacity per unit mass (dust + gas), with an opacity index $\beta=1.5$ typical of cold and dense environment \citep{ossenkopf-henning94}, and $B(T,\,\nu)$ is the Planck function computed at the representative frequency of the observations in Band 6 with the dust temperature $T$ (see Table~\ref{t:alma-imf-fields} for central frequencies and distances, and Tables~\ref{t:core-extraction-smoothed-G008} to~\ref{t:core-extraction-smoothed-W51-IRS2} for source temperature). 

Errors on the masses mostly arise from the uncertainties on the opacity index $\beta$ and on the dust temperature, $T$. Masses would be a factor of two smaller using $\beta=2$ instead of $\beta=1.5$. As for the temperature, a 25\,\% difference leads to a mass shift by $\lesssim$40\,\%. 

To address the boundedness of cores, we computed the ratio $M_{\rm BE}/M_{\rm core}$ where $M_{\rm core}$ is the core mass and $M_{\rm BE}$ is the mass of the critical Bonnor-Ebert sphere \citep{bonnor56} whose size matches that of the core: $M_{\rm BE}=2.4\times R \times \sigma_{\rm th}^2/G$. Here, $R$ is the equivalent core radius, estimated as $R=(a\times b^2)^{\frac{1}{3}}\times D$\footnote{We stress that choosing 3D oblate core shape ($V\propto a^2\times b$) has no effect on the boundedness of the cores.}, where $a$ and $b$ are the major and minor axis of the source ellipse, respectively. Moreover, $\sigma_{\rm th}$ is the thermal broadening of lines, as $\sigma_{\rm th}=\sqrt{\gamma\,k_{\rm B}\,T/(\mu\times m_{\rm p})} $ where $\gamma=1$ is the adiabatic index (isothermal), $k_B$ is the Boltzmann constant, $\mu=2.4$ is the mean molecular weight per free  particle, and $m_{\rm p}$ is the mass of a proton. We consider that the sources are gravitationally bound if $M_{\rm BE}/M_{\rm source}<2$ \citep[see e.g.][]{fkl21}. As listed in Table~\ref{t:core-extraction-getsf}, this filter excludes no sources. We note that this is a first-order check of the boundedness of cores. A more accurate determination would require computing the equilibrium of each source taking into account its turbulence, external pressure, and magnetic field support in addition to the thermal support. Unfortunately, we lack all these pieces of information at the moment. 

In Figs.~\ref{f:contmaps1} we show the thermal-dust cores for each field. The Tables~\ref{t:core-extraction-smoothed-G327} to~\ref{t:core-extraction-smoothed-w51irs2-suite2} list all the sources detected by the \textsl{getsf} extraction algorithm in each field. The first group of sources in each table corresponds to thermal dust cores that are gravitationally bound, and the second group corresponds to sources whose fluxes are arguably contaminated by free-free emission. For each source, the last column indicates whether it was also detected by \textsl{GExt2D}, which is true for 80\,\% of the extracted sources.


\section{Core mass function}
\label{s:ana}

\subsection{Completeness tests}
\label{ss:comptests}

In order to draw a coherent sample gathering cores from the 15 ALMA-IMF protoclusters, we need to ascertain that the core extractions are complete at the selected lower mass limit. To do so, we performed completeness tests in each protocluster: we injected synthetic cores on top of their background emission (i.e. the emission subtracted from each source). The synthetic cores are injected in the form of 2D circular Gaussian with a FWHM corresponding to 3840\,au — the mean core size (see Sect.~\ref{ss:extractsmoo}). We inject these synthetic sources randomly, provided the centre of the cores are separated by more than 2.5$^{\prime\prime}$. We forbid the injection of sources within 5$^{\prime\prime}$ of the border of the maps, where the noise increases due to the primary beam correction (see Fig.~\ref{f:contmaps1}). The flux of the synthetic sources ranges from equivalent masses of 0.5 to 5\,\msun~in all fields for gas at 20\,K, except in  W51-E where we adopted a mass range of 1 to 10\,\msun. The fluxes are equally split into ten bins of flux. From these fluxes, we compute the mass of the synthetic cores when adopting the mean temperature of the cores extracted in the corresponding region. We then extract the synthetic cores with the \textsl{getsf} method (see Sect.~\ref{ss:extraction}). To obtain a good statistics in each bin of mass, we repeated this procedure four times per field. Cumulating the four draws we obtain $\simeq$3800 cores in total in each field (or $\simeq$380 per bin) and we probe $\simeq$80\,\% of the background area. We plot the percentage of synthetic cores extracted as a function of their mass in Fig.~\ref{af:comptest}. Following Paper~\Romannum{3}, we consider reaching the completeness limit when 90\,\% of the synthetic cores are extracted. The mass completeness varies from 0.80 to 1.64\,\msun~with a mean value of 1.34\,\msun~and a standard deviation of 0.2\,\msun. These numbers exclude W51-E, for which we find completeness as high as 3.9\,\msun. We exclude W51-E from the analysis of the global CMF (see Sect.~\ref{ss:globalcmf}) and we restrict the core samples to cores exceeding 1.64\,\msun~to fit the high-mass tail of the CMF.

\subsection{Fitting core mass functions}
\label{ss:cmf-all}

\subsubsection{Method}
\label{sss:fitmethod}

The representation of a distribution in the form of histograms, as in Fig.~\ref{f:cmfallnofree-free}, further fitted through linear regression may give inaccurate results \citep{clauset09}. In addition, the representation in the form of a CMF in log-log scales prohibits estimating the fit uncertainty since the noise is not Gaussian. Moreover, the choice of the bin width induces a free parameter that thwarts any uncertainty estimate. The representation in the form of a cumulative CMF, however, prevents the estimation of the uncertainty as the data are not independent. To circumvent these issues, we choose to fit the high-mass end of the CMF using maximum likelihood estimates (MLE). We follow the procedure presented by \cite{clauset09}, and use its implementation in the python package \texttt{powerlaw} presented by \cite{alstott14}. If we assume that the CMF can be represented by a power-law $p(x)=Cx^{-\alpha}$, where $C$ is a constant, there must be some lower mass value, $x_{\rm min}$, from which the power-law fit is accurate, and prevents the divergence of $p$ when $x\rightarrow 0$. After normalisation (imposing that $\int_{x_{\rm min}}^{+\infty} p(x) \, dx =1$) it comes:
\begin{equation}
\label{e:power}
p(x)=\frac{\alpha-1}{x_{\rm min}}\,\left(\frac{x}{x_{\rm min}}\right)^{-\alpha}
\end{equation}
 
When $x_{\rm min}$ is known, the MLE gives an estimate of the exponent of the power-law as:
\begin{equation}
\alpha=1+n\left[\sum_{i=1}^n \ln\frac{x_i}{x_{\rm min}} \right]^{-1}
\end{equation}
where $x_i$ is the mass of the core $i$ (provided $x_i\geq x_{\rm min} $), and the uncertainty on $\alpha$ reads:
\begin{equation}
\sigma = \frac{\alpha-1}{\sqrt{n}}+O\left(\frac{1}{n}\right)
\label{e:sigma}
\end{equation}
where $O$ is the mathematical notation for: not negligible in front of.

\begin{figure} 
    \centering
\subfloat{\includegraphics[trim=0cm 0cm 0cm 0cm, width=1.0\linewidth]{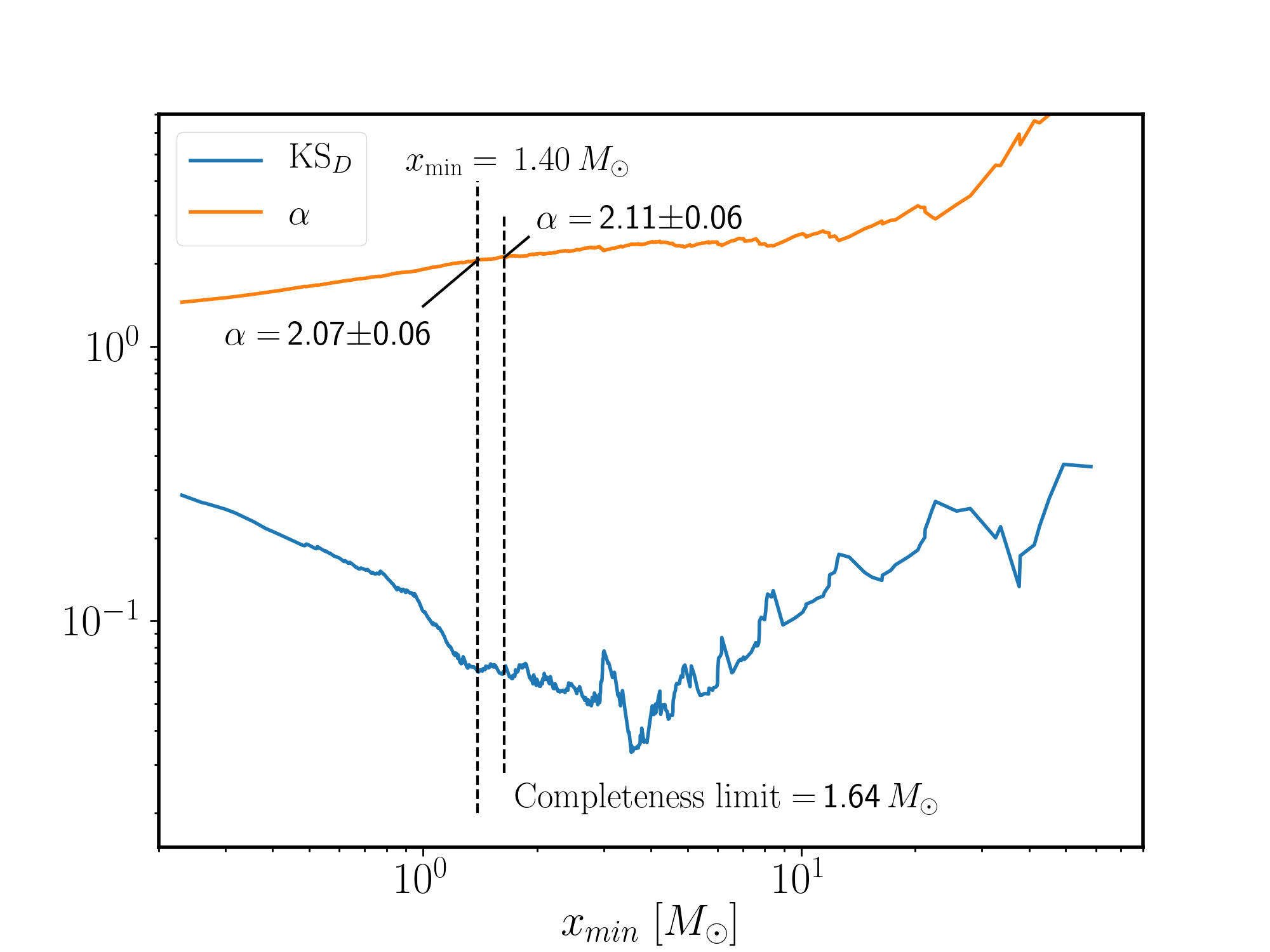}}
\caption
{ 
Convergence towards the best fit of a power-law through MLE and KS$_D$ method for the global \textsl{getsf} bound core catalogue without the free-free sources. The left vertical dashed line highlights the first minimal KS$_D$ (in blue), occurring at $x_{\rm min}= 1.40$\,\msun. The right vertical dashed line highlights the completeness limits at $x_{\rm min}= 1.64$\,\msun. The orange curve shows the exponent of the power law fit of the high-mass tail of the CMF as a function of $x_{\rm min}$. 
} 
\label{f:convmle}
\end{figure}

The originality of the method by \cite{clauset09} is to propose a method to determine the best parameters $x_{\rm min}$ and $\alpha$ using the Kolmogorov-Smirnov (KS) distance. First, the method computes $\alpha$ for $x_{\rm min}$ taking successively each mass value of the core sample. Then it computes the maximal distance — the KS distance, $KS_{\rm D}$ — between the observational CMF for all elements whose mass exceeds $x_{\rm min}$ and the synthetic CMF following a probability distribution function as defined by Eq.~\ref{e:power}. The best value for the parameters $x_{\rm min}$ and $\alpha$ is that which minimizes the $KS_{\rm D}$. Figure~\ref{f:convmle} shows the evolution of $\alpha$ and $KS_{\rm D}$ for all $x_{\rm min}$ values. In our case, we need to select the highest value between the $x_{\rm min}$ that minimizes the $KS_{\rm D}$ and our completeness limit (see Sect.~\ref{ss:comptests}). The first minimum in the $KS_D$ is met for $x_{\rm min}$\,=1.40\,\msun, below our completeness level at 1.64\,\msun. We, therefore, select $x_{\rm min}=1.64$\,\msun, which corresponds to a core sample of 330 elements (excluding W51-E). From that sub-sample, we obtain a power-law index fit $\alpha=2.11\pm0.06$\footnote{We note that $x_{\rm min}=1.40$\,\msun~corresponds to $\alpha=2.07\pm0.06$, compatible with the $\alpha$ associated with $x_{\rm min}=1.64$\,\msun.}. We put forward that 330 cores above the completeness limit convey a statistical conclusive sample. \cite{clauset09} report stable index values when the sample exceeds $\approx$50 elements above the completeness limits. Consistently, \cite{fkl21} report stable index values when the sample exceeds $\approx$40 elements above the completeness limits.

\subsubsection{Uncertainty of the fit}
\label{sss:fituncertainty}

\begin{figure} 
    \centering
\subfloat{\includegraphics[trim=0cm 0cm 0cm 0cm, width=1.0\linewidth]{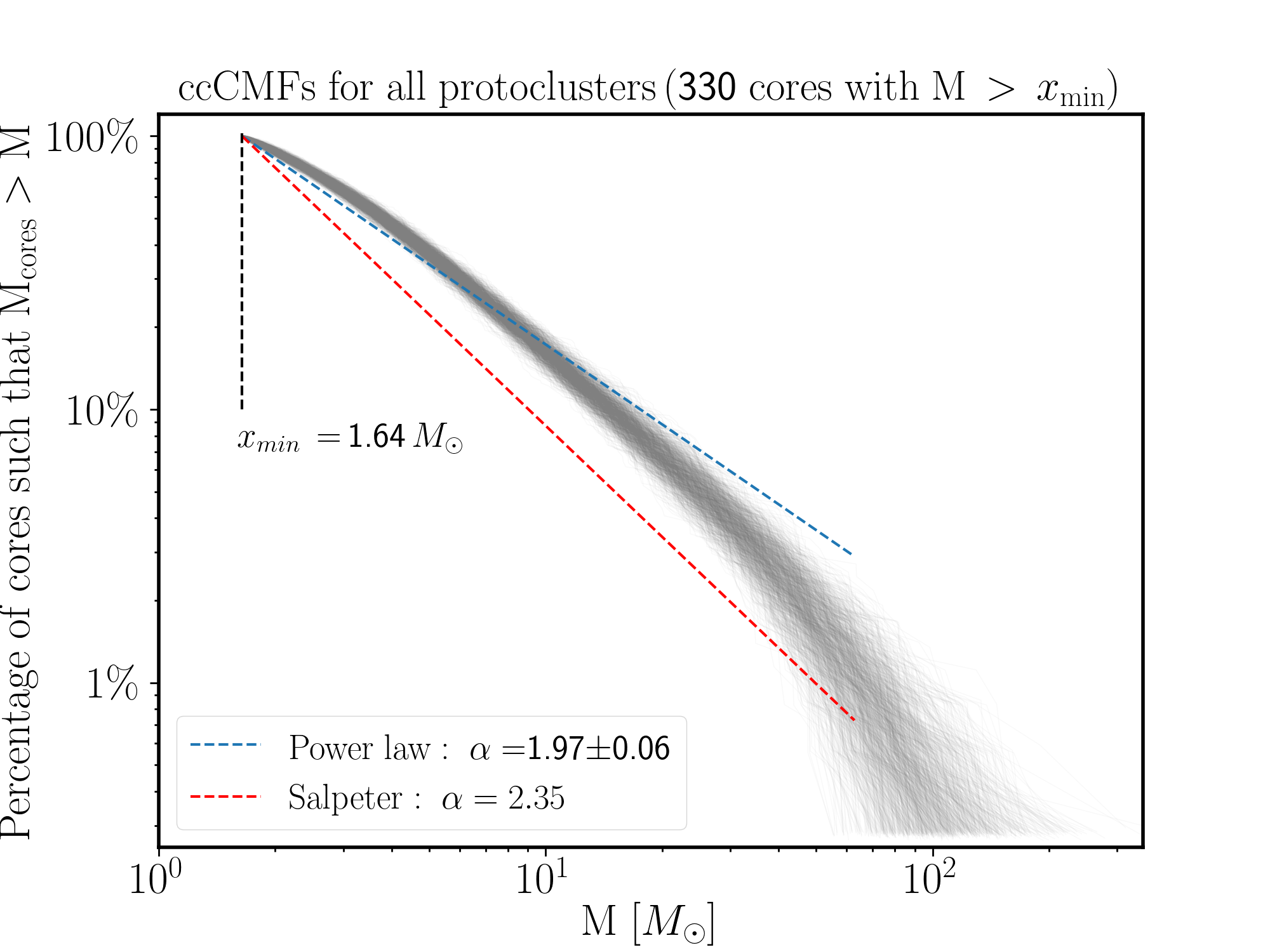}}
\caption
{ 
The grey curves show the complementary cumulative CMFs (ccCMFs) of the $10^3$ core samples obtained by including flux measurement uncertainties, varying core temperatures, and applying opacity index variations on a per-regions basis (see Sect.~\ref{sss:fituncertainty}), overlaid with the average fit by a power-law (in blue) and the best fit with a power-law of index 2.35.
} 
\label{f:mle}
\end{figure}

As pointed out in Sect.~\ref{ss:masses}, errors in the mass estimates arise primarily from the uncertainties on the opacity index and core temperature. To study how these uncertainties affect the slope of the CMF, we performed a Monte Carlo simulation on our 330 cores by allowing their opacity index, their temperature as well as their source flux to vary simultaneously. About the latter, we used the Gaussian error associated with the flux measurement. \cite{koen09} showed that including such errors on the fluxes flattens the distribution. However, since our relative uncertainty diminishes for brighter sources, we expect this effect to be marginal on the high-mass end of our CMF. In our Monte Carlo simulation, the opacity index $\beta$ can take random normal values with a mean value of 1.5 \citep{ossenkopf-henning94} and a standard deviation of 0.2 for each core. In parallel, we allow the core temperature to take random normal values with the mean and standard deviation as described in Sect.~\ref{sss:temperature}. We show the resulting cumulative CMFs generated from $10^3$ trials in Fig.~\ref{f:mle}. Each draw is fitted via MLE (see Sect.~\ref{sss:fitmethod}), fixing $x_{\rm min}$ to 1.64\,\msun. We obtain a mean exponent value for the slope of the CMF of 1.97$\pm$0.06, where the uncertainty is the quadratic sum of the statistical uncertainty from Eq.~\ref{e:sigma} ($\sigma\simeq 0.06$) and from the uncertainties on the temperature, opacity index, and flux measurement uncertainties ($\sigma\simeq 0.02$). We show in Appendix~\ref{a:incertitudes} the results of the $10^3$ trials for a power-law fit.

\subsubsection{Comparison with the Salpeter-slope}
\label{sss:salpeter}

\begin{figure} 
    \centering
\subfloat{\includegraphics[trim=0cm 0cm 0cm 0cm, width=1.0\linewidth]{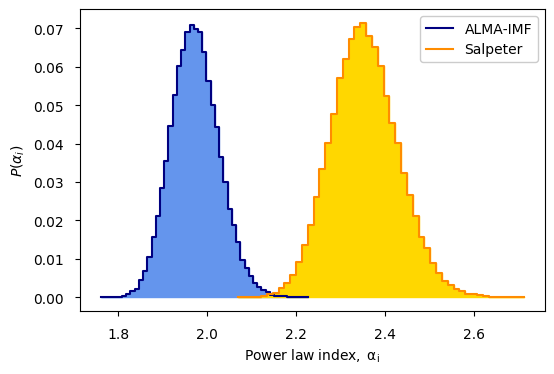}}
\caption
{ 
The figure illustrates the power-law exponents $\alpha$ retrieved when sorting 330 cores from a synthetic core sample whose CMF has a power-law exponent of 1.97 (in blue) and 2.35 (in orange).
} 
\label{f:bootstrapall}
\end{figure}

The ALMA-IMF large program aims to test if the high-mass slope of the CMF differs from the high-mass slope of the canonical IMF ($\alpha=2.35$ when fitted by a power-law, \citealt{salpeter55}). To see if our sample of 330 cores (with $M>x_{\rm min}=$\,1.64\,\msun) permits such a claim, we selected 330 sources from a perfect mass distribution with an exponent of 2.35 and fitted the slope of its high-mass tail through the MLE. We repeated the operation $10^5$ times. Figure~\ref{f:bootstrapall} shows the probability of retrieving an exponent $\alpha$ when the parental distribution has a slope $\alpha$=2.35. With 330 cores, the probability of retrieving a slope compatible with ALMA-IMF, $\alpha=1.97\pm0.06$, is lower than 1\% ($\sigma\approx2.4$). Therefore, we report that the CMF in the protoclusters studied by ALMA-IMF is flatter than, and cannot reasonably be reconciled with the Salpeter-IMF slope to a 2.4$\sigma$ level\footnote{$\sigma$ refers to a confidence interval of a Gaussian distribution.}.

\section{Discussion}
\label{s:disc}

\subsection{A global top-heavy core mass function}
\label{ss:globalcmf}

In Sects.~\ref{sss:fitmethod} and~\ref{sss:fituncertainty}, we presented the best fit to the high-mass tail of the CMF by a power law. The best-fit yields $\alpha=1.97\pm 0.06$. This result, based on 330 cores, confirms the many investigations led in recent years that reported top-heavy CMFs in high-mass protoclusters \citep[e.g.][]{mottenat,cheng18,sanhueza19,kong19,moser20}. We show that this slope cannot be reconciled, at the 2.4\,$\sigma$ level, with the canonical IMF slope ($\alpha=2.35$, see Sect.~\ref{sss:salpeter}) even after the inclusion of all known uncertainties. 

Observing top-heavy CMFs in high-mass protoclusters contrasts with nearby low-mass star-forming regions where a high-mass tail compatible with Salpeter is consistently found (see, e.g., Table~6 of \citealt{fiorellino21} for all CMF exponents from the Herschel Gould Belt Survey).

\subsection{From the CMF to the IMF}
\label{ss:CMF-IMF}

To link the CMF to the IMF, one must assume that cores constitute the mass reservoir for the accretion of protostars. One also assumes a fragmentation cascade and a mass transfer efficiency $\epsilon$ from cores to protostars. Regarding the latter, the common assumption is that one core will give birth to one star — and therefore assumes no sub-fragmentation — and that $\epsilon$ remains constant whatever the mass of the cores. If we adopt these assumptions, since our massive protoclusters display a top-heavy CMF they shall engender a top-heavy IMF with a high-mass slope equals to that of the CMF: $\alpha\simeq 1.97$. Meanwhile, \cite{bontemps10}, \cite{louvet14} and \cite{csengeri17} showed that $\epsilon$ actually depends almost linearly on the local density. As a matter of facts, the massive cores — $>$25\,\msun — spread equally over the observed size range and are therefore, on average, denser than low- and intermediate-mass cores (see red crosses in Figure~\ref{f:coresize}). This suggests that massive cores will convert gas into stars more efficiently than low- and intermediate-mass cores, giving birth to a flatter IMF than the parental CMF. In the past ten years some examples of top-heavy IMFs have been reported in young massive clusters in and out of our Galaxy \citep[see for example][]{lu13,schneider18}. As a consequence, we speculate that the ALMA-IMF protoclusters might the precursors of star-clusters whose IMF is top-heavy. From another perspective, we now observe more regularly star-forming regions with at top-heavy CMF than star clusters with a top-heavy IMF. To reconcile these two observables, we hypothesise that some of the star-forming regions hosting a top-heavy CMF will evolve to produce a Salpeter-like IMF.

\subsection{Limitations and perspectives}
\label{ss:biases}

We recognize two limitations in our study:
\begin{itemize}
\item[$\bullet$] Free-free emission: in Sect.~\ref{ss:fluxmass}, we rejected all sources that are arguably contaminated by free-free emission from our core sample. We ended up rejecting 84 sources (see Table~\ref{t:core-extraction-getsf}). Galv\'{a}n-Madrid et al. subm. gives estimates of the free-free contribution, based on the H$_{\rm  41\alpha}$ recombination lines collected by ALMA-IMF. This characterization will allow us to correct and re-introduce these sources in the core sample. These sources, with free-free emission, arise from massive young stellar objects and would steepen our CMF even further. Therefore, our conclusions will stay unchanged by the inclusion of the free-free contaminated cores.  
\item[$\bullet$] Sub-fragmentation: we probed the cores at a spatial resolution of 2700\,au. Recent observations in the high-mass protostellar core G335 showed that a binary system, and perhaps even a triple system, could take place below 1000\,au \citep{olguin21,olguin22}. Similar results were obtain by \cite{izquierdo18} in W33A. However, other works reported no fragmentation in G336.01-0.82 \citep{olguin23} or HH80-81 \citep{girart18}. At present date, it is thus uncertain how will be the multiplicity in the ALMA-IMF fields. A straightforward path would consist in conducting observations at a yet higher angular resolution to reach a spatial resolution of $\simeq$200\,au, the typical scale of Solar-type proto-planetary disks \citep[e.g.][]{louvet18}. The main difficulty consists in obtaining high sensitivities with an optically thin tracer at the scale of protoclusters.
\end{itemize}

\section{Conclusions}
\label{s:concl}

We present the core catalogues of the 15 high-mass protoclusters observed by the ALMA-IMF large programme. At a homogeneous sensitivity and spatial resolution (2700\,au), we collected about 680 sources. Rejecting sources arguably contaminated by free-free emission, we analysed a core sample of $\simeq$600 elements. We performed completeness tests and found the matched-physical resolution core catalogues to be complete down to 1.64\,\msun, with the exception of the W51-E protocluster whose cores were discarded from the CMF analysis. In total, we analysed 330 cores with masses above the completeness limit of 1.64\,\msun~and with little to no free-free contamination.

We fit the high-mass slope of the core mass function (CMF) with the maximum likelihood estimate technique and found a best-fit power law probability distribution function (PDF) $\frac{{\rm d}N}{{\rm d}M}\propto M^{-\alpha}$ with $\alpha=1.97\pm 0.06$. Such an exponent is flatter than, and cannot be reconciled with, the Salpeter-IMF slope of $\alpha\simeq 2.35$. We confirm that the CMF in a representative sample of high-mass Galactic protoclusters is shallower than the Salpeter slope. We suggest that these massive protoclusters will give birth to top-heavy stellar clusters, or, in order to reconcile with the Universal IMF, the CMF needs to evolve and become Salpeter-like at latter stages of the cluster formation.

Together with this article, we provide the core catalogues for the 15 protoclusters at both the native (1300 to 2700\,au) and smoothed (2700\,au) linear resolutions. These catalogues include, among other, the sources position, size (in arcsec and in au), peak and integrated fluxes at both 1.3 and 3\,mm, their temperature estimates from PPMAP and their estimated mass.

\begin{acknowledgements}
This paper makes use of the following ALMA data: ADS/JAO.ALMA\#2017.1.01355.L. ALMA is a partnership of ESO (representing its member states), NSF (USA) and NINS (Japan), together with NRC (Canada), MOST and ASIAA (Taiwan), and KASI (Republic of Korea), in cooperation with the Republic of Chile. The Joint ALMA Observatory is operated by ESO, AUI/NRAO and NAOJ. 

This paper also use the following ALMA data: ADS/JAO. ALMA\#2013.1.01365.S and ADS/JAO.ALMA\#2015.1.01273.S. 

FL acknowledges support by the Marie Curie Action of the European Union (project \textit{MagiKStar}, Grant agreement number 841276). 

 FM acknowledges the support of the French Agence Nationale de la Recherche (ANR) under reference ANR-20-CE31-009, of the Programme National de Physique Stellaire and Physique et Chimie du Milieu Interstellaire (PNPS and PCMI) of CNRS/INSU (with INC/INP/IN2P3).

AG acknowledges support from the NSF under grants AST 2008101 and CAREER 2142300. 

PS was supported by a Grant-in-Aid for Scientific Research (KAKENHI Number JP22H01271 and JP23H01221) of the Japan Society for the Promotion of Science (JSPS). P.S. and H.-L.L. gratefully acknowledge the support from the NAOJ Visiting Fellow Program to visit the National Astronomical Observatory of Japan in 2019, February. 

SB acknowledges support by the French Agence Nationale de la Recherche (ANR) through the project \textit{GENESIS} (ANR-16-CE92-0035-01). 

T. Cs. has received financial support from the French State in the framework of the IdEx Universit\'e de Bordeaux Investments for the future Program. 

RG-M and TN acknowledge support from UNAM-PAPIIT project IN108822 and from CONACyT Ciencia de Frontera project ID: 86372. 

AS gratefully acknowledges support by the Fondecyt Regular (project code 1220610), and ANID BASAL project FB210003. 

LB gratefully acknowledges support by the ANID BASAL projects ACE210002 and FB210003.

NC acknowledges funding from the ERC under the European Union's Horizon 2020 research and innovation programme (ECOGAL, grant agreement no. 855130). 

GB acknowledges funding from the State Agency for Research (AEI) of the Spanish MCIU through the AYA2017-84390-C2-2-R grant.

GB and ALS acknowledges funding from the European Research Council (ERC) under the European Union's Horizon 2020 research and innovation programme, for the Project "The Dawn of Organic Chemistry" (DOC), grant agreement No 741002.

TB acknowledges the support from S. N. Bose National Centre for Basic Sciences under the Department of Science and Technology, Govt. of India. 

MB is a postdoctoral fellow in the University of Virginia's VICO collaboration and is funded by grants from the NASA Astrophysics Theory Program (grant number 80NSSC18K0558) and the NSF Astronomy \& Astrophysics program (grant number 2206516).

The project leading to this publication has received support from 
ORP, that is funded by the European Union's Horizon 2020 research and innovation programme under grant agreement No 101004719.

\end{acknowledgements}

\bibliographystyle{aa} 
\bibliography{cmf.bib}


\clearpage

\begin{appendix}

\section{Core completeness tests in ALMA-IMF observations}
\label{a:comptests}
We display the completeness tests for each of the 15 ALMA-IMF fields. We define the completeness as the mass limit above which more than 90\,\% of the synthetic cores get extracted.

\begin{figure*}
	\hspace*{-1cm}
	\centering
\subfloat{\includegraphics[trim=0cm 0cm 0cm 0cm, width=0.55\linewidth]{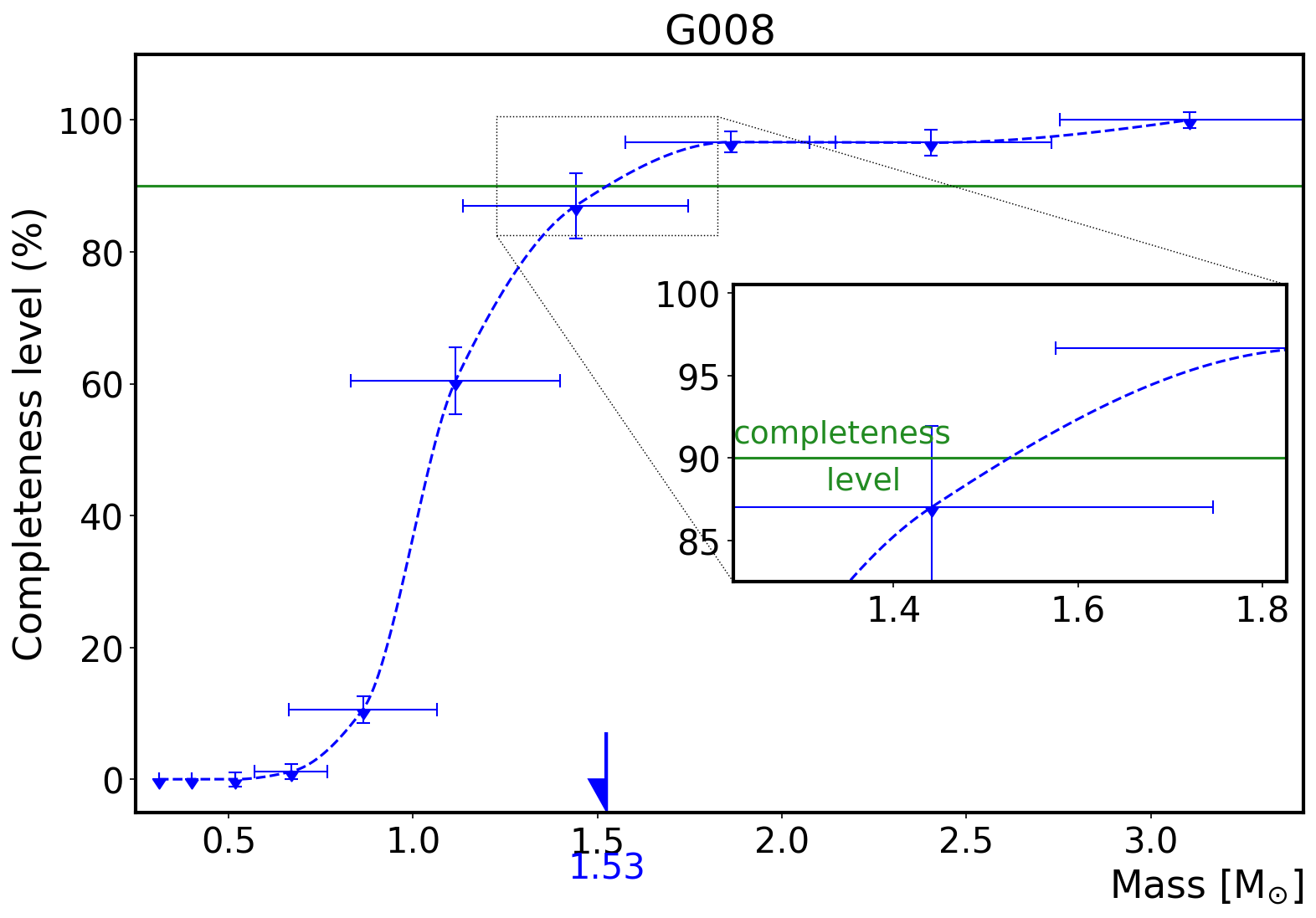}}
\subfloat{\includegraphics[trim=0cm 0cm 0cm 0cm, width=0.55\linewidth]{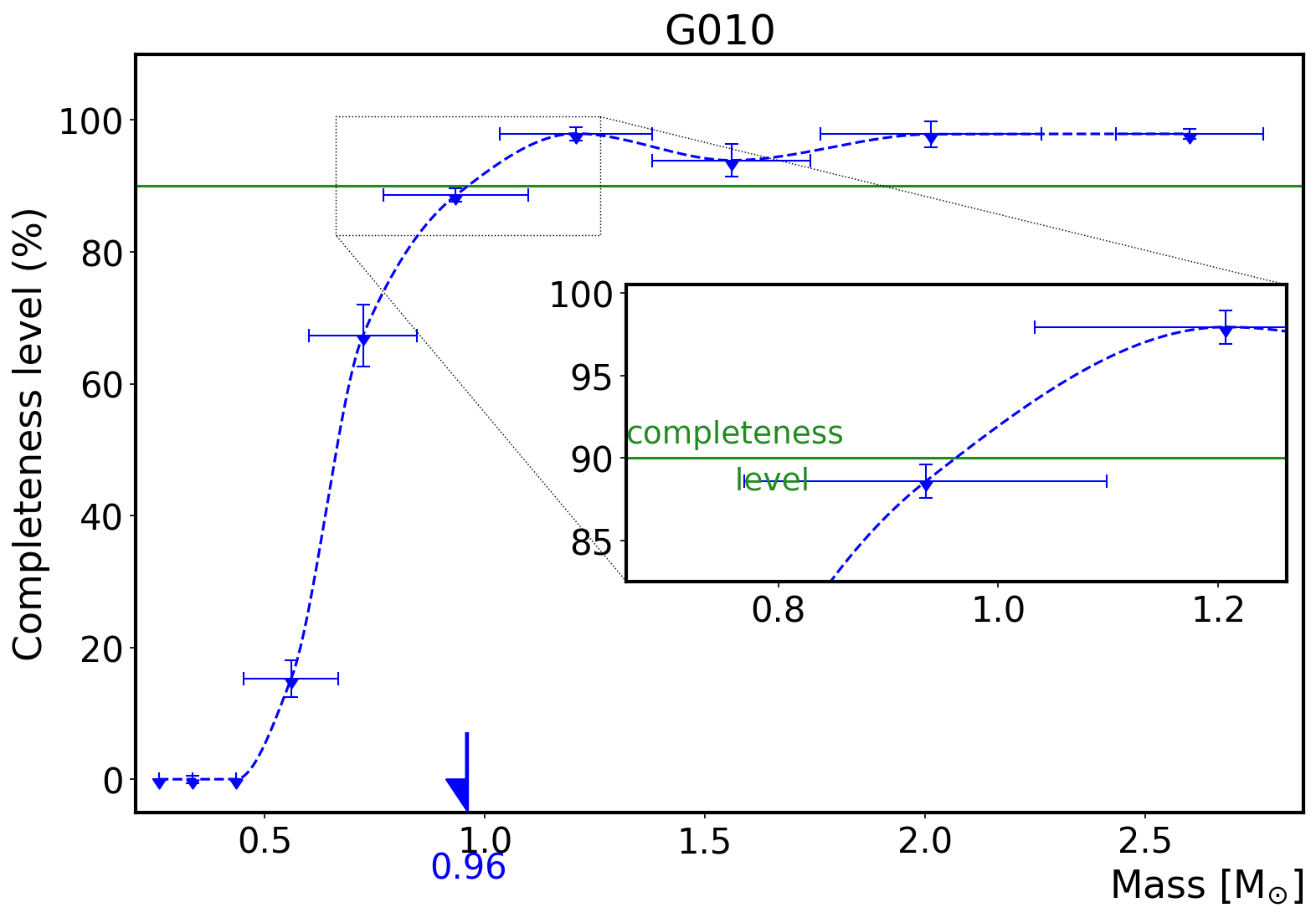}}\\
	\hspace*{-1cm}
\subfloat{\includegraphics[trim=0cm 0cm 0cm 0cm, width=0.55\linewidth]{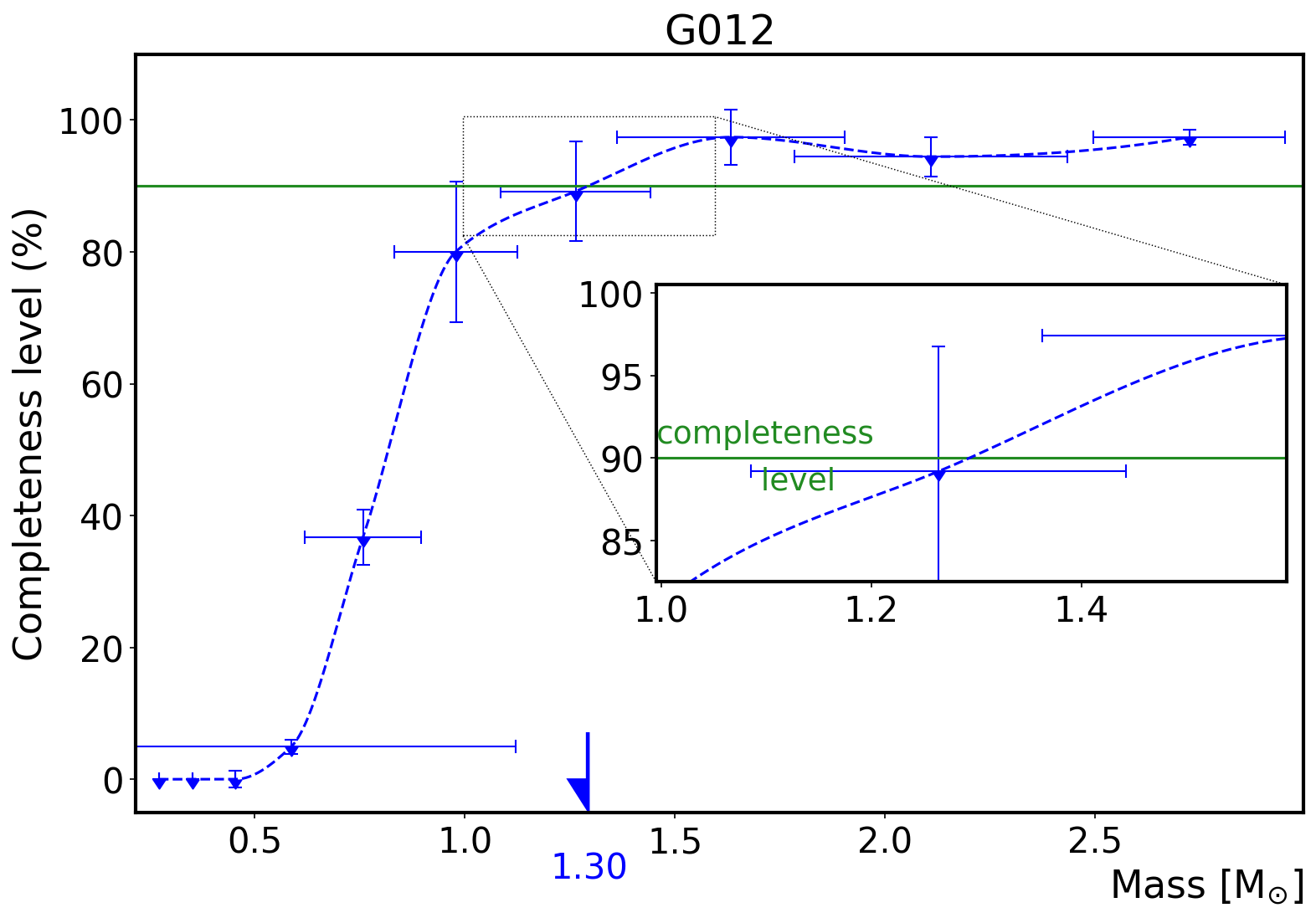}}
\subfloat{\includegraphics[trim=0cm 0cm 0cm 0cm, width=0.55\linewidth]{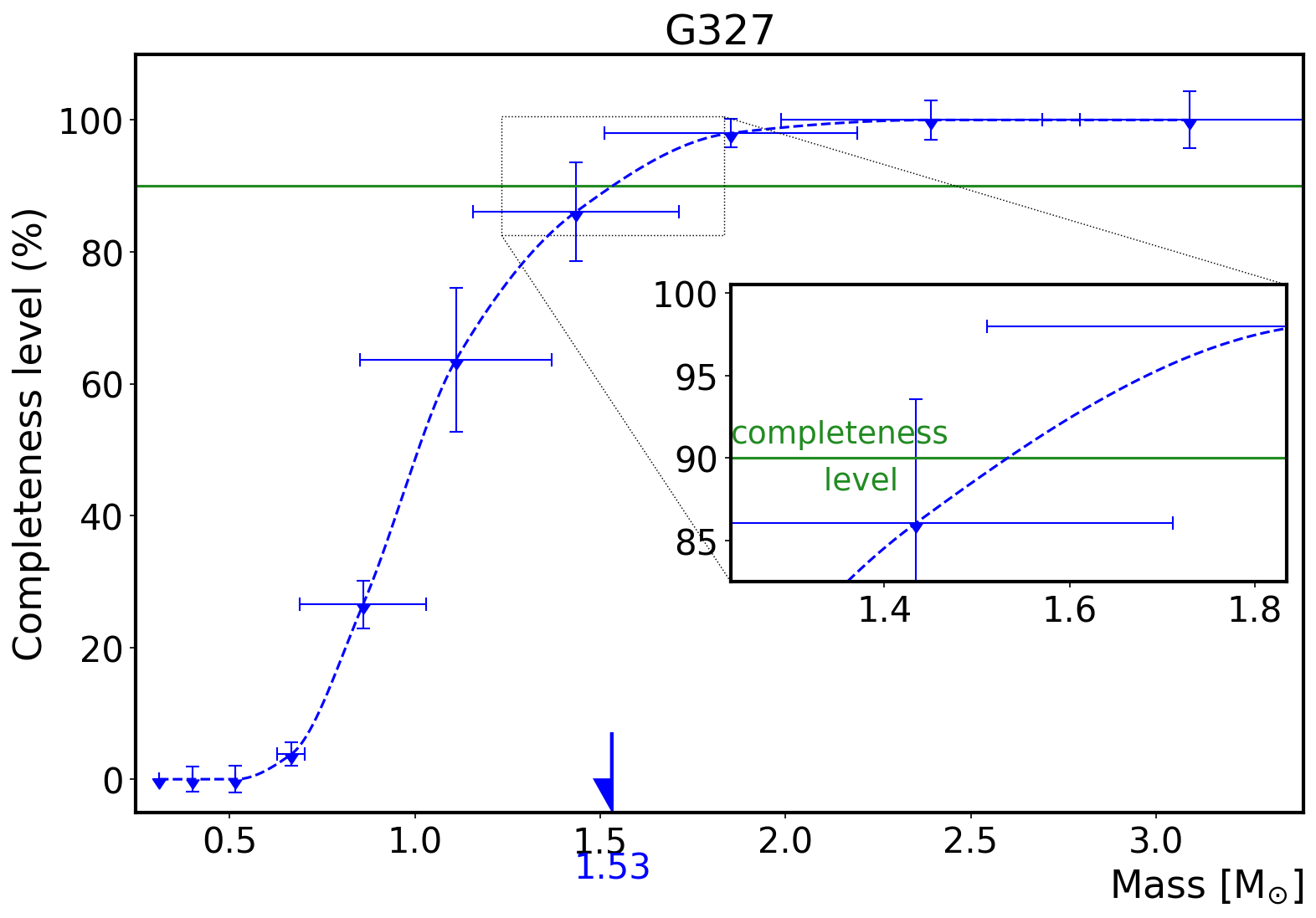}}\\
	\hspace*{-1cm}
\subfloat{\includegraphics[trim=0cm 0cm 0cm 0cm, width=0.55\linewidth]{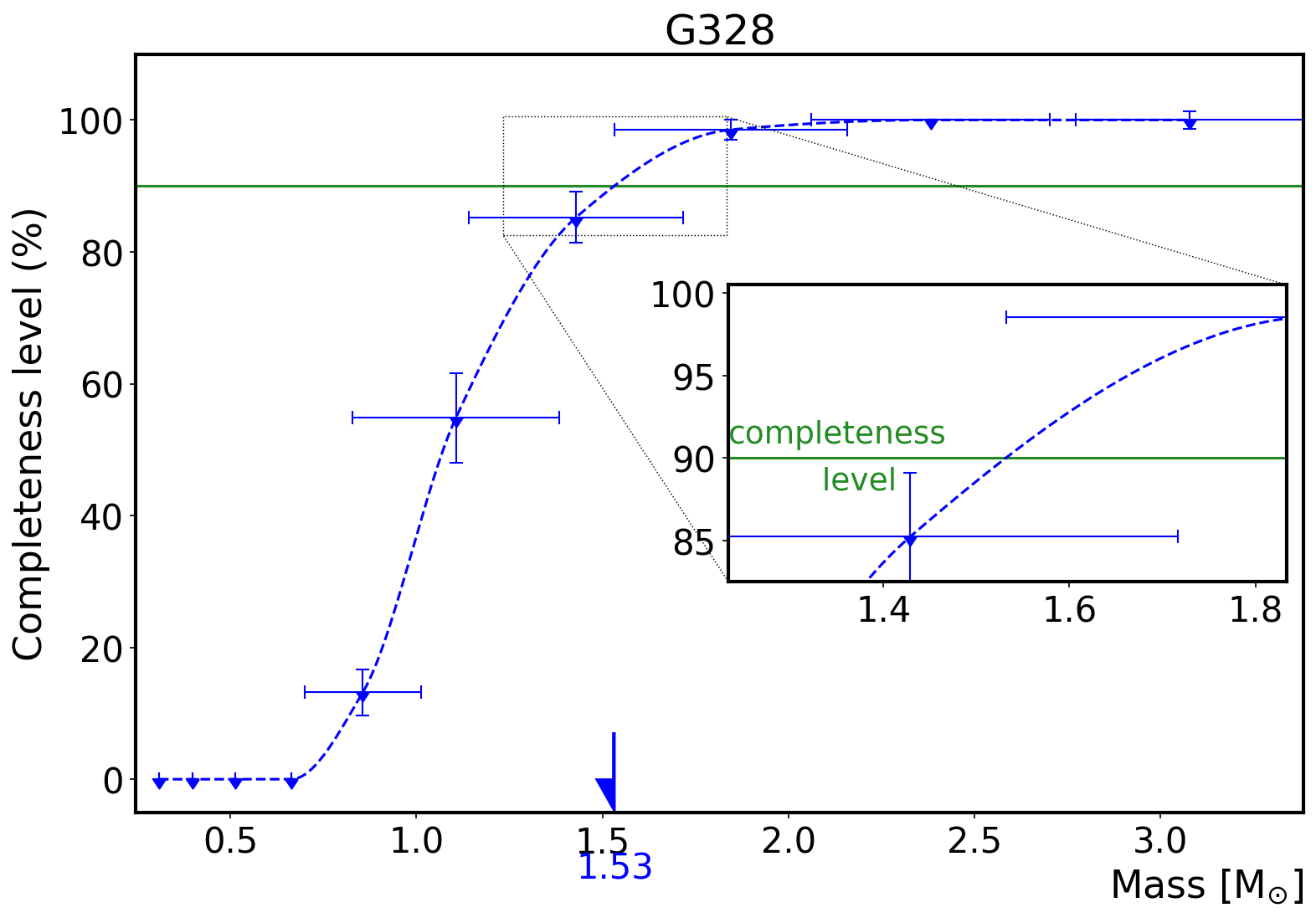}}
\subfloat{\includegraphics[trim=0cm 0cm 0cm 0cm, width=0.55\linewidth]{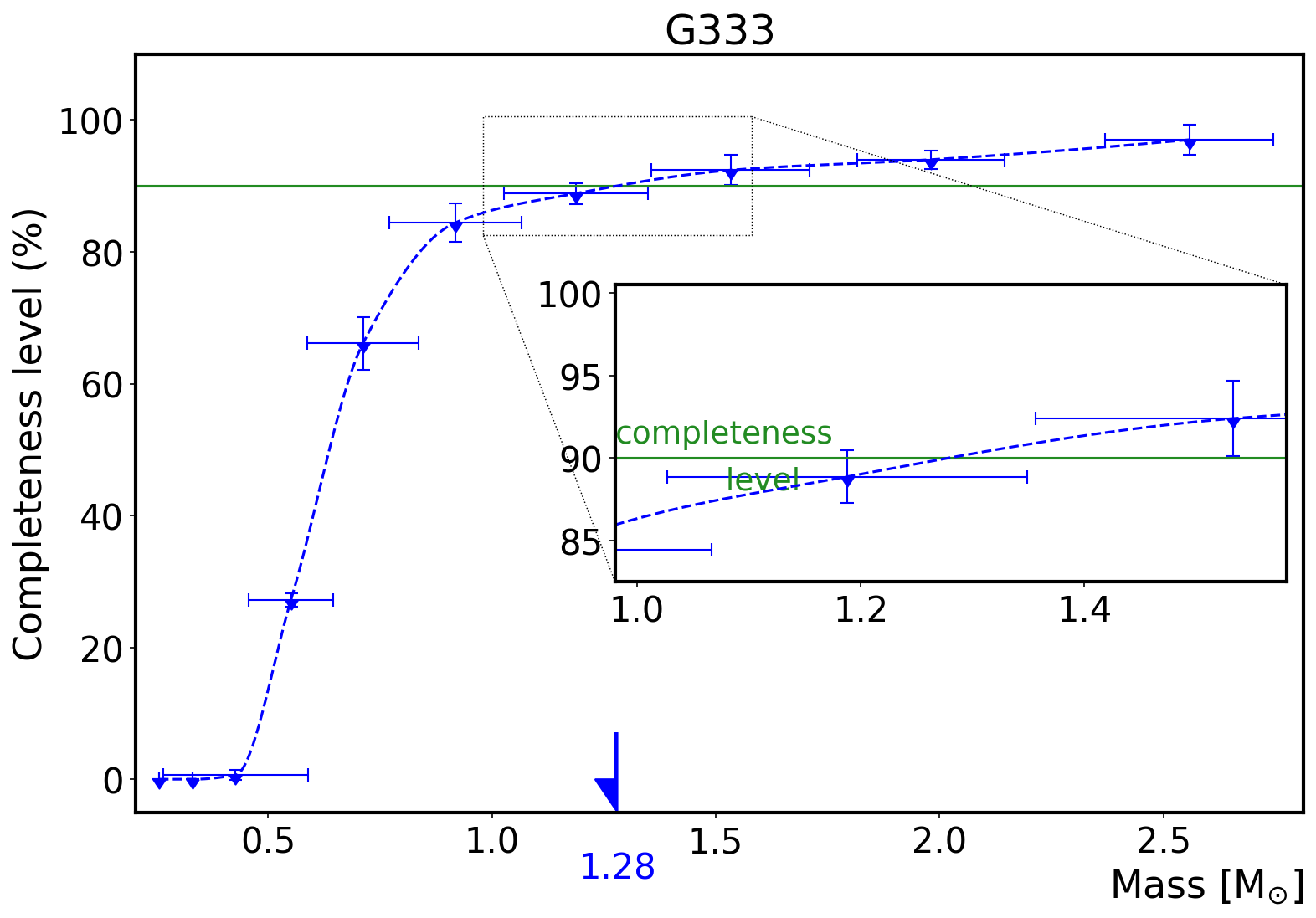}}
\caption{Core completeness tests on the ALMA-IMF fields. The curve shows the percentage of synthetic cores extracted as a function of their mass. The green line shows the 90\,\% threshold. }
\label{af:comptest}
\end{figure*}

\addtocounter{figure}{-1}
\begin{figure*}
	\hspace*{-1cm}
	\centering
\subfloat{\includegraphics[trim=0cm 0cm 0cm 0cm, width=0.55\linewidth]{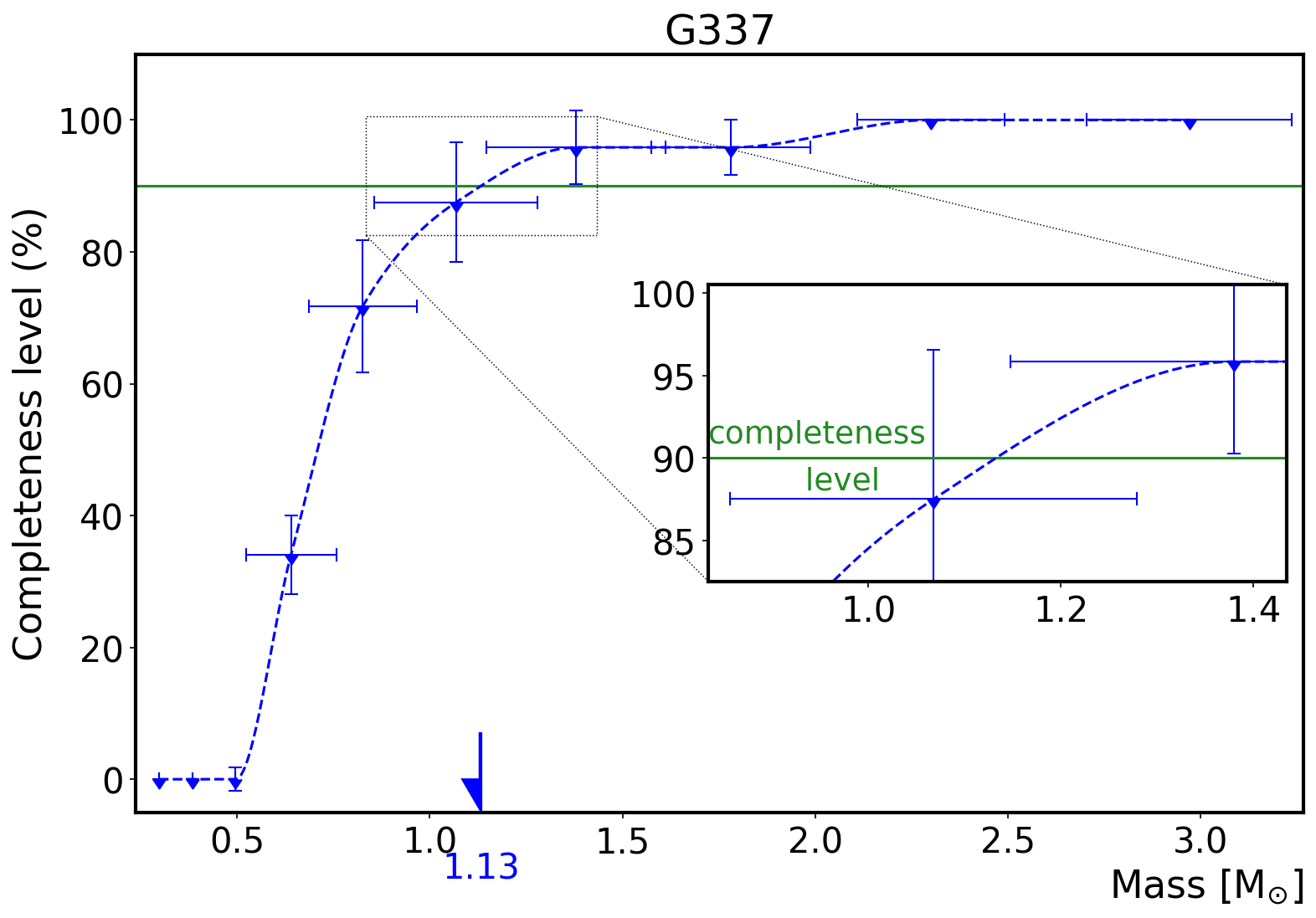}}
\subfloat{\includegraphics[trim=0cm 0cm 0cm 0cm, width=0.55\linewidth]{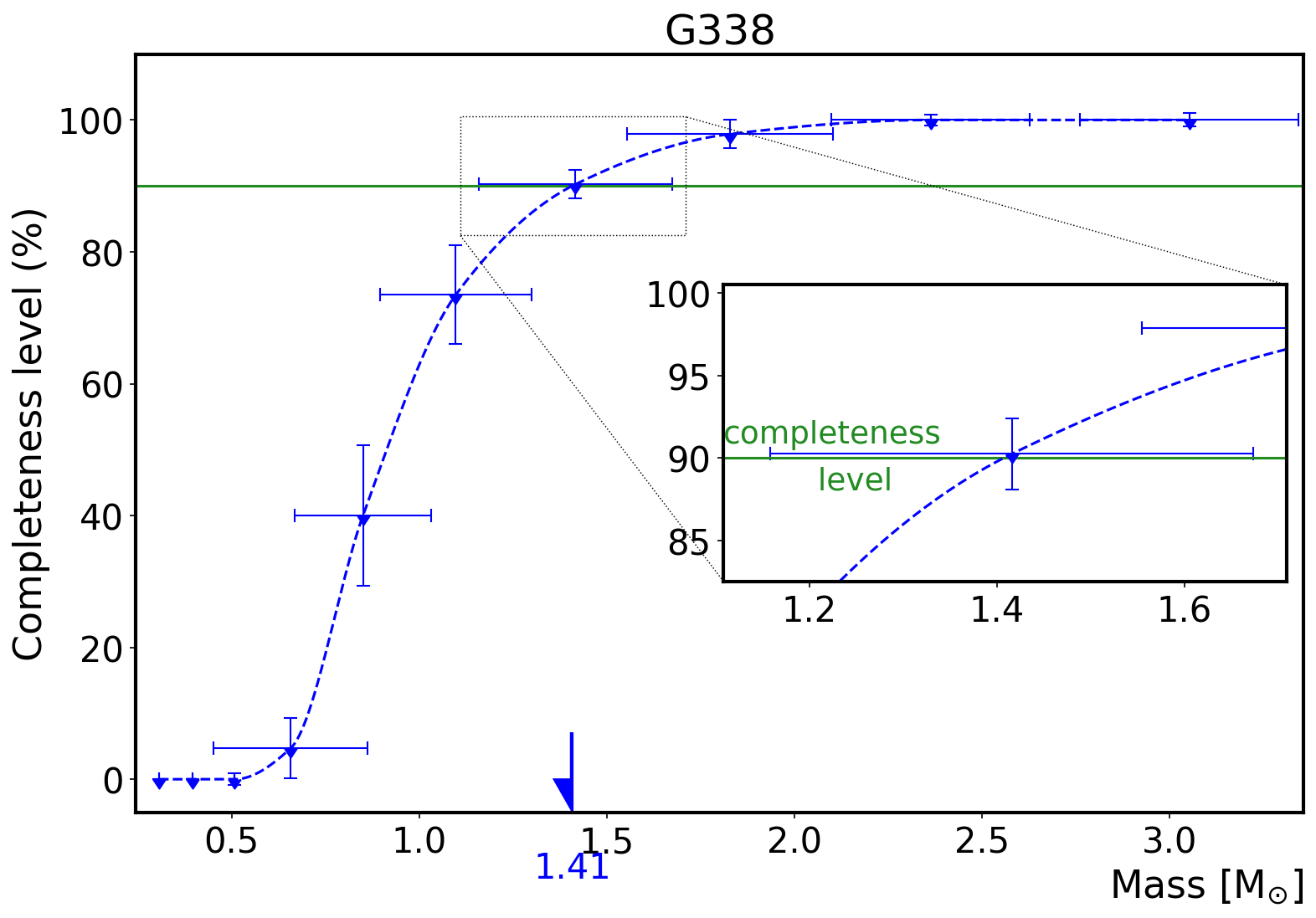}}\\
	\hspace*{-1cm}
\subfloat{\includegraphics[trim=0cm 0cm 0cm 0cm, width=0.55\linewidth]{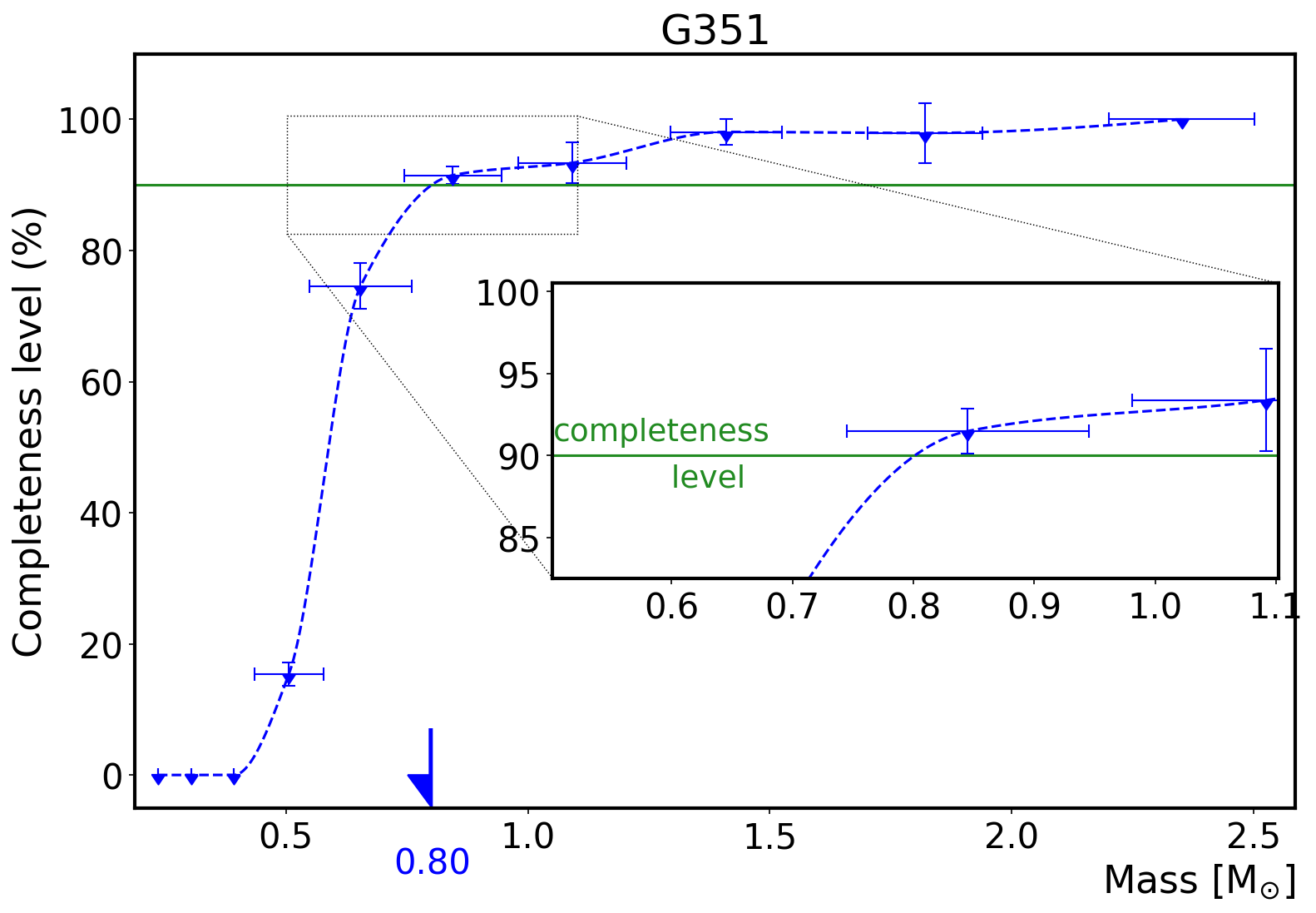}}
\subfloat{\includegraphics[trim=0cm 0cm 0cm 0cm, width=0.55\linewidth]{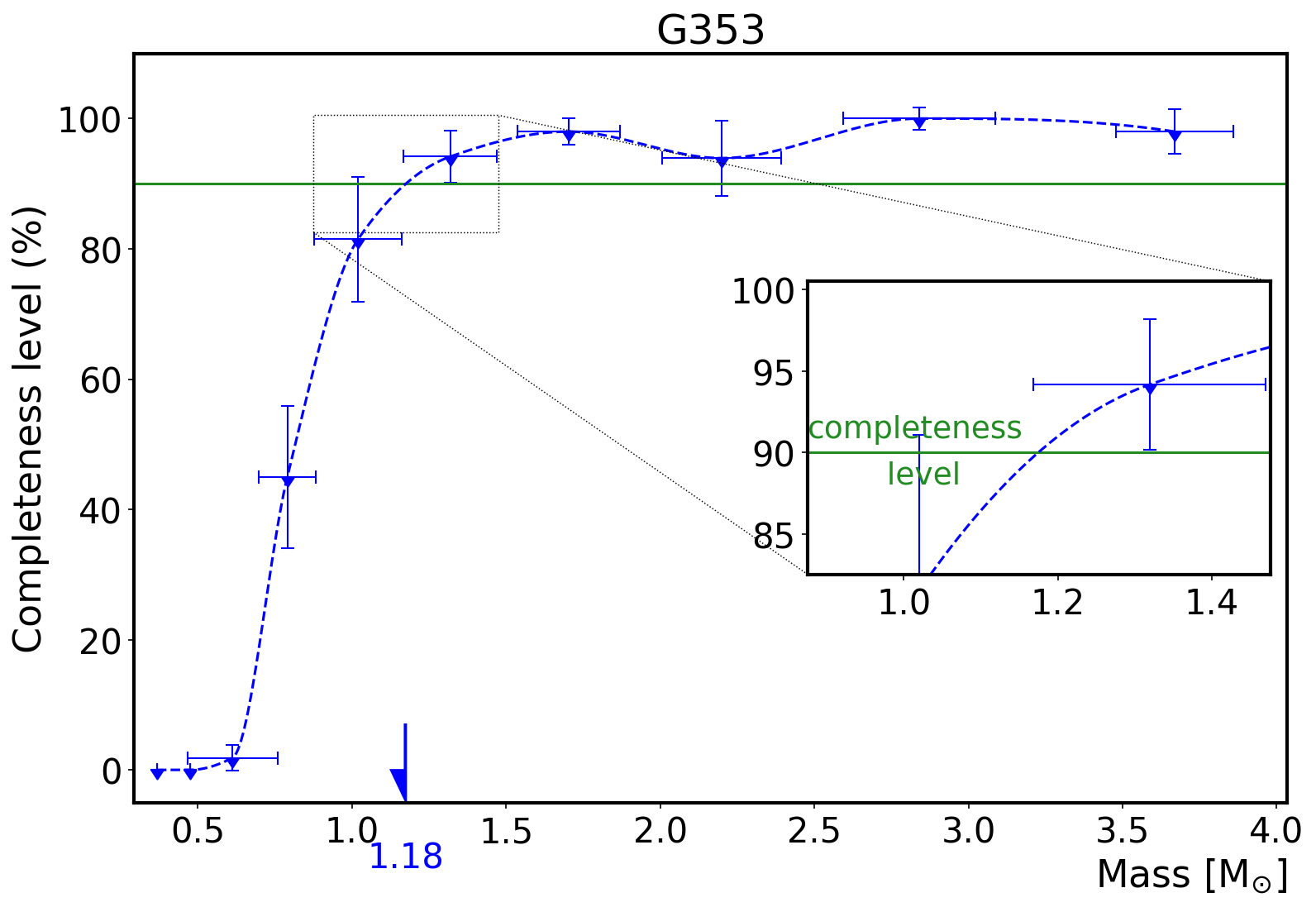}}\\
	\hspace*{-1cm}
\subfloat{\includegraphics[trim=0cm 0cm 0cm 0cm, width=0.55\linewidth]{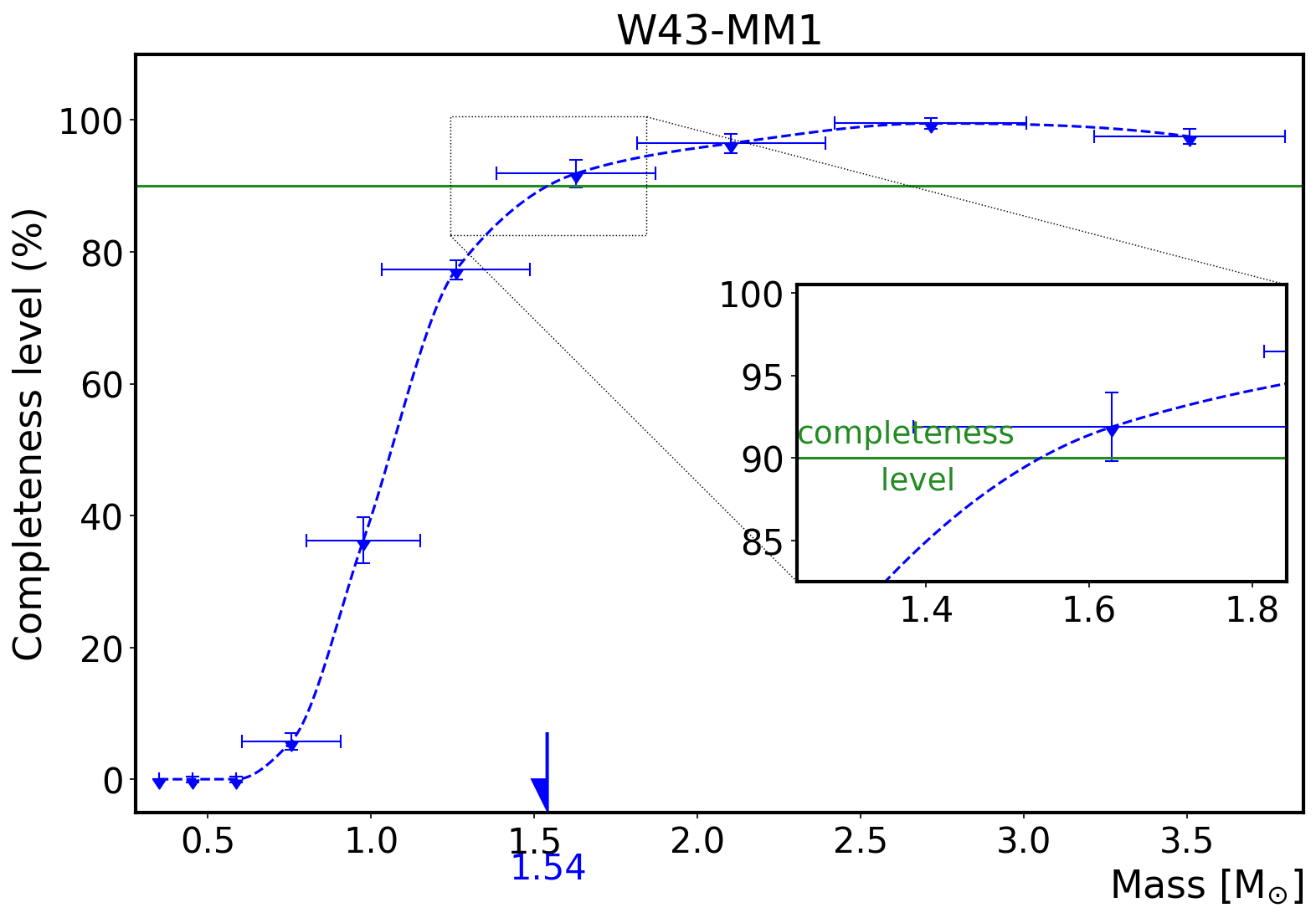}}
\subfloat{\includegraphics[trim=0cm 0cm 0cm 0cm, width=0.55\linewidth]{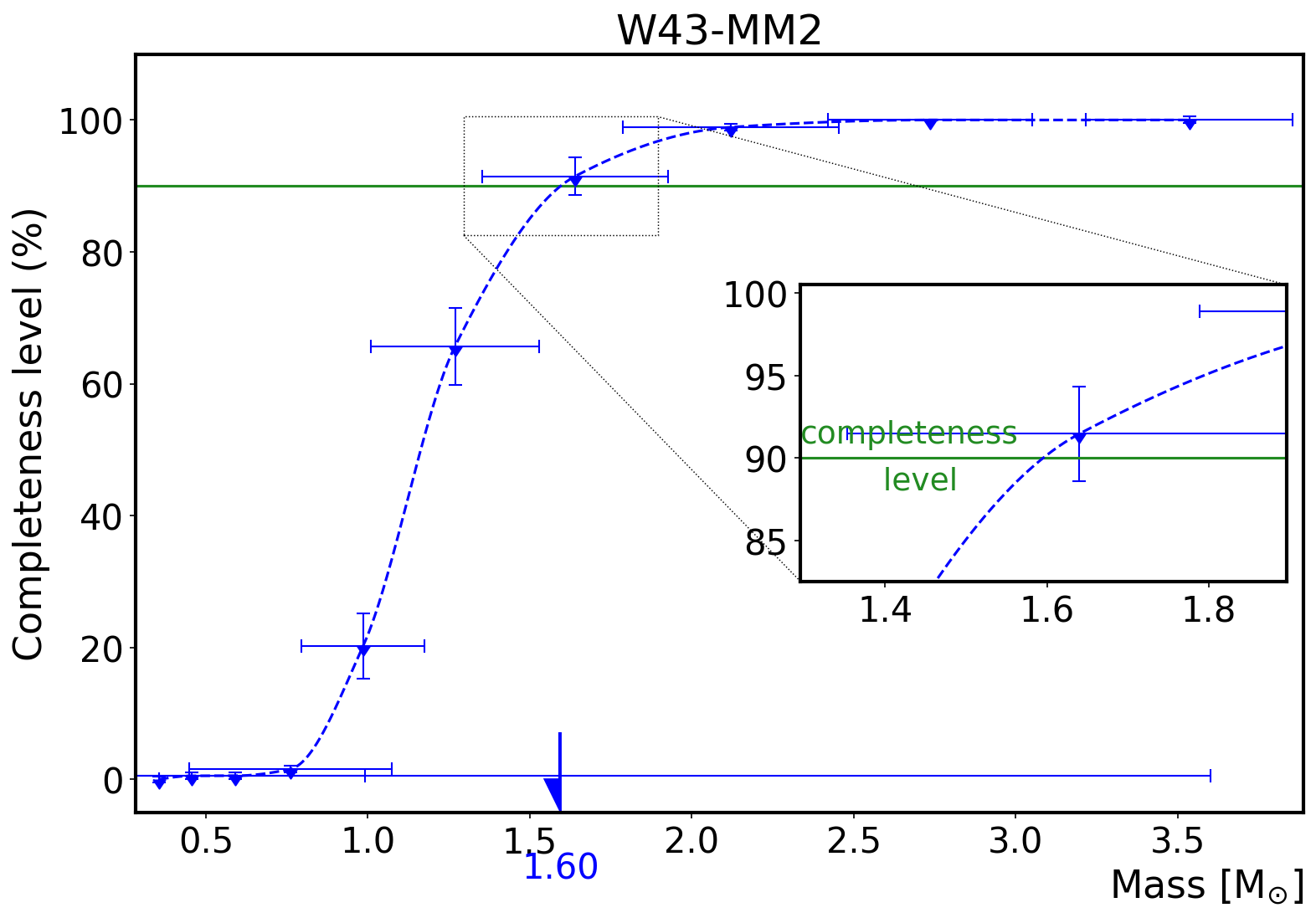}}
\caption{Completeness tests (continued).}
\end{figure*}

\addtocounter{figure}{-1}
\begin{figure*}
	\hspace*{-1cm}
	\centering
\subfloat{\includegraphics[trim=0cm 0cm 0cm 0cm, width=0.55\linewidth]{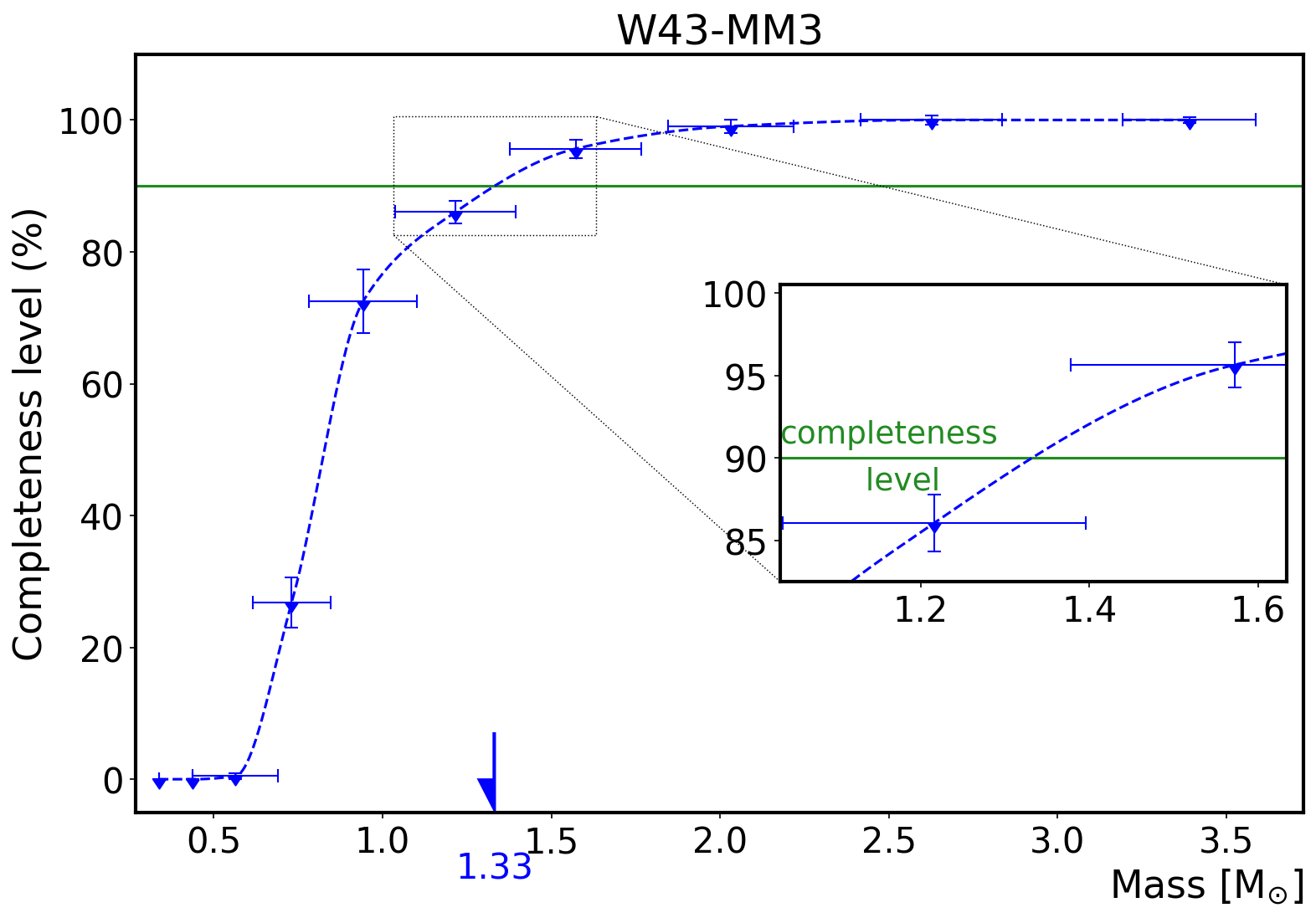}}
\subfloat{\includegraphics[trim=0cm 0cm 0cm 0cm, width=0.55\linewidth]{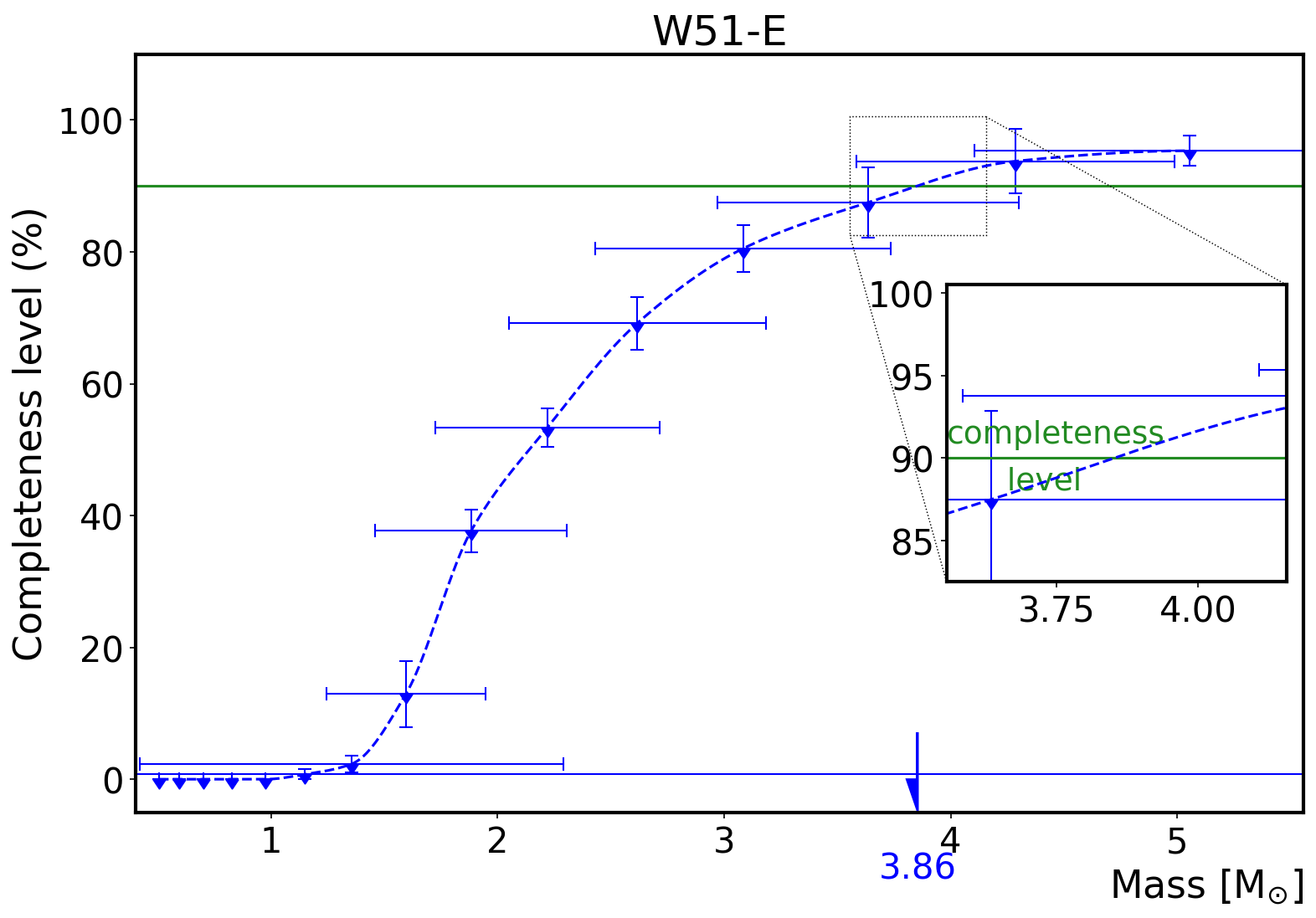}}\\
	\hspace*{-1cm}
\subfloat{\includegraphics[trim=0cm 0cm 0cm 0cm, width=0.55\linewidth]{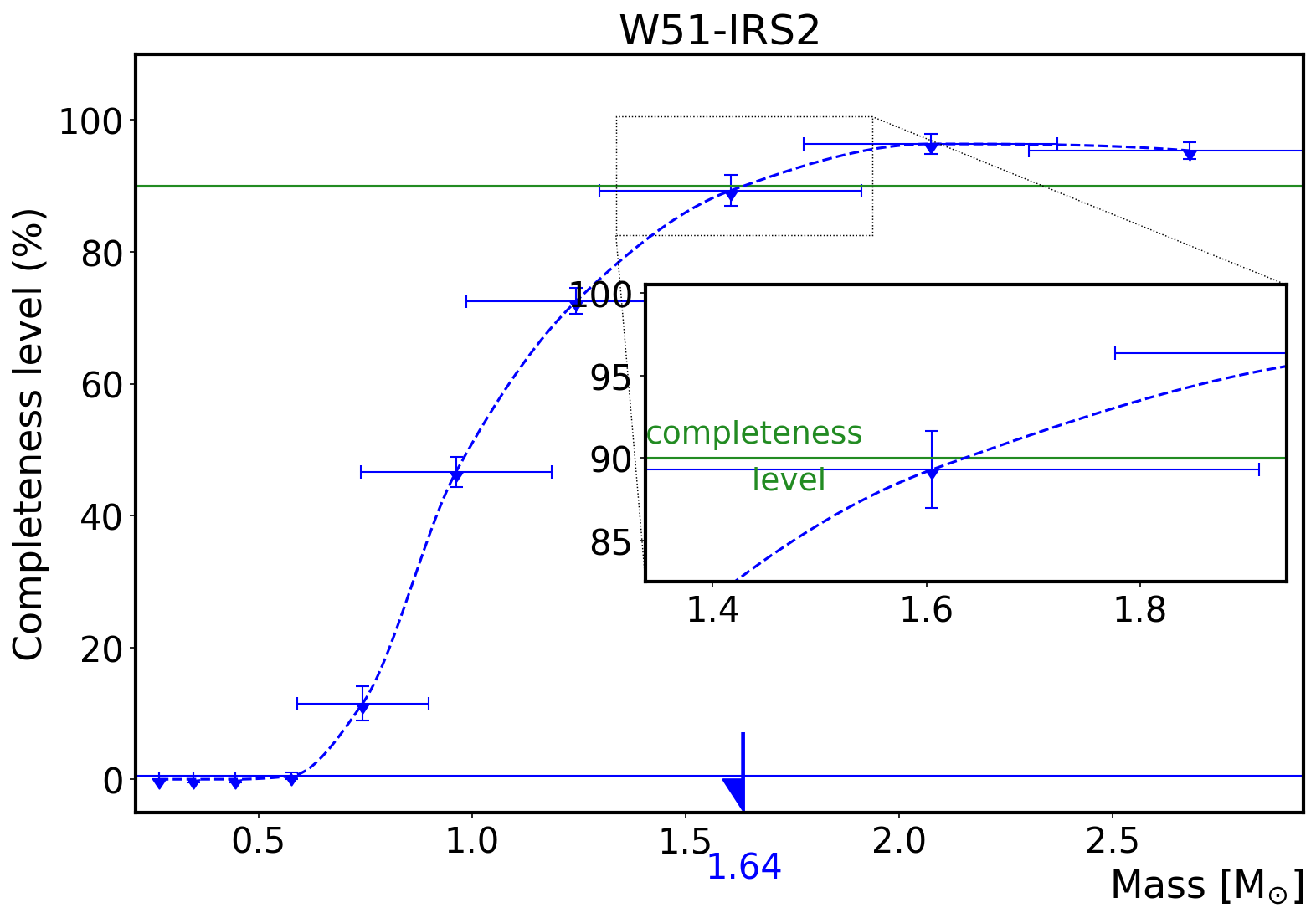}}
\caption{Completeness tests (continued).}
\end{figure*}

\section{Effect of the smoothing on the \textsl{GExt2D} extractions}

\begin{figure*} 
	\hspace*{-1cm}
    \centering
\subfloat{\includegraphics[trim=0cm 0cm 0cm 0cm, width=0.55\linewidth]{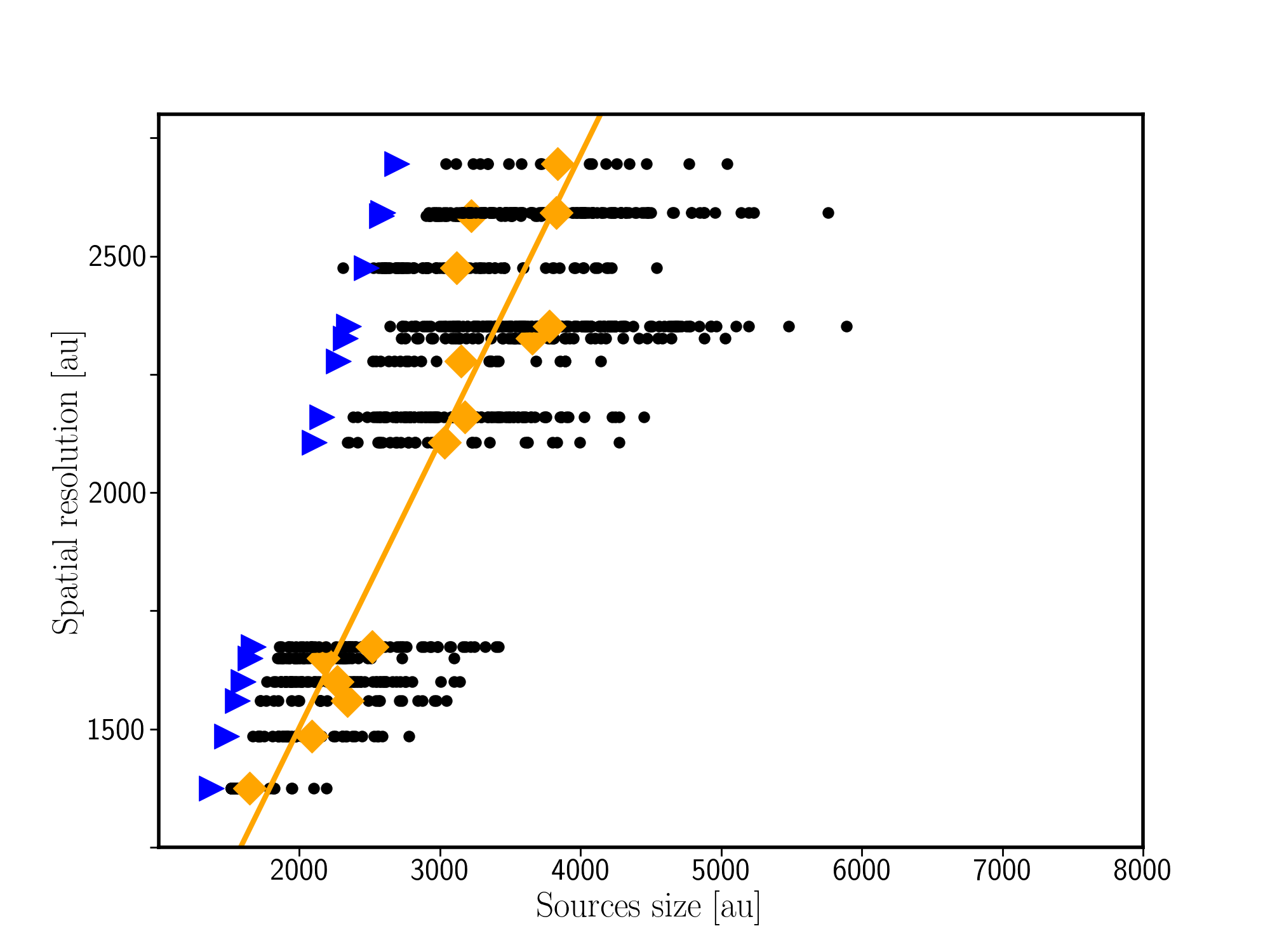}}
\subfloat{\includegraphics[trim=0cm 0cm 0cm 0cm, width=0.55\linewidth]{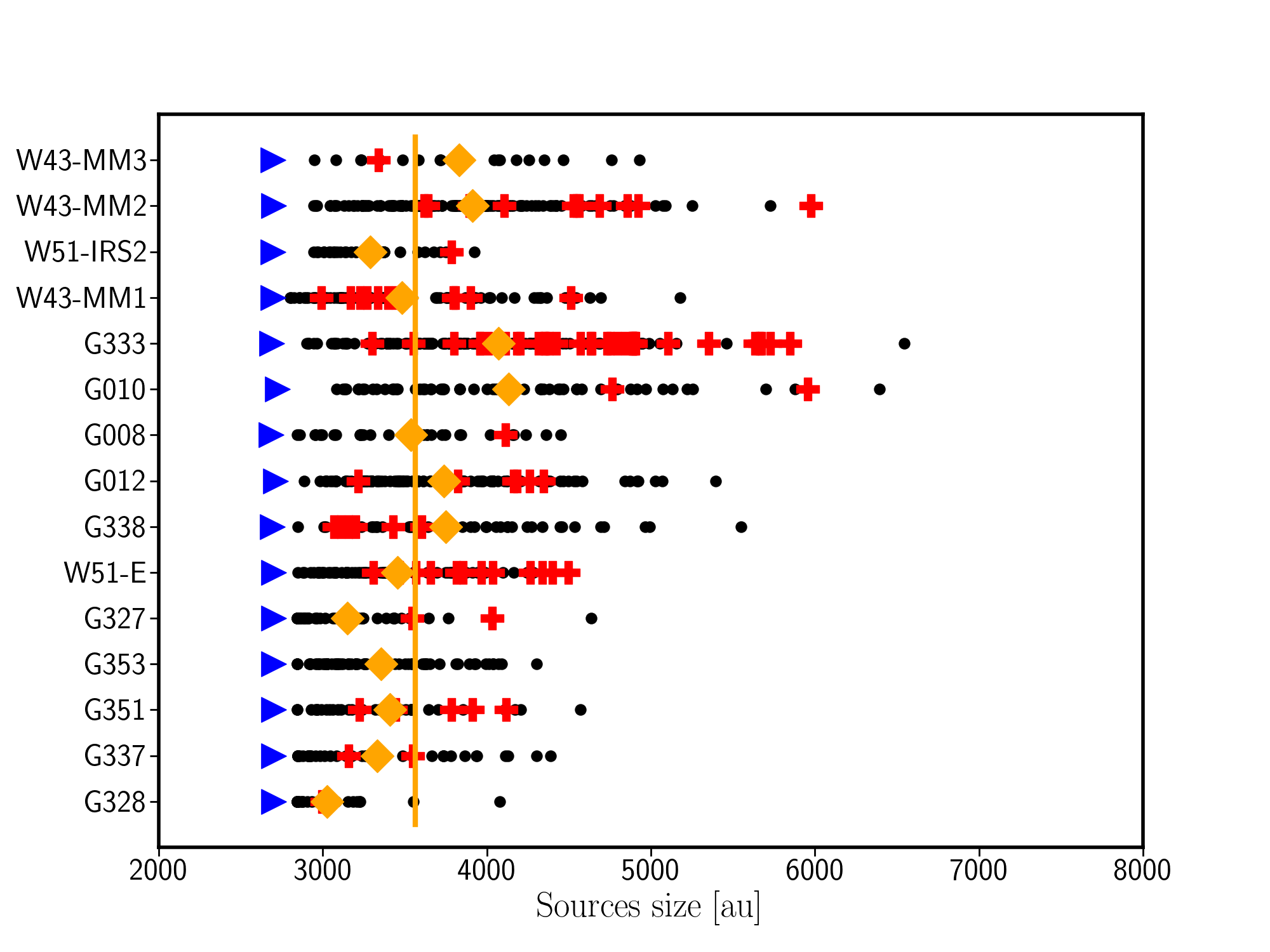}}\caption
{ 
Same as Fig.~\ref{f:coresize} for the sources extracted by GExt2D.
} 
\label{f:coresize-getext2d}
\end{figure*}

We show the effect of smoothing the maps on the source sizes derived by \textsl{GExt2D}. Figure~\ref{f:coresize-getext2d} shows the scatter plot of all source sizes before (left panel) and after (right panel) smoothing the maps to a match spacial scale of 2700\,au.

\section{Continuum maps overlaid by both getsf and GExt2D sources}
\label{s:doubleextract}

\begin{figure*} 
    \centering
\includegraphics[trim=0cm 0cm 0cm 0cm, width=1.0\linewidth]{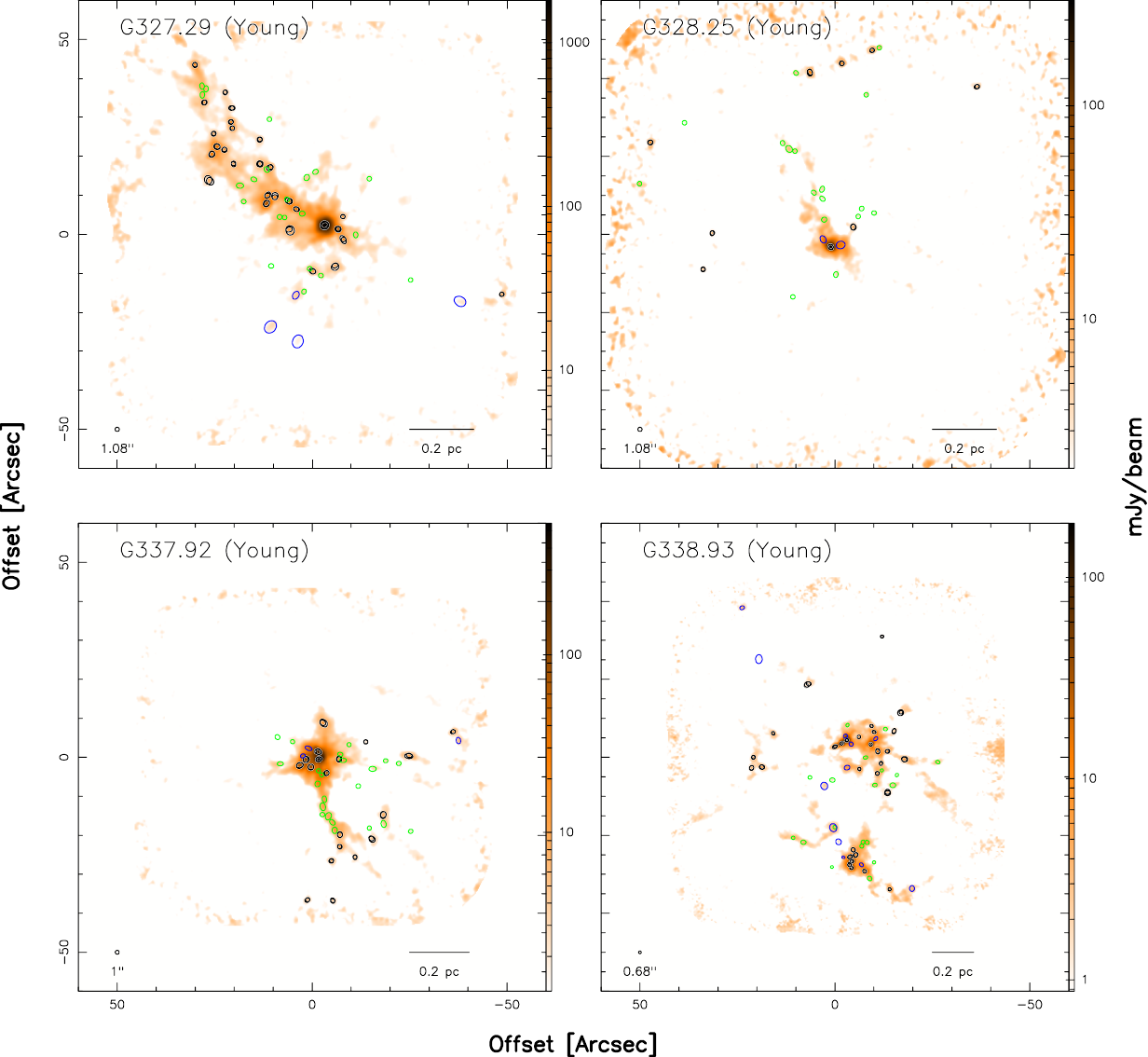}
\caption
{ 
Continuum emission maps at 1.3\,mm for the 15 ALMA-IMF fields with a spatial resolution of 2700\,au. The ellipses locate all the cores found by \textsl{getsf} in blue, by \textsl{GExt2D} in green, and by both algorithms in black. The field name is indicated in the top left corner of each panel, its evolutionary stage next to the field name, and the beam size in the bottom left corner. 
} 
\label{f:contmapsgag1}
\end{figure*}

In this Appendix, we display for each field the sources extracted by \textsl{getsf} and by \textsl{GExt2D}. In Fig.~\ref{f:contmapsgag1}, we show the sources extracted by \textsl{getsf} only with blue ellipses, those extracted by \textsl{GExt2D} only with green ellipses, and those extracted by both algorithms with black ellipses.

\addtocounter{figure}{-1}
\begin{figure*} 
    \centering
\includegraphics[trim=0cm 0cm 0cm 0cm, width=1.0\linewidth]{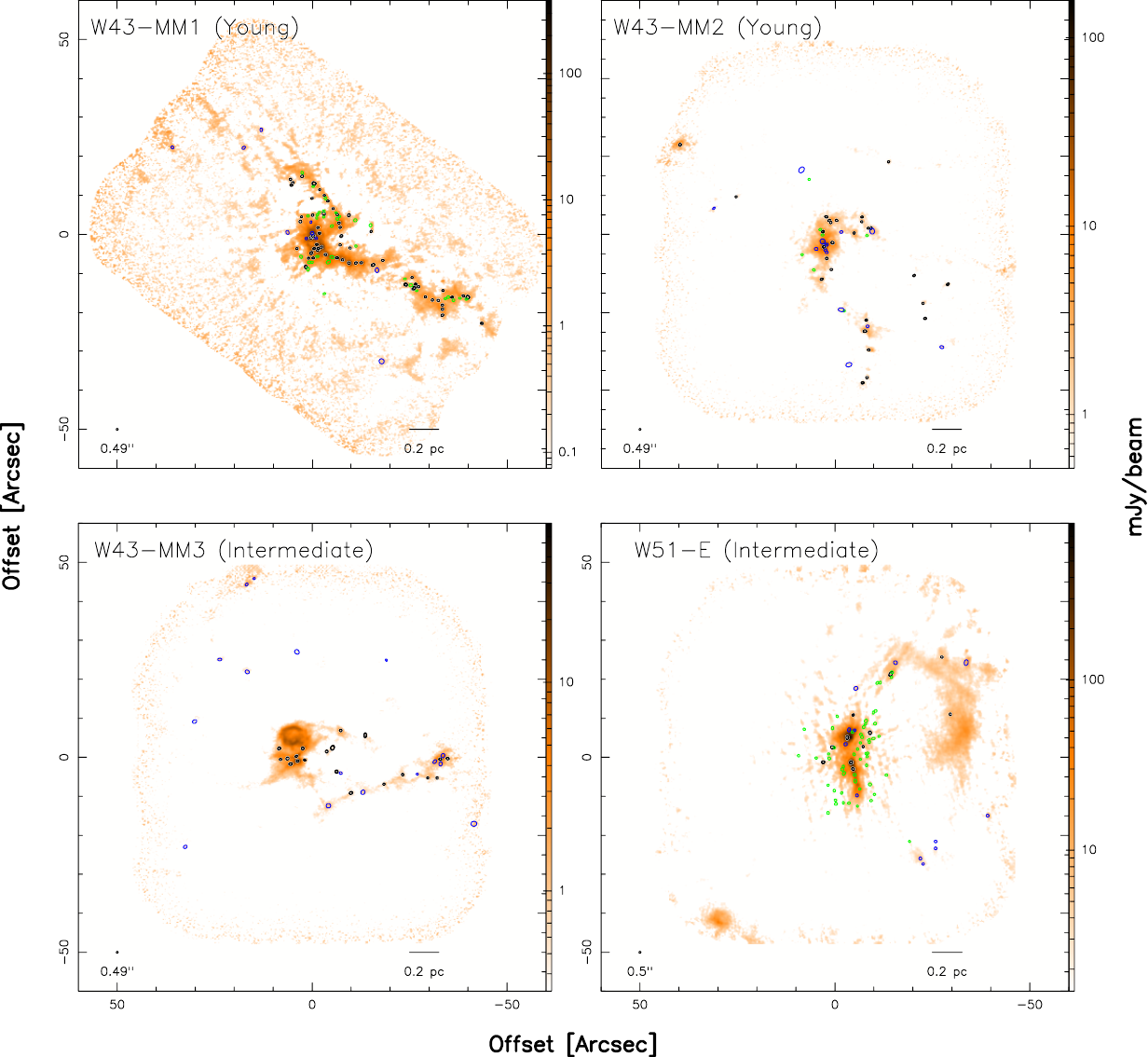}
\caption
{ 
Continuum emission maps at 1.3\,mm. Continued.
} 
\end{figure*}

\addtocounter{figure}{-1}
\begin{figure*} 
    \centering
\includegraphics[trim=0cm 0cm 0cm 0cm, width=1.0\linewidth]{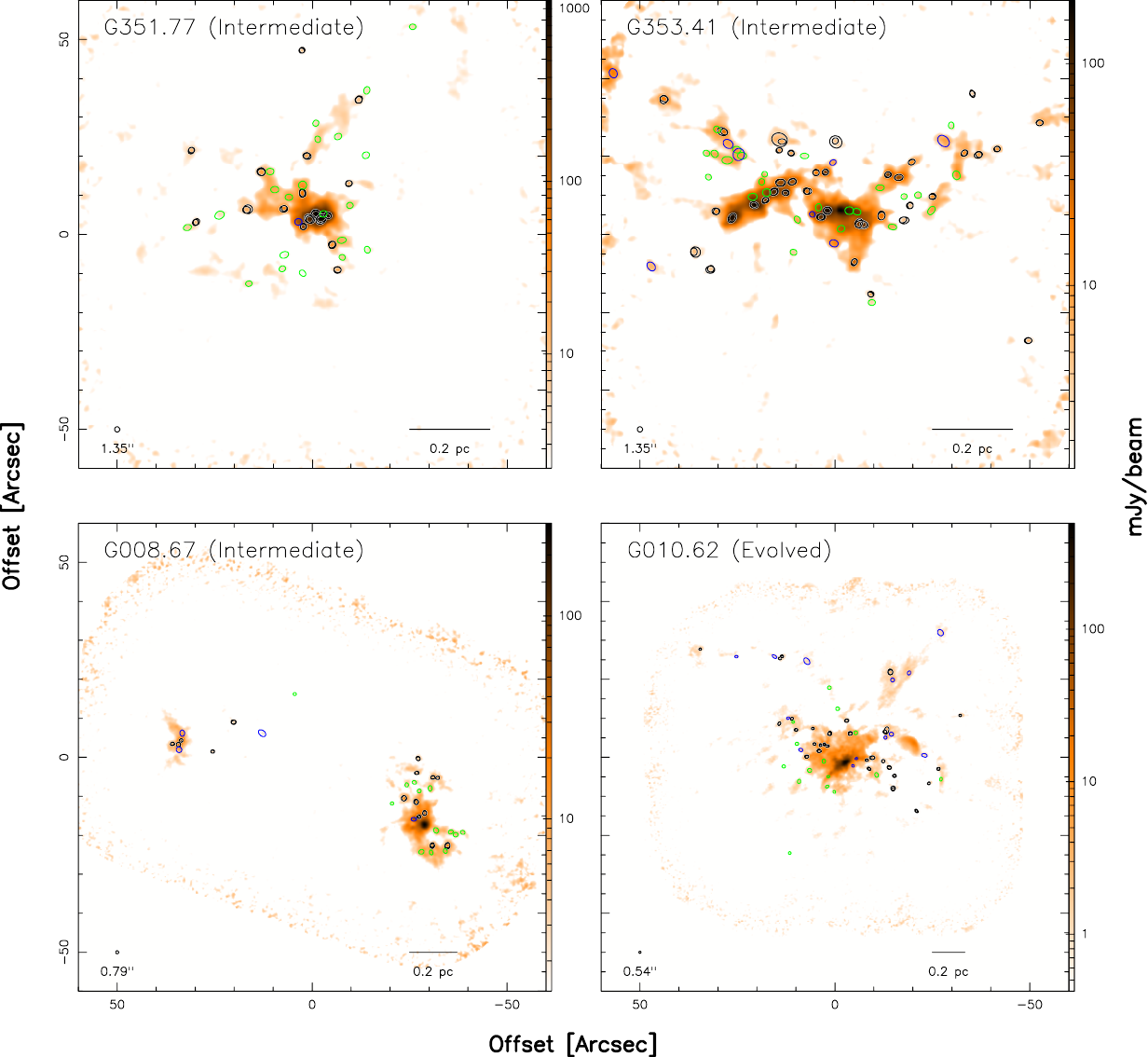}
\caption
{ 
Continuum emission maps at 1.3\,mm. Continued.
} 
\end{figure*}

\addtocounter{figure}{-1}
\begin{figure*} 
    \centering
\includegraphics[trim=0cm 0cm 0cm 0cm, width=1.0\linewidth]{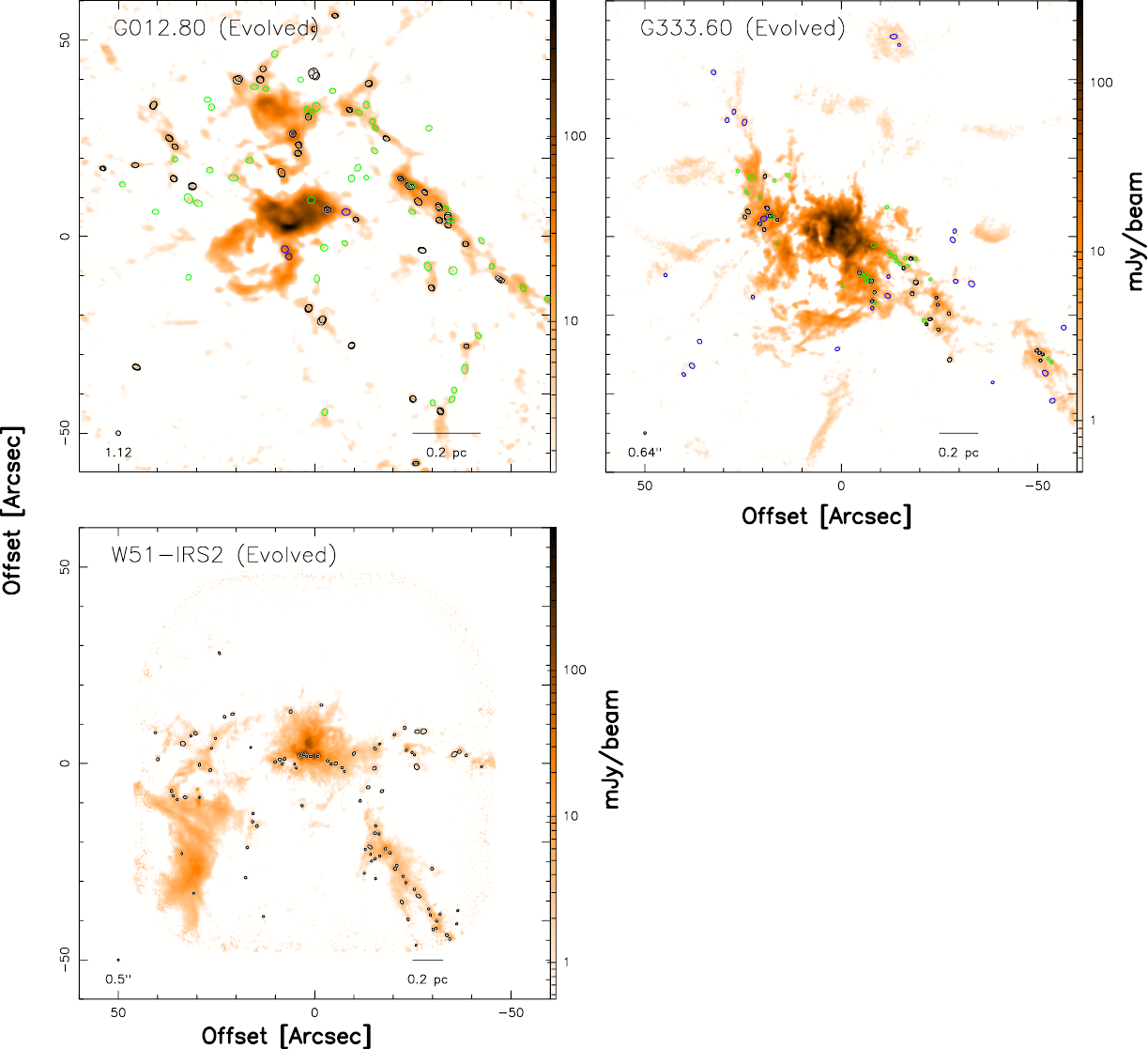}
\caption
{ 
Continuum emission maps at 1.3\,mm. Continued.
} 
\end{figure*}

\section{Table of sources, after smoothing the maps}
\label{s:table-smooth}

In this Appendix, we show the core catalogues for each of the 15 protoclusters. The first group of sources in each table corresponds to thermal dust cores that are gravitationally bound, and the second group corresponds to sources whose fluxes are arguably contaminated by free-free emission. All the tables are available on the ALMA-IMF website (\url{https://www.almaimf.com/}), and on \href{http://cdsportal.u-strasbg.fr/}{CDS}.

\input{G327-catpap.tex}

\input{G328-catpap.tex}

\input{G337-catpap.tex}

\input{G338-catpap.tex}

\input{W43-MM1-catpap.tex}

\input{W43-MM2-catpap.tex}

\input{W43-MM3-catpap.tex}

\input{W51-E-catpap.tex}

\input{G351-catpap.tex}

\input{G353-catpap.tex}

\input{G008-catpap.tex}

\input{G010-catpap.tex}

\input{G012-catpap.tex}

\input{G333-catpap.tex}

\input{W51-IRS2-catpap.tex}

\section{Monte Carlo simulations for fitting the slope of the CMF}

In this Appendix we display in Fig.~{a:incertitudes} the functional parameters retrieved when adjusting a power-law (in blue), an exponential (in green), or a log-normal (in orange) with the MLE technique (see Sect.~{\ref{sss:fitmethod}}.

\begin{figure} 
    \centering
\subfloat{\includegraphics[trim=0cm 0cm 0cm 0cm, width=1.0\linewidth]{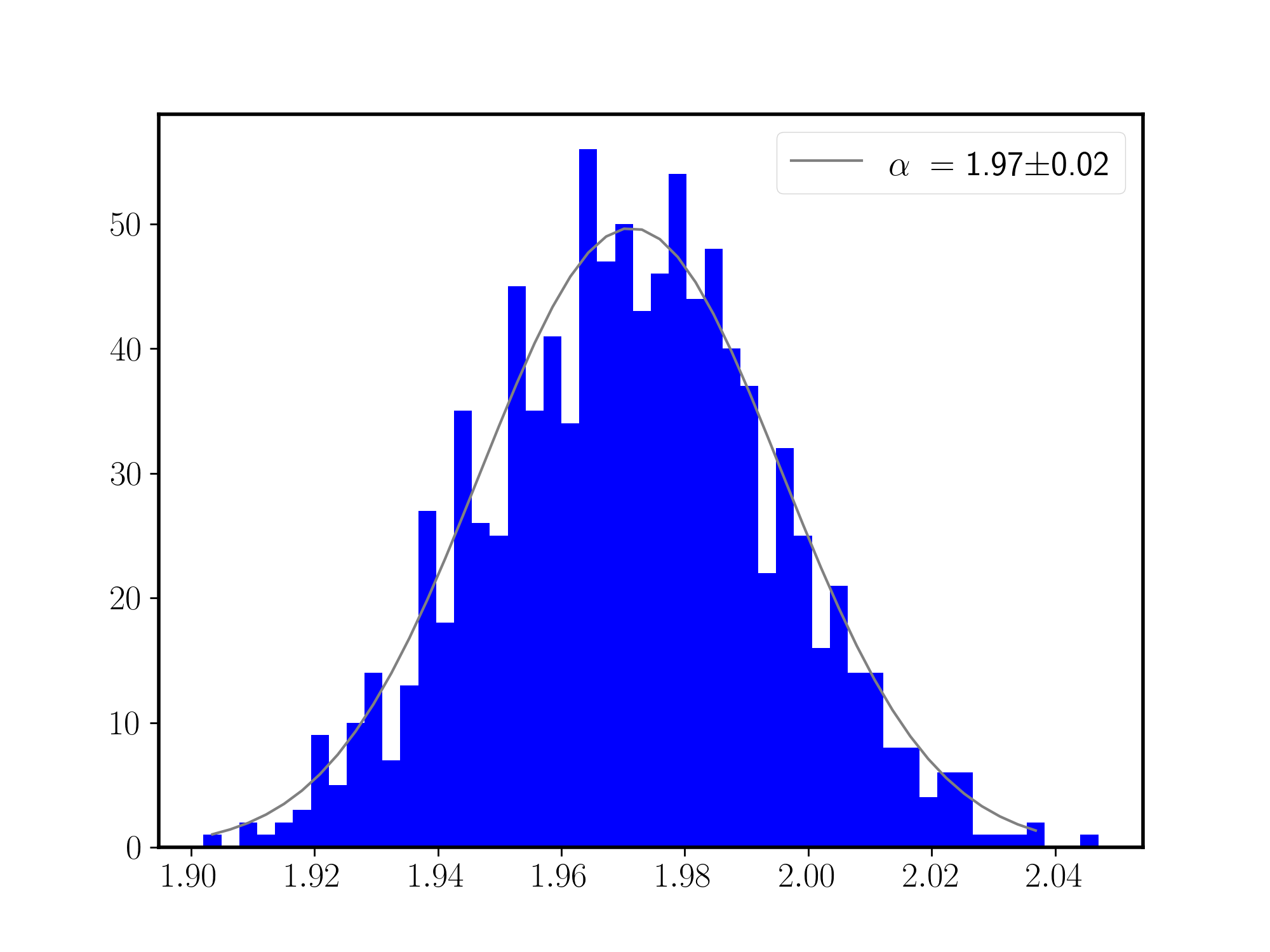}}
\caption
{ 
The plot shows the probability density function for a power-law on the 10$^3$ draws to estimate the fit error on the CMF (see Sect.~\ref{ss:cmf-all}).
} 
\label{a:incertitudes}
\end{figure}

\end{appendix}

\end{document}

%% file: alma-imf-fields.tex
\begin{table*} 
\caption{General properties of the ALMA-IMF fields.}
\addtolength{\tabcolsep}{0pt}
\label{t:alma-imf-fields}      
\begin{tabular}{cccccccccccc} 
\hline\hline
\noalign{\smallskip}
Protocluster& \multicolumn{2}{c}{Phase center} & Evolutionary & $D$ & $\nu_{\rm 1.3\,mm}$  & $\theta$ & Scale & \textbf{${\rm max}$} & $\sigma_{\rm 1.3\,mm}$ & $\nu_{\rm 3\,mm}$ & $\sigma_{\rm 3\,mm}$\\
cloud name  &   RA[J2000] & DEC[J2000] & stage & [pc] & [GHz] & [$\arcsec$] & [au] &  [$\arcsec$]  & [mK] & [GHz] & [mK] \\ 
     &  (1)        & (1)        & (1) & (1) & (2)  & (3) & (4) & (5) & (6) & (2) & (6) \\
\noalign{\smallskip}
\hline
\noalign{\smallskip}
G327.29               &     15:53:08.130   &       $-$54:37:08.60        &  Young        & 2500   &  229.507   & 0.66   & 1650   &    \,\,\,8.3   &       25 &       101.776 &  \,\,\,5  \\
G328.25               &     15:57:59.680   &       $-$53:57:57.43        &  Young        & 2500   &  227.575   & 0.55   & 1365   &    \,\,\,7.1   &       35 &       101.500 &  \,\,\,3  \\
G337.92               &     16:41:10.620   &       $-$47:08:02.90        &  Young        & 2700   &  227.503   & 0.55   & 1475   &    \,\,\,6.6   &       15 &       101.602 &       23  \\
G338.93               &     16:40:34.420   &       $-$45:41:40.60        &  Young        & 3900   &  229.226   & 0.54   & 2100   &    \,\,\,5.8   &       15 &       100.882 &       10  \\
W43-MM1               &     18:47:47.000   &       $-$01:54:26.00        &  Young        & 5500   &  229.680   & 0.45   & 2470   &         12.0   &       26 &  \,\,\,99.759 &       17  \\
W43-MM2               &     18:47:36.610   &       $-$02:00:51.10        &  Young        & 5500   &  227.597   & 0.47   & 2580   &    \,\,\,8.1   &       15 &       101.017 &       23  \\
W43-MM3               &     18:47:41.460   &       $-$02:00:27.60        &  Intermediate & 5500   &  228.931   & 0.49   & 2685   &         11.5   &  \,\,\,9 &       100.911 &       17  \\
W51-E                 &     19:23:44.180   &    \,\,\,14:30:29.50        &  Intermediate & 5400   &  228.918   & 0.31   & 1650   &    \,\,\,4.5   &       60 &       101.426 &       26  \\
G351.77               &     17:26:42.620   &       $-$36:09:22.47        &  Intermediate & 2000   &  227.991   & 0.78   & 1560   &    \,\,\,8.9   &       21 &       100.228 &       16  \\
G353.41               &     17:30:26.280   &       $-$34:41:51.67        &  Intermediate & 2000   &  229.431   & 0.80   & 1600   &    \,\,\,9.1   &       14 &       100.547 &  \,\,\,9  \\
G008.67               &     18:06:21.072   &       $-$21:37:14.84        &  Intermediate & 3400   &  228.732   & 0.67   & 2270   &    \,\,\,9.1   &       22 &       100.526 &       14  \\
G010.62               &     18:10:28.840   &       $-$19:55:48.30        &  Evolved      & 4950   &  229.268   & 0.47   & 2310   &    \,\,\,6.4   &       14 &       100.704 &       23  \\
G012.80               &     18:14:13.370   &       $-$17:55:47.17        &  Evolved      & 2400   &  229.080   & 0.90   & 2155   &    \,\,\,9.5   &       20 &       100.680 &       21  \\
G333.60               &     16:22:09.360   &       $-$50:06:00.87        &  Evolved      & 4200   &  229.062   & 0.56   & 2335   &         11.8   &       22 &       100.756 &       24  \\
W51-IRS2              &     19:23:39.810   &    \,\,\,14:31:03.50        &  Evolved      & 5400   &  228.530   & 0.48   & 2575   &    \,\,\,7.0   &       13 &       101.263 &       24  \\
\noalign{\smallskip}
\hline
\noalign{\smallskip}
\end{tabular}
\tablefoot
{ 
(1) Phase center, evolutionary stage and distance to the region taken from Tables 1 and 4 of Paper~\Romannum{1}.
(2) Central frequencies in Band 6 ($\simeq$1.3\,mm) and in Band 3 ($\simeq$3.0\,mm) taken from Table~D.1 of Paper~\Romannum{2}.
(3) Angular resolution in Band 6 (geometric average over the major and minor axis of the beam).
(4) Spatial resolution on source in Band 6.
(5) Largest angular scale recovered in Band 6.
(6) Standard deviations of the noise level in Band 6 and in Band 3.
}
\end{table*} 

%% file: alma-imf-smoothing.tex
\begin{table*} 
\caption{Resolutions, noise levels, and sources extracted after smoothing to a linear scale of 2700\,au. }
\addtolength{\tabcolsep}{0pt}
\label{t:field-smoothing}      
\begin{tabular}{ccccccccc} 
\hline\hline
\noalign{\smallskip}
Protocluster & $\theta^{\,(1)}$& $\sigma^{\,(2)}$ & $\sigma^{\,(3)}$ & \textsl{getsf} & \textsl{GExt2D} & Common$^{(4)}$  \\
cloud name & [$\arcsec$] & [mJy.beam$^{-1}$] & [$M_{\odot}$] \\ 
\noalign{\smallskip}
\hline
\noalign{\smallskip}
G327.29               &   1.08  & 0.30  & 0.03 & \,\,\,32 &  \,\,\,\,\,\,\,\,\,49 &       \,\,\,28 & Table~\ref{t:core-extraction-smoothed-G327}\\
G328.25               &   1.08  & 0.19  & 0.02 & \,\,\,11 &  \,\,\,\,\,\,\,\,\,26 &  \,\,\,\,\,\,9 & Table~\ref{t:core-extraction-smoothed-G328}\\
G337.92               &   1.00  & 0.09  & 0.01 & \,\,\,22 &  \,\,\,\,\,\,\,\,\,42 &       \,\,\,19 & Table~\ref{t:core-extraction-smoothed-G337}\\
G338.93               &   0.69  & 0.12  & 0.03 & \,\,\,42 &  \,\,\,\,\,\,\,\,\,50 &       \,\,\,31 & Table~\ref{t:core-extraction-smoothed-G338}\\
W43-MM1               &   0.49  & 0.09  & 0.01 & \,\,\,71 &  \,\,\,\,\,\,\,\,\,94 &       \,\,\,62 & Table~\ref{t:core-extraction-smoothed-W43-MM1}\\
W43-MM2               &   0.49  & 0.11  & 0.06 & \,\,\,40 &  \,\,\,\,\,\,\,\,\,34 &       \,\,\,27 & Table~\ref{t:core-extraction-smoothed-W43-MM3}\\
W43-MM3               &   0.49  & 0.08  & 0.04 & \,\,\,38 &  \,\,\,\,\,\,\,\,\,21 &       \,\,\,20 & Table~\ref{t:core-extraction-smoothed-W43-MM3}\\
W51-E                 &   0.50  & 0.06  & 0.03 & \,\,\,31 &  \,\,\,\,\,\,\,\,\,75 &       \,\,\,22 & Table~\ref{t:core-extraction-smoothed-W51-E}\\
G351.77               &   1.35  & 0.28  & 0.02 & \,\,\,19 &  \,\,\,\,\,\,\,\,\,40 &       \,\,\,18 & Table~\ref{t:core-extraction-smoothed-G351}\\
G353.41               &   1.35  & 0.19  & 0.02 & \,\,\,46 &  \,\,\,\,\,\,\,\,\,68 &       \,\,\,39 & Table~\ref{t:core-extraction-smoothed-G353}\\
G008.67               &   0.79  & 0.25  & 0.05 & \,\,\,20 &  \,\,\,\,\,\,\,\,\,30 &       \,\,\,16 & Table~\ref{t:core-extraction-smoothed-G008}\\
G010.62               &   0.55  & 0.12  & 0.03 & \,\,\,50 &  \,\,\,\,\,\,\,\,\,59 &       \,\,\,37 & Table~\ref{t:core-extraction-smoothed-G010}\\
G012.80               &   1.13  & 0.52  & 0.05 & \,\,\,57 &       \,\,\,\,\,\,125 &       \,\,\,57 & Table~\ref{t:core-extraction-smoothed-G012}\\
G333.60               &   0.64  & 0.24  & 0.07 & \,\,\,95 &       \,\,\,\,\,\,175 &       \,\,\,64 & Table~\ref{t:core-extraction-smoothed-G333}\\
W51-IRS2              &   0.50  & 0.12  & 0.03 &      109 &       \,\,\,\,\,\,136 &       \,\,\,94 & Table~\ref{t:core-extraction-smoothed-W51-IRS2}\\
\noalign{\smallskip}
\hline
\noalign{\smallskip}
Total$^{(5)}$         &         &       &      &      677 &                  1020 &            539 \\
\end{tabular}
\tablefoot
{ 
(1) Angular resolution after smoothing Band 3 and 6 images to a matched spatial resolution of 2700\,au.
(2) Standard deviation of the noise level in Band 6.
(3) Equivalent mass to the standard deviation in Band 6, using Eq.~\ref{e:mass}, assuming $S^{\rm peak}_{\rm 1.3\,mm}/\left(\Omega^{\rm 1.3\,mm}\times B(20\,{\rm K},\,\nu)\right)\rightarrow 1$.
(4) Number of sources in common between the \textsl{getsf} and \textsl{gext2d} extractions.
(5) The total number of sources differs from the direct sum per field because two pairs of fields host sources in common. In the \textsl{getsf} extractions, sources number 3, 20, 30, and 31 in W51-E correspond to sources 3, 22, 38, and 108 in W51-IRS2, respectively; sources number 10 and 46 in W43-MM2 correspond to sources 2 and 37 in W43-MM3, respectively. In the \textsl{GExt2D} extractions, sources number 29, 32, and 46 in W51-E correspond to sources 21, 11, and 28 in W51-IRS2, respectively; source number 14 in W43-MM2 corresponds to source number 4 in W43-MM3. 
}
\end{table*} 

%% file: Stat-extractions-getsf.tex
\begin{table*} 
\caption{Evolution of the core sample through the selection process}
\addtolength{\tabcolsep}{0pt}
\label{t:core-extraction-getsf}      
\begin{tabular}{ccccccccc} 
\hline\hline
\noalign{\smallskip}
Protocluster  & All$^{(1)}$ & $\gamma<2^{(2)}$ & $S^{\rm theo}_{\rm 3\,mm}<\sigma_{\rm 3\,mm}^{(3)}$ & Unbound$^{(4)}$ & Bound$^{(5)}$ & $M_{low}^{(6)}$ & $M\,>\,M_{low}^{(7)}$  & $M\,>\,1.64$\,\msun$^{(8)}$ \\
              &             &                  &                                                     &                 &               &    [\msun]        \\
\noalign{\smallskip}
\hline
\noalign{\smallskip}
G327.29               & \,\,\,32 &  \,\,\,0  &  0 &  0  & \,\,\,32   &   1.53  &       \,\,\,26 &      \,\,\,25            \\
G328.25               & \,\,\,11 &  \,\,\,0  &  0 &  0  & \,\,\,11   &   1.53  & \,\,\,\,\,\, 7 & \,\,\,\,\,\,7            \\
G337.92               & \,\,\,22 &  \,\,\,0  &  0 &  0  & \,\,\,22   &   1.13  &       \,\,\,14 &      \,\,\,11            \\
G338.93               & \,\,\,42 &  \,\,\,0  &  0 &  0  & \,\,\,42   &   1.41  &       \,\,\,31 &      \,\,\,29            \\
W43-MM1               & \,\,\,71 &  \,\,\,1  &  0 &  0  & \,\,\,70   &   1.54  &       \,\,\,53 &      \,\,\,49            \\
W43-MM2               & \,\,\,40 &  \,\,\,0  &  0 &  0  & \,\,\,40   &   1.60  &       \,\,\,24 &      \,\,\,24            \\
W43-MM3               & \,\,\,36 &  \,\,\,1  &  0 &  0  & \,\,\,37   &   1.33  &       \,\,\,16 &      \,\,\,12            \\
W51-E                 & \,\,\,31 &  \,\,\,7  &  1 &  0  & \,\,\,23   &   3.86  &       \,\,\,20 &      \,\,\,20            \\
G351.77               & \,\,\,19 &  \,\,\,1  &  0 &  0  & \,\,\,18   &   0.80  &       \,\,\,11 & \,\,\,\,\,\,6            \\
G353.41               & \,\,\,46 &  \,\,\,1  &  0 &  0  & \,\,\,45   &   1.18  &       \,\,\,23 &      \,\,\,17            \\
G008.67               & \,\,\,20 &  \,\,\,1  &  0 &  0  & \,\,\,19   &   1.53  &       \,\,\,15 &      \,\,\,14            \\
G010.62               & \,\,\,50 &  \,\,\,8  &  0 &  0  & \,\,\,42   &   0.96  &       \,\,\,29 &      \,\,\,22            \\
G012.80               & \,\,\,57 &  \,\,\,9  &  4 &  0  & \,\,\,44   &   1.30  &       \,\,\,30 &      \,\,\,25            \\
G333.60               & \,\,\,95 &       34  &  7 &  0  & \,\,\,54   &   1.28  &       \,\,\,30 &      \,\,\,23            \\
W51-IRS2              &      109 &  \,\,\,6  &  4 &  0  & \,\,\,99   &   1.64  &       \,\,\,67 &      \,\,\,66            \\
\noalign{\smallskip}
\hline
\noalign{\smallskip}
Total$^{(9)}$         &      677 &       68  & 16 &  0  &      593   &         &            393 &           350$^{(10)}$   \\
\end{tabular}
\tablefoot
{ 
(1) Number of sources extracted by \textsl{getsf}.
(2) Number of sources contaminated by freefree.
(3) Number of sources whose emission could be contaminated by freefree.
(4) Number of sources not gravitationnally bound.
(5) Number of sources gravatiationnally bound.
(6) Mass completeness limit in the field.
(7) Number of gravitationnally bound cores exceeding the field's completeness level (see Sect.~\ref{ss:comptests}).
(8) Number of gravitationnally bound cores exceeding the global completeness level (see Sect.~\ref{ss:comptests}).
(9) The total number of sources differs from the direct sum per field because two pairs of fields host sources in common (see Table~\ref{t:field-smoothing}).
(10) The total number of sources reduces to 330 when discarding W51-E.
}
\end{table*}

%% file: G327-catpap.tex
\begin{landscape}
\begin{table}
\caption{Compact sources extracted by \textsl{getsf} in G327, after smoothing.}
\label{t:core-extraction-smoothed-G327}
\centerline{ 
\addtolength{\tabcolsep}{-2pt}
\begin{tabular}{cccc|cccc|cccc|ccccccc}
\hline\hline 
\noalign{\smallskip}
  &           &             &              &      \multicolumn{4}{c}{1.3\,mm}                                                               &  \multicolumn{4}{c}{3\,mm}                            \\ 
n & Core name & RA          & DEC          & $a  \times b $                                  & PA    &$S^{\rm Peak}$                   &$S^{\rm int}$                 & $a \times b$               & PA    &$S^{\rm Peak}$                 &$S^{\rm int}$            & T    & Mass$^{\,(1)}$& $Size^{\,(1)}$ & $\alpha_{\rm BE}^{\,(1)}$ & $\alpha$ & $\textsl{GExt2D}^{\,(2)}$\\ 
& & [IRCS]                & [ICRS]                 & [$\arcsec\times\arcsec$]            & [deg]              & [mJy.Beam$^{-1}$]           & [mJy]      & [$\arcsec\times\arcsec$]            & [deg]              & [mJy.Beam$^{-1}$]           & [mJy]          & [K]            & [$M_{\odot}$] & [au]   &                   &                      &          &                          \\ 
\noalign{\smallskip}
\hline 
\noalign{\smallskip} 
  1 & 238.28234-54.61840 &  15:53:7.760 & -54:37:6.250 & 2.14$\times$2.09 & 164 & 1781.0$\pm$10.1 & 7896.0$\pm$24.0 & 2.22$\times$2.10 & \,\,\,24 & \,\,\,90.7$\pm$0.8\,\,\, & 392.9$\pm$1.7 & 300$\pm$100      &    \,\,\,49.5$\pm$0.1\,\,\, & 5264 & 0.13 & 3.7 & T \\ 
  2 & 238.28948-54.61686 &  15:53:9.480 & -54:37:0.680 & 1.78$\times$1.50 & 152 & \,\,\,169.8$\pm$6.7 & 383.0$\pm$8.3 & 1.75$\times$1.52 & 128 & \,\,\,17.2$\pm$0.7\,\,\, & \,\,\,34.9$\pm$0.8 & 100$\pm$50      &    \,\,\,\,\,\,7.0$\pm$0.1\,\,\, & 3972 & 0.23 & 2.9 & T \\ 
  3 & 238.28066-54.61868 &  15:53:7.360 & -54:37:7.250 & 1.62$\times$1.47 & \,\,\,62 & \,\,\,118.9$\pm$7.7 & 195.6$\pm$6.0 & 1.52$\times$1.28 & \,\,\,79 & \,\,\,\,\,\,9.8$\pm$1.1\,\,\, & \,\,\,17.8$\pm$1.0 & \,\,\,42$\pm$11      &    \,\,\,\,\,\,9.3$\pm$0.3\,\,\, & 3790 & 0.07 & 2.9 & T \\
\end{tabular}}
\tablefoot
{
The coordinates are given at the J2000 Epoch. 
(1) Estimated from the measures at 1.3\,mm. 
(2) Indicates if the source found by \textsl{getsf} was also found by \textsl{GExt2D} (True or False). 
The $^*$ next to the uncertainty indicates that the true value is $<$0.05. 
}
\end{table}
\end{landscape}

%% file: G328-catpap.tex
\begin{landscape}
\begin{table}
\caption{Compact sources extracted by \textsl{getsf} in G328, after smoothing.}
\label{t:core-extraction-smoothed-G328}
\centerline{ 
\addtolength{\tabcolsep}{-2pt}
\begin{tabular}{cccc|cccc|cccc|ccccccc}
\hline\hline 
\noalign{\smallskip}
  &           &             &              &      \multicolumn{4}{c}{1.3\,mm}                                                               &  \multicolumn{4}{c}{3\,mm}                            \\ 
n & Core name & RA          & DEC          & $a  \times b $                                  & PA    &$S^{\rm Peak}$                   &$S^{\rm int}$                 & $a \times b$               & PA    &$S^{\rm Peak}$                 &$S^{\rm int}$            & T    & Mass$^{\,(1)}$& $Size^{\,(1)}$ & $\alpha_{\rm BE}^{\,(1)}$ & $\alpha$ & $\textsl{GExt2D}^{\,(2)}$\\ 
& & [IRCS]                & [ICRS]                 & [$\arcsec\times\arcsec$]            & [deg]              & [mJy.Beam$^{-1}$]           & [mJy]      & [$\arcsec\times\arcsec$]            & [deg]              & [mJy.Beam$^{-1}$]           & [mJy]          & [K]            & [$M_{\odot}$] & [au]   &                   &                      &          &                          \\ 
\noalign{\smallskip}
\hline 
\noalign{\smallskip} 
  1 & 239.49918-53.96685 &  15:57:59.80 & -53:58:0.660 & 1.49$\times$1.32 & \,\,\,29 & \,\,\,300.3$\pm$2.5 & 599.2$\pm$4.1 & 1.54$\times$1.42 & \,\,\,10 & \,\,\,20.2$\pm$0.3\,\,\, & \,\,\,37.3$\pm$0.4 & 100$\pm$50      &    \,\,\,11.5$\pm$0.1\,\,\, & 3442 & 0.12 & 3.4 & T \\ 
  2 & 239.49784-53.95377 &  15:57:59.48 & -53:57:13.58 & 1.28$\times$1.16 & \,\,\,12 & \,\,\,\,\,\,25.3$\pm$1.6 & \,\,\,30.5$\pm$1.4 & 1.25$\times$1.17 & \,\,\,95 & \,\,\,\,\,\,3.0$\pm$0.1\,\,\, & \,\,\,\,\,\,3.4$\pm$0.1 & \,\,\,24$\pm$06      &    \,\,\,\,\,\,2.8$\pm$0.1\,\,\, & 2998 & 0.10 & 2.7 & T \\ 
  3 & 239.50170-53.95443 &  15:58:0.410 & -53:57:15.96 & 1.81$\times$1.30 & \,\,\,14 & \,\,\,\,\,\,21.2$\pm$1.2 & \,\,\,39.6$\pm$1.3 & 1.52$\times$1.36 & \,\,\,52 & \,\,\,\,\,\,2.4$\pm$0.2\,\,\, & \,\,\,\,\,\,4.5$\pm$0.2 & \,\,\,24$\pm$06      &    \,\,\,\,\,\,3.6$\pm$0.1\,\,\, & 3632 & 0.10 & 2.7 & T \\
\end{tabular}}
\tablefoot
{
The coordinates are given at the J2000 Epoch. 
(1) Estimated from the measures at 1.3\,mm. 
(2) Indicates if the source found by \textsl{getsf} was also found by \textsl{GExt2D} (True or False). 
The $^*$ next to the uncertainty indicates that the true value is $<$0.05. 
}
\end{table}
\end{landscape}

%% file: G337-catpap.tex
\begin{landscape}
\begin{table}
\caption{Compact sources extracted by \textsl{getsf} in G337, after smoothing.}
\label{t:core-extraction-smoothed-G337}
\centerline{ 
\addtolength{\tabcolsep}{-2pt}
\begin{tabular}{cccc|cccc|cccc|ccccccc}
\hline\hline 
\noalign{\smallskip}
  &           &             &              &      \multicolumn{4}{c}{1.3\,mm}                                                               &  \multicolumn{4}{c}{3\,mm}                            \\ 
n & Core name & RA          & DEC          & $a  \times b $                                  & PA    &$S^{\rm Peak}$                   &$S^{\rm int}$                 & $a \times b$               & PA    &$S^{\rm Peak}$                 &$S^{\rm int}$            & T    & Mass$^{\,(1)}$& $Size^{\,(1)}$ & $\alpha_{\rm BE}^{\,(1)}$ & $\alpha$ & $\textsl{GExt2D}^{\,(2)}$\\ 
& & [IRCS]                & [ICRS]                 & [$\arcsec\times\arcsec$]            & [deg]              & [mJy.Beam$^{-1}$]           & [mJy]      & [$\arcsec\times\arcsec$]            & [deg]              & [mJy.Beam$^{-1}$]           & [mJy]          & [K]            & [$M_{\odot}$] & [au]   &                   &                      &          &                          \\ 
\noalign{\smallskip}
\hline 
\noalign{\smallskip} 
  1 & 250.29355-47.13427 &  16:41:10.45 & -47:08:3.380 & 1.61$\times$1.23 & 105 & \,\,\,491.9$\pm$13.3 & 1007.0$\pm$16.9 & 1.43$\times$1.12 & 101 & \,\,\,41.5$\pm$1.1\,\,\, & \,\,\,86.8$\pm$1.6 & 100$\pm$50      &    \,\,\,23.7$\pm$0.4\,\,\, & 3622 & 0.06 & 3.0 & T \\ 
  2 & 250.29362-47.13375 &  16:41:10.47 & -47:08:1.490 & 1.85$\times$1.27 & \,\,\,69 & \,\,\,181.6$\pm$13.6 & 392.8$\pm$14.7 & 1.83$\times$1.29 & \,\,\,64 & \,\,\,17.4$\pm$1.1\,\,\, & \,\,\,37.9$\pm$1.2 & 100$\pm$50      &    \,\,\,\,\,\,8.7$\pm$0.3\,\,\, & 3884 & 0.18 & 2.9 & T \\ 
  3 & 250.29490-47.13430 &  16:41:10.78 & -47:08:3.480 & 1.51$\times$1.35 & 170 & \,\,\,\,\,\,56.4$\pm$13.0 & 105.1$\pm$12.0 & 1.52$\times$1.34 & \,\,\,\,\,\,7 & \,\,\,\,\,\,4.4$\pm$1.2\,\,\, & \,\,\,\,\,\,8.0$\pm$1.1 & \,\,\,40$\pm$10      &    \,\,\,\,\,\,6.2$\pm$0.7\,\,\, & 3786 & 0.10 & 3.2 & T \\
\end{tabular}}
\tablefoot
{
The coordinates are given at the J2000 Epoch. 
(1) Estimated from the measures at 1.3\,mm. 
(2) Indicates if the source found by \textsl{getsf} was also found by \textsl{GExt2D} (True or False). 
The $^*$ next to the uncertainty indicates that the true value is $<$0.05. 
}
\end{table}
\end{landscape}

%% file: G338-catpap.tex
\begin{landscape}
\begin{table}
\caption{Compact sources extracted by \textsl{getsf} in G338, after smoothing.}
\label{t:core-extraction-smoothed-G338}
\centerline{ 
\addtolength{\tabcolsep}{-2pt}
\begin{tabular}{cccc|cccc|cccc|ccccccc}
\hline\hline 
\noalign{\smallskip}
  &           &             &              &      \multicolumn{4}{c}{1.3\,mm}                                                               &  \multicolumn{4}{c}{3\,mm}                            \\ 
n & Core name & RA          & DEC          & $a  \times b $                                  & PA    &$S^{\rm Peak}$                   &$S^{\rm int}$                 & $a \times b$               & PA    &$S^{\rm Peak}$                 &$S^{\rm int}$            & T    & Mass$^{\,(1)}$& $Size^{\,(1)}$ & $\alpha_{\rm BE}^{\,(1)}$ & $\alpha$ & $\textsl{GExt2D}^{\,(2)}$\\ 
& & [IRCS]                & [ICRS]                 & [$\arcsec\times\arcsec$]            & [deg]              & [mJy.Beam$^{-1}$]           & [mJy]      & [$\arcsec\times\arcsec$]            & [deg]              & [mJy.Beam$^{-1}$]           & [mJy]          & [K]            & [$M_{\odot}$] & [au]   &                   &                      &          &                          \\ 
\noalign{\smallskip}
\hline 
\noalign{\smallskip} 
  1 & 250.14220-45.69340 &  16:40:34.13 & -45:41:36.25 & 0.89$\times$0.79 & \,\,\,84 & \,\,\,177.1$\pm$2.1 & 284.3$\pm$2.6 & 0.94$\times$0.85 & \,\,\,52 & \,\,\,10.0$\pm$0.2\,\,\, & \,\,\,17.0$\pm$0.3 & 100$\pm$50      &    \,\,\,13.2$\pm$0.1\,\,\, & 3214 & 0.10 & 3.4 & T \\ 
  2 & 250.14273-45.69362 &  16:40:34.25 & -45:41:37.04 & 0.97$\times$0.73 & 124 & \,\,\,167.8$\pm$2.4 & 271.5$\pm$2.5 & 1.03$\times$0.77 & 124 & \,\,\,14.9$\pm$0.3\,\,\, & \,\,\,24.9$\pm$0.3 & 100$\pm$50      &    \,\,\,12.6$\pm$0.1\,\,\, & 3130 & 0.10 & 2.9 & T \\ 
  3 & 250.14170-45.70202 &  16:40:34.01 & -45:42:7.290 & 0.86$\times$0.79 & \,\,\,92 & \,\,\,119.8$\pm$2.5 & 168.6$\pm$2.5 & 0.88$\times$0.81 & 110 & \,\,\,\,\,\,5.1$\pm$0.3\,\,\, & \,\,\,\,\,\,6.9$\pm$0.3 & 100$\pm$50      &    \,\,\,\,\,\,7.7$\pm$0.1\,\,\, & 3166 & 0.16 & 3.9 & T \\
\end{tabular}}
\end{table}
\end{landscape}

\begin{landscape}
\begin{table}
\caption{Continuation of Table~\ref{t:core-extraction-smoothed-G338}}
\label{t:core-extraction-smoothed-w51irs2-suite2}
\centerline{
\addtolength{\tabcolsep}{-2pt}
\begin{tabular}{cccc|cccc|cccc|ccccccc}
  &           &             &              &      \multicolumn{4}{c}{1.3\,mm}                                                               &  \multicolumn{4}{c}{3\,mm}                            \\ 
n & Core name & RA          & DEC          & $a  \times b $                                  & PA    &$S^{\rm Peak}$                   &$S^{\rm int}$                 & $a \times b$               & PA    &$S^{\rm Peak}$                 &$S^{\rm int}$            & T    & Mass$^{\,(1)}$& $Size^{\,(1)}$ & $\alpha_{\rm BE}^{\,(1)}$ & $\alpha$ & $\textsl{GExt2D}^{\,(2)}$\\ 
\noalign{\smallskip}
\hline 
\noalign{\smallskip} 
 47 & 250.14219-45.69534 &  16:40:34.12 & -45:41:43.23 & 1.47$\times$1.17 & 106 & \,\,\,\,\,\,\,\,\,3.4$\pm$1.3 & \,\,\,\,\,\,9.7$\pm$1.4 &        -         &      -      &        \,\,\,-      &       \,\,\,\,\,\,-        & \,\,\,27$\pm$07      &    \,\,\,\,\,\,1.8$\pm$0.3\,\,\, & 4930 & 0.30 &  -  & F \\ 
\noalign{\smallskip}
\hline
\noalign{\smallskip}
\end{tabular}}
\tablefoot
{
The coordinates are given at the J2000 Epoch. 
(1) Estimated from the measures at 1.3\,mm. 
(2) Indicates if the source found by \textsl{getsf} was also found by \textsl{GExt2D} (True or False). 
The $^*$ next to the uncertainty indicates that the true value is $<$0.05. 
}
\end{table}
\end{landscape}

%% file: W43-MM1-catpap.tex
\begin{landscape}
\begin{table}
\caption{Compact sources extracted by \textsl{getsf} in W43-MM1, after smoothing.}
\label{t:core-extraction-smoothed-W43-MM1}
\centerline{ 
\addtolength{\tabcolsep}{-2pt}
\begin{tabular}{cccc|cccc|cccc|ccccccc}
\hline\hline 
\noalign{\smallskip}
  &           &             &              &      \multicolumn{4}{c}{1.3\,mm}                                                               &  \multicolumn{4}{c}{3\,mm}                            \\ 
n & Core name & RA          & DEC          & $a  \times b $                                  & PA    &$S^{\rm Peak}$                   &$S^{\rm int}$                 & $a \times b$               & PA    &$S^{\rm Peak}$                 &$S^{\rm int}$            & T    & Mass$^{\,(1)}$& $Size^{\,(1)}$ & $\alpha_{\rm BE}^{\,(1)}$ & $\alpha$ & $\textsl{GExt2D}^{\,(2)}$\\ 
& & [IRCS]                & [ICRS]                 & [$\arcsec\times\arcsec$]            & [deg]              & [mJy.Beam$^{-1}$]           & [mJy]      & [$\arcsec\times\arcsec$]            & [deg]              & [mJy.Beam$^{-1}$]           & [mJy]          & [K]            & [$M_{\odot}$] & [au]   &                   &                      &          &                          \\ 
\noalign{\smallskip}
\hline 
\noalign{\smallskip} 
  1 & 281.94593-1.907470 &  18:47:47.02 & -01:54:26.90 & 0.71$\times$0.62 & \,\,\,88 & \,\,\,360.5$\pm$4.0 & 676.9$\pm$4.8 & 0.72$\times$0.62 & \,\,\,99 & \,\,\,15.6$\pm$0.3\,\,\, & \,\,\,29.4$\pm$0.4 & 100$\pm$50      &    \,\,\,82.8$\pm$0.4\,\,\, & 3572 & 0.02 & 3.8 & T \\ 
  2 & 281.94516-1.908130 &  18:47:46.84 & -01:54:29.28 & 0.61$\times$0.56 & 143 & \,\,\,202.3$\pm$6.8 & 299.3$\pm$7.2 & 0.67$\times$0.54 & 121 & \,\,\,13.1$\pm$0.5\,\,\, & \,\,\,19.2$\pm$0.5 & 100$\pm$50      &    \,\,\,30.5$\pm$0.6\,\,\, & 3180 & 0.04 & 3.3 & T \\ 
  3 & 281.94321-1.909270 &  18:47:46.37 & -01:54:33.38 & 0.72$\times$0.63 & \,\,\,37 & \,\,\,108.9$\pm$2.5 & 223.3$\pm$3.2 & 0.72$\times$0.65 & \,\,\,75 & \,\,\,\,\,\,8.1$\pm$0.2\,\,\, & \,\,\,16.1$\pm$0.2 & 100$\pm$50      &    \,\,\,20.9$\pm$0.3\,\,\, & 3624 & 0.07 & 3.1 & T \\
\end{tabular}}
\end{table}
\end{landscape}

\begin{landscape}
\begin{table}
\caption{Continuation of Table~\ref{t:core-extraction-smoothed-W43-MM1}}
\label{t:core-extraction-smoothed-w51irs2-suite2}
\centerline{
\addtolength{\tabcolsep}{-2pt}
\begin{tabular}{cccc|cccc|cccc|ccccccc}
  &           &             &              &      \multicolumn{4}{c}{1.3\,mm}                                                               &  \multicolumn{4}{c}{3\,mm}                            \\ 
n & Core name & RA          & DEC          & $a  \times b $                                  & PA    &$S^{\rm Peak}$                   &$S^{\rm int}$                 & $a \times b$               & PA    &$S^{\rm Peak}$                 &$S^{\rm int}$            & T    & Mass$^{\,(1)}$& $Size^{\,(1)}$ & $\alpha_{\rm BE}^{\,(1)}$ & $\alpha$ & $\textsl{GExt2D}^{\,(2)}$\\ 
\noalign{\smallskip}
\hline 
\noalign{\smallskip} 
 43 & 281.95581-1.901030 &  18:47:49.39 & -01:54:3.710 & 0.76$\times$0.69 & \,\,\,28 & \,\,\,\,\,\,\,\,\,2.3$\pm$0.3 & \,\,\,\,\,\,5.0$\pm$0.4 &        -         &      -      &        \,\,\,-      &       \,\,\,\,\,\,-        & \,\,\,25$\pm$06      &    \,\,\,\,\,\,2.0$\pm$0.2\,\,\, & 3928 & 0.20 &  -  & F \\ 
 44 & 281.94660-1.905970 &  18:47:47.18 & -01:54:21.49 & 0.56$\times$0.54 & 153 & \,\,\,\,\,\,\,\,\,4.4$\pm$1.3 & \,\,\,\,\,\,5.1$\pm$1.0 &        -         &      -      &        \,\,\,-      &       \,\,\,\,\,\,-        & \,\,\,25$\pm$06      &    \,\,\,\,\,\,2.1$\pm$0.4\,\,\, & 2984 & 0.14 &  -  & T \\ 
 45 & 281.94470-1.904850 &  18:47:46.73 & -01:54:17.47 & 0.62$\times$0.44 & \,\,\,92 & \,\,\,\,\,\,\,\,\,5.5$\pm$1.2 & \,\,\,\,\,\,6.0$\pm$0.9 &        -         &      -      &        \,\,\,-      &       \,\,\,\,\,\,-        & \,\,\,25$\pm$06      &    \,\,\,\,\,\,2.5$\pm$0.4\,\,\, & 2722 & 0.11 &  -  & T \\ 
 46 & 281.94732-1.903710 &  18:47:47.36 & -01:54:13.36 & 0.57$\times$0.54 & 120 & \,\,\,\,\,\,\,\,\,2.7$\pm$0.4 & \,\,\,\,\,\,3.3$\pm$0.3 &        -         &      -      &        \,\,\,-      &       \,\,\,\,\,\,-        & \,\,\,23$\pm$06      &    \,\,\,\,\,\,1.4$\pm$0.1\,\,\, & 3030 & 0.20 &  -  & T \\
\end{tabular}}
\tablefoot
{
The coordinates are given at the J2000 Epoch. 
(1) Estimated from the measures at 1.3\,mm. 
(2) Indicates if the source found by \textsl{getsf} was also found by \textsl{GExt2D} (True or False). 
The $^*$ next to the uncertainty indicates that the true value is $<$0.05. 
}
\end{table}
\end{landscape}

%% file: W43-MM2-catpap.tex
\begin{landscape}
\begin{table}
\caption{Compact sources extracted by \textsl{getsf} in W43-MM2, after smoothing.}
\label{t:core-extraction-smoothed-W43-MM2}
\centerline{ 
\addtolength{\tabcolsep}{-2pt}
\begin{tabular}{cccc|cccc|cccc|ccccccc}
\hline\hline 
\noalign{\smallskip}
  &           &             &              &      \multicolumn{4}{c}{1.3\,mm}                                                               &  \multicolumn{4}{c}{3\,mm}                            \\ 
n & Core name & RA          & DEC          & $a  \times b $                                  & PA    &$S^{\rm Peak}$                   &$S^{\rm int}$                 & $a \times b$               & PA    &$S^{\rm Peak}$                 &$S^{\rm int}$            & T    & Mass$^{\,(1)}$& $Size^{\,(1)}$ & $\alpha_{\rm BE}^{\,(1)}$ & $\alpha$ & $\textsl{GExt2D}^{\,(2)}$\\ 
& & [IRCS]                & [ICRS]                 & [$\arcsec\times\arcsec$]            & [deg]              & [mJy.Beam$^{-1}$]           & [mJy]      & [$\arcsec\times\arcsec$]            & [deg]              & [mJy.Beam$^{-1}$]           & [mJy]          & [K]            & [$M_{\odot}$] & [au]   &                   &                      &          &                          \\ 
\noalign{\smallskip}
\hline 
\noalign{\smallskip} 
  1 & 281.90332-2.015080 &  18:47:36.80 & -02:00:54.27 & 0.76$\times$0.70 & 147 & \,\,\,152.4$\pm$1.1 & 423.6$\pm$3.2 & 1.04$\times$0.52 & \,\,\,\,\,\,9 & \,\,\,\,\,\,6.6$\pm$0.1\,\,\, & \,\,\,24.8$\pm$0.2 & 100$\pm$50      &    \,\,\,42.6$\pm$0.3\,\,\, & 3980 & 0.04 & 3.5 & T \\ 
  2 & 281.90013-2.022430 &  18:47:36.03 & -02:01:20.74 & 0.78$\times$0.48 & \,\,\,84 & \,\,\,\,\,\,13.3$\pm$0.2 & \,\,\,24.5$\pm$0.3 & 0.69$\times$0.51 & \,\,\,84 & \,\,\,\,\,\,1.4$\pm$0.1\,\,\, & \,\,\,\,\,\,2.4$\pm$0.1 & \,\,\,25$\pm$06      &    \,\,\,10.8$\pm$0.1\,\,\, & 3118 & 0.03 & 2.9 & T \\ 
  3 & 281.90042-2.021090 &  18:47:36.10 & -02:01:15.94 & 0.90$\times$0.70 & \,\,\,81 & \,\,\,\,\,\,14.9$\pm$0.5 & \,\,\,41.4$\pm$0.8 & 0.78$\times$0.68 & \,\,\,81 & \,\,\,\,\,\,1.0$\pm$0.1\,\,\, & \,\,\,\,\,\,2.5$\pm$0.1 & \,\,\,24$\pm$06      &    \,\,\,18.7$\pm$0.3\,\,\, & 4200 & 0.02 & 3.5 & T \\
\end{tabular}}
\tablefoot
{
The coordinates are given at the J2000 Epoch. 
(1) Estimated from the measures at 1.3\,mm. 
(2) Indicates if the source found by \textsl{getsf} was also found by \textsl{GExt2D} (True or False). 
The $^*$ next to the uncertainty indicates that the true value is $<$0.05. 
}
\end{table}
\end{landscape}

%% file: W43-MM3-catpap.tex
\begin{landscape}
\begin{table}
\caption{Compact sources extracted by \textsl{getsf} in W43-MM3, after smoothing.}
\label{t:core-extraction-smoothed-W43-MM3}
\centerline{ 
\addtolength{\tabcolsep}{-2pt}
\begin{tabular}{cccc|cccc|cccc|ccccccc}
\hline\hline 
\noalign{\smallskip}
  &           &             &              &      \multicolumn{4}{c}{1.3\,mm}                                                               &  \multicolumn{4}{c}{3\,mm}                            \\ 
n & Core name & RA          & DEC          & $a  \times b $                                  & PA    &$S^{\rm Peak}$                   &$S^{\rm int}$                 & $a \times b$               & PA    &$S^{\rm Peak}$                 &$S^{\rm int}$            & T    & Mass$^{\,(1)}$& $Size^{\,(1)}$ & $\alpha_{\rm BE}^{\,(1)}$ & $\alpha$ & $\textsl{GExt2D}^{\,(2)}$\\ 
& & [IRCS]                & [ICRS]                 & [$\arcsec\times\arcsec$]            & [deg]              & [mJy.Beam$^{-1}$]           & [mJy]      & [$\arcsec\times\arcsec$]            & [deg]              & [mJy.Beam$^{-1}$]           & [mJy]          & [K]            & [$M_{\odot}$] & [au]   &                   &                      &          &                          \\ 
\noalign{\smallskip}
\hline 
\noalign{\smallskip} 
  1 & 281.92380-2.007940 &  18:47:41.71 & -02:00:28.60 & 0.65$\times$0.50 & \,\,\,90 & \,\,\,\,\,\,52.0$\pm$2.0 & \,\,\,82.2$\pm$2.6 & 0.79$\times$0.56 & 103 & \,\,\,\,\,\,2.8$\pm$0.3\,\,\, & \,\,\,\,\,\,4.0$\pm$0.3 & 100$\pm$50      &    \,\,\,\,\,\,7.4$\pm$0.2\,\,\, & 2984 & 0.16 & 3.7 & T \\ 
  2 & 281.91360-2.007800 &  18:47:39.26 & -02:00:28.09 & 0.73$\times$0.57 & \,\,\,57 & \,\,\,\,\,\,17.1$\pm$0.4 & \,\,\,36.1$\pm$0.6 & 1.05$\times$0.87 & 106 & \,\,\,\,\,\,0.6$\pm$0.1$^*$ & \,\,\,\,\,\,1.1$\pm$0.0 & 100$\pm$50      &    \,\,\,\,\,\,3.2$\pm$0.1\,\,\, & 3388 & 0.43 & 4.2 & T \\ 
  3 & 281.92072-2.005760 &  18:47:40.97 & -02:00:20.74 & 0.67$\times$0.65 & \,\,\,20 & \,\,\,\,\,\,\,\,\,7.3$\pm$0.2 & \,\,\,21.8$\pm$0.4 & 0.68$\times$0.59 & 136 & \,\,\,\,\,\,0.4$\pm$0.1$^*$ & \,\,\,\,\,\,0.8$\pm$0.0 & \,\,\,27$\pm$07      &    \,\,\,\,\,\,8.2$\pm$0.1\,\,\, & 3602 & 0.05 & 4.1 & T \\
\end{tabular}}
\tablefoot
{
The coordinates are given at the J2000 Epoch. 
(1) Estimated from the measures at 1.3\,mm. 
(2) Indicates if the source found by \textsl{getsf} was also found by \textsl{GExt2D} (True or False). 
The $^*$ next to the uncertainty indicates that the true value is $<$0.05. 
}
\end{table}
\end{landscape}

%% file: W51-E-catpap.tex
\begin{landscape}
\begin{table}
\caption{Compact sources extracted by \textsl{getsf} in W51-E, after smoothing.}
\label{t:core-extraction-smoothed-W51-E}
\centerline{ 
\addtolength{\tabcolsep}{-2pt}
\begin{tabular}{cccc|cccc|cccc|ccccccc}
\hline\hline 
\noalign{\smallskip}
  &           &             &              &      \multicolumn{4}{c}{1.3\,mm}                                                               &  \multicolumn{4}{c}{3\,mm}                            \\ 
n & Core name & RA          & DEC          & $a  \times b $                                  & PA    &$S^{\rm Peak}$                   &$S^{\rm int}$                 & $a \times b$               & PA    &$S^{\rm Peak}$                 &$S^{\rm int}$            & T    & Mass$^{\,(1)}$& $Size^{\,(1)}$ & $\alpha_{\rm BE}^{\,(1)}$ & $\alpha$ & $\textsl{GExt2D}^{\,(2)}$\\ 
& & [IRCS]                & [ICRS]                 & [$\arcsec\times\arcsec$]            & [deg]              & [mJy.Beam$^{-1}$]           & [mJy]      & [$\arcsec\times\arcsec$]            & [deg]              & [mJy.Beam$^{-1}$]           & [mJy]          & [K]            & [$M_{\odot}$] & [au]   &                   &                      &          &                          \\ 
\noalign{\smallskip}
\hline 
\noalign{\smallskip} 
  2 & 290.93319+14.509590 &  19:23:43.96 &  14:30:34.52 & 1.11$\times$0.66 & 177 & \,\,\,756.4$\pm$9.4 & 2856.0$\pm$18.3 & 0.81$\times$0.51 & 178 & 112.5$\pm$0.8\,\,\, & 419.6$\pm$2.8 & 300$\pm$100      &    \,\,\,94.2$\pm$0.5\,\,\, & 4228 & 0.05 & 2.4 & T \\ 
  4 & 290.93291+14.507830 &  19:23:43.90 &  14:30:28.18 & 0.85$\times$0.73 & \,\,\,\,\,\,6 & \,\,\,531.1$\pm$35.3 & 1238.0$\pm$40.6 & 0.72$\times$0.55 & \,\,\,\,\,\,7 & \,\,\,81.8$\pm$3.5\,\,\, & 211.1$\pm$6.4 & 100$\pm$50      &    193.1$\pm$3.5\,\,\, & 4138 & 0.01 & 2.2 & T \\ 
  8 & 290.93304+14.510150 &  19:23:43.93 &  14:30:36.55 & 1.07$\times$0.71 & 167 & \,\,\,105.7$\pm$16.5 & 307.1$\pm$17.0 & 1.00$\times$0.68 & 167 & \,\,\,13.2$\pm$1.0\,\,\, & \,\,\,34.5$\pm$1.2 & \,\,\,54$\pm$14      &    \,\,\,57.7$\pm$2.8\,\,\, & 4378 & 0.02 & 2.7 & F \\
\end{tabular}}
\tablefoot
{
The coordinates are given at the J2000 Epoch. 
(1) Estimated from the measures at 1.3\,mm. 
(2) Indicates if the source found by \textsl{getsf} was also found by \textsl{GExt2D} (True or False). 
The $^*$ next to the uncertainty indicates that the true value is $<$0.05. 
}
\end{table}
\end{landscape}

%% file: G351-catpap.tex
\begin{landscape}
\begin{table}
\caption{Compact sources extracted by \textsl{getsf} in G351, after smoothing.}
\label{t:core-extraction-smoothed-G351}
\centerline{ 
\addtolength{\tabcolsep}{-2pt}
\begin{tabular}{cccc|cccc|cccc|ccccccc}
\hline\hline 
\noalign{\smallskip}
  &           &             &              &      \multicolumn{4}{c}{1.3\,mm}                                                               &  \multicolumn{4}{c}{3\,mm}                            \\ 
n & Core name & RA          & DEC          & $a  \times b $                                  & PA    &$S^{\rm Peak}$                   &$S^{\rm int}$                 & $a \times b$               & PA    &$S^{\rm Peak}$                 &$S^{\rm int}$            & T    & Mass$^{\,(1)}$& $Size^{\,(1)}$ & $\alpha_{\rm BE}^{\,(1)}$ & $\alpha$ & $\textsl{GExt2D}^{\,(2)}$\\ 
& & [IRCS]                & [ICRS]                 & [$\arcsec\times\arcsec$]            & [deg]              & [mJy.Beam$^{-1}$]           & [mJy]      & [$\arcsec\times\arcsec$]            & [deg]              & [mJy.Beam$^{-1}$]           & [mJy]          & [K]            & [$M_{\odot}$] & [au]   &                   &                      &          &                          \\ 
\noalign{\smallskip}
\hline 
\noalign{\smallskip} 
  1 & 261.67725-36.15480 &  17:26:42.54 & -36:09:17.28 & 2.40$\times$1.79 & \,\,\,91 & \,\,\,743.6$\pm$13.6 & 1817.0$\pm$18.6 & 2.24$\times$1.68 & \,\,\,98 & \,\,\,43.8$\pm$3.8\,\,\, & \,\,\,92.7$\pm$4.3 & 100$\pm$50      &    \,\,\,22.9$\pm$0.2\,\,\, & 3952 & 0.07 & 3.6 & T \\ 
  2 & 261.67677-36.15517 &  17:26:42.43 & -36:09:18.62 & 2.88$\times$2.22 & 126 & \,\,\,824.7$\pm$23.1 & 2877.0$\pm$34.7 & 3.13$\times$2.46 & 120 & \,\,\,59.3$\pm$1.1\,\,\, & 191.7$\pm$1.6 & 100$\pm$50      &    \,\,\,36.5$\pm$0.4\,\,\, & 4836 & 0.05 & 3.3 & T \\ 
  4 & 261.67616-36.15493 &  17:26:42.28 & -36:09:17.76 & 1.62$\times$1.33 & \,\,\,78 & \,\,\,431.4$\pm$31.0 & 490.6$\pm$24.5 & 1.71$\times$1.54 & \,\,\,79 & \,\,\,55.2$\pm$2.0\,\,\, & \,\,\,55.8$\pm$1.3 & \,\,\,45$\pm$11      &    \,\,\,14.9$\pm$0.7\,\,\, & 2840 & 0.03 & 2.6 & T \\
\end{tabular}}
\tablefoot
{
The coordinates are given at the J2000 Epoch. 
(1) Estimated from the measures at 1.3\,mm. 
(2) Indicates if the source found by \textsl{getsf} was also found by \textsl{GExt2D} (True or False). 
The $^*$ next to the uncertainty indicates that the true value is $<$0.05. 
}
\end{table}
\end{landscape}

%% file: G353-catpap.tex
\begin{landscape}
\begin{table}
\caption{Compact sources extracted by \textsl{getsf} in G353, after smoothing.}
\label{t:core-extraction-smoothed-G353}
\centerline{ 
\addtolength{\tabcolsep}{-2pt}
\begin{tabular}{cccc|cccc|cccc|ccccccc}
\hline\hline 
\noalign{\smallskip}
  &           &             &              &      \multicolumn{4}{c}{1.3\,mm}                                                               &  \multicolumn{4}{c}{3\,mm}                            \\ 
n & Core name & RA          & DEC          & $a  \times b $                                  & PA    &$S^{\rm Peak}$                   &$S^{\rm int}$                 & $a \times b$               & PA    &$S^{\rm Peak}$                 &$S^{\rm int}$            & T    & Mass$^{\,(1)}$& $Size^{\,(1)}$ & $\alpha_{\rm BE}^{\,(1)}$ & $\alpha$ & $\textsl{GExt2D}^{\,(2)}$\\ 
& & [IRCS]                & [ICRS]                 & [$\arcsec\times\arcsec$]            & [deg]              & [mJy.Beam$^{-1}$]           & [mJy]      & [$\arcsec\times\arcsec$]            & [deg]              & [mJy.Beam$^{-1}$]           & [mJy]          & [K]            & [$M_{\odot}$] & [au]   &                   &                      &          &                          \\ 
\noalign{\smallskip}
\hline 
\noalign{\smallskip} 
  2 & 262.61650-34.69559 &  17:30:27.96 & -34:41:44.11 & 1.98$\times$1.59 & \,\,\,65 & \,\,\,140.8$\pm$13.0 & 262.9$\pm$14.3 & 1.75$\times$1.51 & \,\,\,48 & \,\,\,18.2$\pm$1.4\,\,\, & \,\,\,33.6$\pm$1.7 & \,\,\,29$\pm$07      &    \,\,\,12.5$\pm$0.6\,\,\, & 3424 & 0.03 & 2.5 & T \\ 
  3 & 262.61842-34.69652 &  17:30:28.42 & -34:41:47.49 & 2.59$\times$1.79 & 146 & \,\,\,136.1$\pm$10.5 & 379.1$\pm$16.0 & 3.64$\times$1.85 & 147 & \,\,\,\,\,\,6.7$\pm$2.2\,\,\, & \,\,\,\,\,\,5.1$\pm$1.8 & 100$\pm$50      &    \,\,\,\,\,\,4.4$\pm$0.2\,\,\, & 4044 & 0.37 & 5.2 & T \\ 
  4 & 262.61032-34.69327 &  17:30:26.48 & -34:41:35.76 & 1.56$\times$1.46 & 104 & \,\,\,\,\,\,55.6$\pm$2.5 & \,\,\,69.9$\pm$2.4 & 1.46$\times$1.30 & \,\,\,97 & \,\,\,\,\,\,7.0$\pm$0.6\,\,\, & \,\,\,\,\,\,8.5$\pm$0.6 & \,\,\,30$\pm$08      &    \,\,\,\,\,\,3.0$\pm$0.1\,\,\, & 2984 & 0.12 & 2.6 & T \\
\end{tabular}}
\end{table}
\end{landscape}

\begin{landscape}
\begin{table}
\caption{Continuation of Table~\ref{t:core-extraction-smoothed-G353}}
\label{t:core-extraction-smoothed-w51irs2-suite2}
\centerline{
\addtolength{\tabcolsep}{-2pt}
\begin{tabular}{cccc|cccc|cccc|ccccccc}
  &           &             &              &      \multicolumn{4}{c}{1.3\,mm}                                                               &  \multicolumn{4}{c}{3\,mm}                            \\ 
n & Core name & RA          & DEC          & $a  \times b $                                  & PA    &$S^{\rm Peak}$                   &$S^{\rm int}$                 & $a \times b$               & PA    &$S^{\rm Peak}$                 &$S^{\rm int}$            & T    & Mass$^{\,(1)}$& $Size^{\,(1)}$ & $\alpha_{\rm BE}^{\,(1)}$ & $\alpha$ & $\textsl{GExt2D}^{\,(2)}$\\ 
\noalign{\smallskip}
\hline 
\noalign{\smallskip} 
 44 & 262.60301-34.69564 &  17:30:24.72 & -34:41:44.31 & 1.76$\times$1.55 & \,\,\,97 & \,\,\,\,\,\,\,\,\,5.9$\pm$1.9 & \,\,\,\,\,\,6.9$\pm$1.9 &        -         &      -      &        \,\,\,-      &       \,\,\,\,\,\,-        & \,\,\,24$\pm$06      &    \,\,\,\,\,\,0.4$\pm$0.1\,\,\, & 3236 & 0.83 &  -  & T \\ 
 45 & 262.61430-34.69094 &  17:30:27.43 & -34:41:27.38 & 4.09$\times$3.38 & \,\,\,67 & \,\,\,\,\,\,\,\,\,5.4$\pm$1.2 & \,\,\,36.5$\pm$2.2 &        -         &      -      &        \,\,\,-      &       \,\,\,\,\,\,-        & \,\,\,23$\pm$06      &    \,\,\,\,\,\,2.1$\pm$0.1\,\,\, & 7204 & 0.33 &  -  & T \\ 
 46 & 262.61876-34.69124 &  17:30:28.50 & -34:41:28.46 & 2.76$\times$2.12 & \,\,\,53 & \,\,\,\,\,\,\,\,\,5.4$\pm$2.1 & \,\,\,12.3$\pm$2.8 &        -         &      -      &        \,\,\,-      &       \,\,\,\,\,\,-        & \,\,\,23$\pm$06      &    \,\,\,\,\,\,0.7$\pm$0.2\,\,\, & 4626 & 0.62 &  -  & F \\ 
 47 & 262.61785-34.69199 &  17:30:28.28 & -34:41:31.18 & 3.17$\times$2.90 & 168 & \,\,\,\,\,\,\,\,\,6.0$\pm$2.6 & \,\,\,22.4$\pm$3.5 &        -         &      -      &        \,\,\,-      &       \,\,\,\,\,\,-        & \,\,\,24$\pm$06      &    \,\,\,\,\,\,1.3$\pm$0.2\,\,\, & 5982 & 0.45 &  -  & F \\
\end{tabular}}
\tablefoot
{
The coordinates are given at the J2000 Epoch. 
(1) Estimated from the measures at 1.3\,mm. 
(2) Indicates if the source found by \textsl{getsf} was also found by \textsl{GExt2D} (True or False). 
The $^*$ next to the uncertainty indicates that the true value is $<$0.05. 
}
\end{table}
\end{landscape}

%% file: G008-catpap.tex
\begin{landscape}
\begin{table}
\caption{Compact sources extracted by \textsl{getsf} in G008, after smoothing.}
\label{t:core-extraction-smoothed-G008}
\centerline{ 
\addtolength{\tabcolsep}{-2pt}
\begin{tabular}{cccc|cccc|cccc|ccccccc}
\hline\hline 
\noalign{\smallskip}
  &           &             &              &      \multicolumn{4}{c}{1.3\,mm}                                                               &  \multicolumn{4}{c}{3\,mm}                            \\ 
n & Core name & RA          & DEC          & $a  \times b $                                  & PA    &$S^{\rm Peak}$                   &$S^{\rm int}$                 & $a \times b$               & PA    &$S^{\rm Peak}$                 &$S^{\rm int}$            & T    & Mass$^{\,(1)}$& $Size^{\,(1)}$ & $\alpha_{\rm BE}^{\,(1)}$ & $\alpha$ & $\textsl{GExt2D}^{\,(2)}$\\ 
& & [IRCS]                & [ICRS]                 & [$\arcsec\times\arcsec$]            & [deg]              & [mJy.Beam$^{-1}$]           & [mJy]      & [$\arcsec\times\arcsec$]            & [deg]              & [mJy.Beam$^{-1}$]           & [mJy]          & [K]            & [$M_{\odot}$] & [au]   &                   &                      &          &                          \\ 
\noalign{\smallskip}
\hline 
\noalign{\smallskip} 
  2 & 271.59783-21.61959 &  18:06:23.48 & -21:37:10.53 & 0.96$\times$0.87 & \,\,\,92 & \,\,\,\,\,\,70.4$\pm$2.1 & \,\,\,98.5$\pm$2.6 & 1.06$\times$0.98 & \,\,\,93 & \,\,\,\,\,\,5.4$\pm$0.2\,\,\, & \,\,\,\,\,\,7.5$\pm$0.2 & 100$\pm$50      &    \,\,\,\,\,\,3.3$\pm$0.1\,\,\, & 3044 & 0.36 & 3.1 & T \\ 
  3 & 271.57961-21.62503 &  18:06:19.11 & -21:37:30.13 & 0.90$\times$0.79 & \,\,\,71 & \,\,\,\,\,\,71.9$\pm$8.4 & \,\,\,76.7$\pm$6.7 & 0.91$\times$0.81 & \,\,\,52 & \,\,\,\,\,\,7.3$\pm$1.0\,\,\, & \,\,\,\,\,\,7.6$\pm$0.7 & \,\,\,34$\pm$09      &    \,\,\,\,\,\,8.7$\pm$0.7\,\,\, & 2804 & 0.04 & 2.8 & T \\ 
  4 & 271.58076-21.62371 &  18:06:19.38 & -21:37:25.35 & 1.42$\times$1.36 & 168 & \,\,\,\,\,\,15.4$\pm$1.0 & \,\,\,41.7$\pm$1.2 &        -         &      -      &        \,\,\,-      &       \,\,\,\,\,\,-        & \,\,\,28$\pm$07      &    \,\,\,\,\,\,5.7$\pm$0.2\,\,\, & 4682 & 0.09 &  -  & T \\
\end{tabular}}
\tablefoot
{
The coordinates are given at the J2000 Epoch. 
(1) Estimated from the measures at 1.3\,mm. 
(2) Indicates if the source found by \textsl{getsf} was also found by \textsl{GExt2D} (True or False). 
The $^*$ next to the uncertainty indicates that the true value is $<$0.05. 
}
\end{table}
\end{landscape}

%% file: G010-catpap.tex
\begin{landscape}
\begin{table}
\caption{Compact sources extracted by \textsl{getsf} in G010, after smoothing.}
\label{t:core-extraction-smoothed-G010}
\centerline{ 
\addtolength{\tabcolsep}{-2pt}
\begin{tabular}{cccc|cccc|cccc|ccccccc}
\hline\hline 
\noalign{\smallskip}
  &           &             &              &      \multicolumn{4}{c}{1.3\,mm}                                                               &  \multicolumn{4}{c}{3\,mm}                            \\ 
n & Core name & RA          & DEC          & $a  \times b $                                  & PA    &$S^{\rm Peak}$                   &$S^{\rm int}$                 & $a \times b$               & PA    &$S^{\rm Peak}$                 &$S^{\rm int}$            & T    & Mass$^{\,(1)}$& $Size^{\,(1)}$ & $\alpha_{\rm BE}^{\,(1)}$ & $\alpha$ & $\textsl{GExt2D}^{\,(2)}$\\ 
& & [IRCS]                & [ICRS]                 & [$\arcsec\times\arcsec$]            & [deg]              & [mJy.Beam$^{-1}$]           & [mJy]      & [$\arcsec\times\arcsec$]            & [deg]              & [mJy.Beam$^{-1}$]           & [mJy]          & [K]            & [$M_{\odot}$] & [au]   &                   &                      &          &                          \\ 
\noalign{\smallskip}
\hline 
\noalign{\smallskip} 
  2 & 272.61336-19.92994 &  18:10:27.21 & -19:55:47.78 & 1.44$\times$0.88 & \,\,\,75 & \,\,\,\,\,\,\,\,\,0.7$\pm$0.3 & \,\,\,\,\,\,1.8$\pm$0.5 &        -         &      -      &        \,\,\,-      &       \,\,\,\,\,\,-        & \,\,\,29$\pm$07      &    \,\,\,\,\,\,0.5$\pm$0.1\,\,\, & 5136 & 1.23 &  -  & F \\ 
  3 & 272.61586-19.92843 &  18:10:27.81 & -19:55:42.35 & 1.09$\times$1.03 & \,\,\,66 & \,\,\,\,\,\,\,\,\,1.2$\pm$0.5 & \,\,\,\,\,\,3.2$\pm$0.7 &        -         &      -      &        \,\,\,-      &       \,\,\,\,\,\,-        & \,\,\,33$\pm$08      &    \,\,\,\,\,\,0.7$\pm$0.2\,\,\, & 5198 & 0.92 &  -  & F \\ 
  4 & 272.62169-19.92914 &  18:10:29.21 & -19:55:44.92 & 0.71$\times$0.62 & \,\,\,97 & \,\,\,\,\,\,22.5$\pm$2.0 & \,\,\,31.3$\pm$1.7 & 0.74$\times$0.60 & \,\,\,66 & \,\,\,\,\,\,2.3$\pm$0.5\,\,\, & \,\,\,\,\,\,3.3$\pm$0.4 & \,\,\,36$\pm$09      &    \,\,\,\,\,\,6.8$\pm$0.4\,\,\, & 3206 & 0.07 & 2.7 & T \\
\end{tabular}}
\end{table}
\end{landscape}

\begin{landscape}
\begin{table}
\caption{Continuation of Table~\ref{t:core-extraction-smoothed-G010}}
\label{t:core-extraction-smoothed-w51irs2-suite2}
\centerline{
\addtolength{\tabcolsep}{-2pt}
\begin{tabular}{cccc|cccc|cccc|ccccccc}
  &           &             &              &      \multicolumn{4}{c}{1.3\,mm}                                                               &  \multicolumn{4}{c}{3\,mm}                            \\ 
n & Core name & RA          & DEC          & $a  \times b $                                  & PA    &$S^{\rm Peak}$                   &$S^{\rm int}$                 & $a \times b$               & PA    &$S^{\rm Peak}$                 &$S^{\rm int}$            & T    & Mass$^{\,(1)}$& $Size^{\,(1)}$ & $\alpha_{\rm BE}^{\,(1)}$ & $\alpha$ & $\textsl{GExt2D}^{\,(2)}$\\ 
\noalign{\smallskip}
\hline 
\noalign{\smallskip} 
 62 & 272.61455-19.92407 &  18:10:27.49 & -19:55:26.65 & 1.04$\times$0.83 & 156 & \,\,\,\,\,\,\,\,\,1.3$\pm$0.5 & \,\,\,\,\,\,2.9$\pm$0.6 &        -         &      -      &        \,\,\,-      &       \,\,\,\,\,\,-        & \,\,\,25$\pm$06      &    \,\,\,\,\,\,0.9$\pm$0.2\,\,\, & 4432 & 0.49 &  -  & T \\ 
\noalign{\smallskip} 
\hline 
\noalign{\smallskip} 
  1 & 272.61942-19.93045 &  18:10:28.66 & -19:55:49.63 & 1.51$\times$0.94 & 128 & \,\,\,448.8$\pm$9.5 & 2115.0$\pm$17.4 & 1.27$\times$0.86 & 132 & 498.4$\pm$4.5\,\,\, & 2603.5$\pm$10.7 &  -      &    -   & 5444 & -  & -0.3 & T \\ 
 10 & 272.62004-19.93072 &  18:10:28.81 & -19:55:50.59 & 1.17$\times$0.87 & 109 & \,\,\,\,\,\,30.9$\pm$15.0 & \,\,\,83.9$\pm$14.7 & 1.11$\times$0.97 & 107 & \,\,\,26.5$\pm$6.0\,\,\, & \,\,\,72.4$\pm$5.6 &  -      &    -   & 4768 & -  & 0.2 & T \\ 
 12 & 272.62021-19.93024 &  18:10:28.85 & -19:55:48.88 & 0.79$\times$0.71 & 161 & \,\,\,\,\,\,15.7$\pm$4.7 & \,\,\,20.8$\pm$4.0 & 0.81$\times$0.74 & 148 & \,\,\,20.7$\pm$4.1\,\,\, & \,\,\,30.8$\pm$3.3 &  -      &    -   & 3644 & -  & -0.5 & T \\ 
 13 & 272.61756-19.92881 &  18:10:28.21 & -19:55:43.73 & 1.59$\times$1.01 & \,\,\,44 & \,\,\,\,\,\,\,\,\,7.9$\pm$1.6 & \,\,\,35.3$\pm$2.2 & 1.51$\times$0.99 & \,\,\,51 & \,\,\,\,\,\,8.4$\pm$0.6\,\,\, & \,\,\,38.3$\pm$1.0 &  -      &    -   & 5824 & -  & -0.1 & T \\
\end{tabular}}
\tablefoot
{
The coordinates are given at the J2000 Epoch. 
(1) Estimated from the measures at 1.3\,mm. 
(2) Indicates if the source found by \textsl{getsf} was also found by \textsl{GExt2D} (True or False). 
The $^*$ next to the uncertainty indicates that the true value is $<$0.05. 
}
\end{table}
\end{landscape}

%% file: G012-catpap.tex
\begin{landscape}
\begin{table}
\caption{Compact sources extracted by \textsl{getsf} in G012, after smoothing.}
\label{t:core-extraction-smoothed-G012}
\centerline{ 
\addtolength{\tabcolsep}{-2pt}
\begin{tabular}{cccc|cccc|cccc|ccccccc}
\hline\hline 
\noalign{\smallskip}
  &           &             &              &      \multicolumn{4}{c}{1.3\,mm}                                                               &  \multicolumn{4}{c}{3\,mm}                            \\ 
n & Core name & RA          & DEC          & $a  \times b $                                  & PA    &$S^{\rm Peak}$                   &$S^{\rm int}$                 & $a \times b$               & PA    &$S^{\rm Peak}$                 &$S^{\rm int}$            & T    & Mass$^{\,(1)}$& $Size^{\,(1)}$ & $\alpha_{\rm BE}^{\,(1)}$ & $\alpha$ & $\textsl{GExt2D}^{\,(2)}$\\ 
& & [IRCS]                & [ICRS]                 & [$\arcsec\times\arcsec$]            & [deg]              & [mJy.Beam$^{-1}$]           & [mJy]      & [$\arcsec\times\arcsec$]            & [deg]              & [mJy.Beam$^{-1}$]           & [mJy]          & [K]            & [$M_{\odot}$] & [au]   &                   &                      &          &                          \\ 
\noalign{\smallskip}
\hline 
\noalign{\smallskip} 
  1 & 273.54933-17.92568 &  18:14:11.84 & -17:55:32.45 & 1.60$\times$1.17 & \,\,\,57 & \,\,\,243.9$\pm$4.5 & 363.3$\pm$4.2 & 1.96$\times$1.61 & \,\,\,59 & \,\,\,27.5$\pm$0.7\,\,\, & \,\,\,25.4$\pm$0.4 & 100$\pm$50      &    \,\,\,\,\,\,6.2$\pm$0.1\,\,\, & 3122 & 0.20 & 3.2 & T \\ 
  3 & 273.55735-17.92251 &  18:14:13.76 & -17:55:21.04 & 1.84$\times$1.52 & \,\,\,\,\,\,8 & \,\,\,135.6$\pm$4.8 & 315.4$\pm$6.2 & 1.57$\times$1.37 & \,\,\,\,\,\,3 & \,\,\,11.8$\pm$4.8\,\,\, & \,\,\,13.3$\pm$5.8 & 100$\pm$50      &    \,\,\,\,\,\,5.3$\pm$0.1\,\,\, & 3898 & 0.29 & 3.9 & T \\ 
  4 & 273.54443-17.93753 &  18:14:10.66 & -17:56:15.12 & 1.41$\times$1.09 & \,\,\,92 & \,\,\,\,\,\,37.6$\pm$1.7 & \,\,\,43.7$\pm$1.7 & 1.59$\times$1.31 & \,\,\,77 & \,\,\,\,\,\,3.4$\pm$0.2\,\,\, & \,\,\,\,\,\,2.7$\pm$0.1 & \,\,\,27$\pm$07      &    \,\,\,\,\,\,3.1$\pm$0.1\,\,\, & 2850 & 0.10 & 3.4 & T \\
\end{tabular}}
\end{table}
\end{landscape}

\begin{landscape}
\begin{table}
\caption{Continuation of Table~\ref{t:core-extraction-smoothed-G012}}
\label{t:core-extraction-smoothed-w51irs2-suite2}
\centerline{
\addtolength{\tabcolsep}{-2pt}
\begin{tabular}{cccc|cccc|cccc|ccccccc}
  &           &             &              &      \multicolumn{4}{c}{1.3\,mm}                                                               &  \multicolumn{4}{c}{3\,mm}                            \\ 
n & Core name & RA          & DEC          & $a  \times b $                                  & PA    &$S^{\rm Peak}$                   &$S^{\rm int}$                 & $a \times b$               & PA    &$S^{\rm Peak}$                 &$S^{\rm int}$            & T    & Mass$^{\,(1)}$& $Size^{\,(1)}$ & $\alpha_{\rm BE}^{\,(1)}$ & $\alpha$ & $\textsl{GExt2D}^{\,(2)}$\\ 
\noalign{\smallskip}
\hline 
\noalign{\smallskip} 
 61 & 273.54755-17.92664 &  18:14:11.41 & -17:55:35.90 & 1.69$\times$1.04 & \,\,\,60 & \,\,\,\,\,\,16.5$\pm$7.7 & \,\,\,19.6$\pm$6.0 &        -         &      -      &        \,\,\,-      &       \,\,\,\,\,\,-        & \,\,\,29$\pm$07      &    \,\,\,\,\,\,1.2$\pm$0.4\,\,\, & 2938 & 0.28 &  -  & T \\ 
 65 & 273.55977-17.91869 &  18:14:14.34 & -17:55:7.270 & 1.70$\times$1.52 & \,\,\,51 & \,\,\,\,\,\,\,\,\,6.9$\pm$2.7 & \,\,\,12.6$\pm$2.7 &        -         &      -      &        \,\,\,-      &       \,\,\,\,\,\,-        & \,\,\,34$\pm$09      &    \,\,\,\,\,\,0.7$\pm$0.1\,\,\, & 3792 & 0.75 &  -  & T \\ 
 66 & 273.54772-17.93073 &  18:14:11.45 & -17:55:50.62 & 2.13$\times$1.44 & \,\,\,75 & \,\,\,\,\,\,\,\,\,5.0$\pm$1.3 & \,\,\,11.4$\pm$1.7 &        -         &      -      &        \,\,\,-      &       \,\,\,\,\,\,-        & \,\,\,30$\pm$08      &    \,\,\,\,\,\,0.7$\pm$0.1\,\,\, & 3934 & 0.68 &  -  & T \\ 
\noalign{\smallskip} 
\hline 
\noalign{\smallskip} 
  2 & 273.55762-17.92916 &  18:14:13.83 & -17:55:44.97 & 2.74$\times$1.40 & \,\,\,95 & \,\,\,335.2$\pm$43.5 & 737.2$\pm$48.2 & 2.95$\times$1.62 & \,\,\,92 & 582.1$\pm$54.0 & 1054.4$\pm$47.6 &  -      &    -   & 4202 & -  & -0.4 & T \\
\end{tabular}}
\tablefoot
{
The coordinates are given at the J2000 Epoch. 
(1) Estimated from the measures at 1.3\,mm. 
(2) Indicates if the source found by \textsl{getsf} was also found by \textsl{GExt2D} (True or False). 
The $^*$ next to the uncertainty indicates that the true value is $<$0.05. 
}
\end{table}
\end{landscape}

%% file: G333-catpap.tex
\begin{landscape}
\begin{table}
\caption{Compact sources extracted by \textsl{getsf} in G333, after smoothing.}
\label{t:core-extraction-smoothed-G333}
\centerline{ 
\addtolength{\tabcolsep}{-2pt}
\begin{tabular}{cccc|cccc|cccc|ccccccc}
\hline\hline 
\noalign{\smallskip}
  &           &             &              &      \multicolumn{4}{c}{1.3\,mm}                                                               &  \multicolumn{4}{c}{3\,mm}                            \\ 
n & Core name & RA          & DEC          & $a  \times b $                                  & PA    &$S^{\rm Peak}$                   &$S^{\rm int}$                 & $a \times b$               & PA    &$S^{\rm Peak}$                 &$S^{\rm int}$            & T    & Mass$^{\,(1)}$& $Size^{\,(1)}$ & $\alpha_{\rm BE}^{\,(1)}$ & $\alpha$ & $\textsl{GExt2D}^{\,(2)}$\\ 
& & [IRCS]                & [ICRS]                 & [$\arcsec\times\arcsec$]            & [deg]              & [mJy.Beam$^{-1}$]           & [mJy]      & [$\arcsec\times\arcsec$]            & [deg]              & [mJy.Beam$^{-1}$]           & [mJy]          & [K]            & [$M_{\odot}$] & [au]   &                   &                      &          &                          \\ 
\noalign{\smallskip}
\hline 
\noalign{\smallskip} 
  1 & 245.54680-50.09881 &  16:22:11.23 & -50:05:55.71 & 1.41$\times$0.90 & \,\,\,67 & \,\,\,117.0$\pm$3.4 & 350.6$\pm$5.1 & 1.09$\times$0.94 & 147 & \,\,\,12.4$\pm$3.8\,\,\, & \,\,\,15.7$\pm$6.8 & \,\,\,49$\pm$12      &    \,\,\,42.5$\pm$0.5\,\,\, & 4402 & 0.02 & 3.8 & T \\ 
  2 & 245.53562-50.10481 &  16:22:8.550 & -50:06:17.33 & 0.94$\times$0.75 & \,\,\,67 & \,\,\,\,\,\,50.6$\pm$2.1 & \,\,\,91.3$\pm$2.3 & 1.22$\times$0.77 & \,\,\,51 & \,\,\,\,\,\,2.9$\pm$2.2\,\,\, & \,\,\,\,\,\,2.8$\pm$1.7 & \,\,\,41$\pm$10      &    \,\,\,13.0$\pm$0.3\,\,\, & 3392 & 0.04 & 4.3 & T \\ 
  3 & 245.52839-50.10507 &  16:22:6.810 & -50:06:18.24 & 0.83$\times$0.79 & \,\,\,88 & \,\,\,\,\,\,23.3$\pm$2.1 & \,\,\,37.1$\pm$2.1 & 0.81$\times$0.71 & 110 & \,\,\,\,\,\,1.8$\pm$0.2\,\,\, & \,\,\,\,\,\,2.7$\pm$0.2 & \,\,\,36$\pm$09      &    \,\,\,\,\,\,5.8$\pm$0.3\,\,\, & 3368 & 0.08 & 3.2 & T \\
\end{tabular}}
\end{table}
\end{landscape}

\begin{landscape}
\begin{table}
\caption{Continuation of Table~\ref{t:core-extraction-smoothed-G333}}
\label{t:core-extraction-smoothed-w51irs2-suite2}
\centerline{
\addtolength{\tabcolsep}{-2pt}
\begin{tabular}{cccc|cccc|cccc|ccccccc}
  &           &             &              &      \multicolumn{4}{c}{1.3\,mm}                                                               &  \multicolumn{4}{c}{3\,mm}                            \\ 
n & Core name & RA          & DEC          & $a  \times b $                                  & PA    &$S^{\rm Peak}$                   &$S^{\rm int}$                 & $a \times b$               & PA    &$S^{\rm Peak}$                 &$S^{\rm int}$            & T    & Mass$^{\,(1)}$& $Size^{\,(1)}$ & $\alpha_{\rm BE}^{\,(1)}$ & $\alpha$ & $\textsl{GExt2D}^{\,(2)}$\\ 
\noalign{\smallskip}
\hline 
\noalign{\smallskip} 
 85 & 245.51446-50.10667 &  16:22:3.470 & -50:06:24.02 & 1.27$\times$1.18 & \,\,\,97 & \,\,\,\,\,\,\,\,\,1.6$\pm$0.3 & \,\,\,\,\,\,5.5$\pm$0.3 &        -         &      -      &        \,\,\,-      &       \,\,\,\,\,\,-        & \,\,\,25$\pm$06      &    \,\,\,\,\,\,1.3$\pm$0.1\,\,\, & 5068 & 0.39 &  -  & F \\ 
 88 & 245.54876-50.10451 &  16:22:11.70 & -50:06:16.24 & 0.94$\times$0.78 & \,\,\,17 & \,\,\,\,\,\,\,\,\,1.7$\pm$0.4 & \,\,\,\,\,\,2.8$\pm$0.4 &        -         &      -      &        \,\,\,-      &       \,\,\,\,\,\,-        & \,\,\,32$\pm$08      &    \,\,\,\,\,\,0.5$\pm$0.1\,\,\, & 3500 & 0.91 &  -  & T \\ 
 91 & 245.55637-50.10999 &  16:22:13.53 & -50:06:35.96 & 1.05$\times$0.80 & \,\,\,42 & \,\,\,\,\,\,\,\,\,1.0$\pm$0.2 & \,\,\,\,\,\,2.3$\pm$0.3 &        -         &      -      &        \,\,\,-      &       \,\,\,\,\,\,-        & \,\,\,26$\pm$07      &    \,\,\,\,\,\,0.5$\pm$0.1\,\,\, & 3668 & 0.73 &  -  & F \\ 
 94 & 245.55462-50.10766 &  16:22:13.11 & -50:06:27.56 & 1.17$\times$1.08 & \,\,\,49 & \,\,\,\,\,\,\,\,\,0.9$\pm$0.3 & \,\,\,\,\,\,2.8$\pm$0.4 &        -         &      -      &        \,\,\,-      &       \,\,\,\,\,\,-        & \,\,\,26$\pm$07      &    \,\,\,\,\,\,0.6$\pm$0.1\,\,\, & 4664 & 0.80 &  -  & F \\
\end{tabular}}
\end{table}
\end{landscape}

\begin{landscape}
\begin{table}
\caption{Continuation of Table~\ref{t:core-extraction-smoothed-G333}}
\label{t:core-extraction-smoothed-w51irs2-suite2}
\centerline{
\addtolength{\tabcolsep}{-2pt}
\begin{tabular}{cccc|cccc|cccc|cccccc}
  &           &             &              &      \multicolumn{4}{c}{1.3\,mm}                                                               &  \multicolumn{4}{c}{3\,mm}                            \\ 
n & Core name & RA          & DEC          & $a  \times b $                                  & PA    &$S^{\rm Peak}$                   &$S^{\rm int}$                 & $a \times b$               & PA    &$S^{\rm Peak}$                 &$S^{\rm int}$            & T    & Mass$^{\,(1)}$& $Size^{\,(1)}$ & $\alpha_{\rm BE}^{\,(1)}$ & $\alpha$ & $\textsl{GExt2D}^{\,(2)}$\\ 
\noalign{\smallskip}
\hline 
\noalign{\smallskip} 
 86 & 245.54286-50.09718 &  16:22:10.29 & -50:05:49.83 & 0.94$\times$0.84 & \,\,\,74 & \,\,\,\,\,\,\,\,\,4.9$\pm$2.2 & \,\,\,\,\,\,8.5$\pm$1.9 & 1.07$\times$0.99 & 151 & \,\,\,\,\,\,5.0$\pm$2.3\,\,\, & \,\,\,\,\,\,7.5$\pm$1.5 &  -      &    -   & 3662 & -  & 0.1 & F \\ 
 90 & 245.53378-50.10306 &  16:22:8.110 & -50:06:11.02 & 0.96$\times$0.80 & 173 & \,\,\,\,\,\,\,\,\,3.4$\pm$1.0 & \,\,\,\,\,\,5.2$\pm$0.9 &        -         &      -      &        \,\,\,-      &       \,\,\,\,\,\,-        &      -      &    -  & 3554 & -  &  -  & F \\ 
 93 & 245.54375-50.10809 &  16:22:10.50 & -50:06:29.12 & 1.29$\times$0.94 & \,\,\,84 & \,\,\,\,\,\,\,\,\,1.8$\pm$0.9 & \,\,\,\,\,\,2.8$\pm$0.9 & 1.23$\times$0.92 & \,\,\,74 & \,\,\,\,\,\,2.5$\pm$1.1\,\,\, & \,\,\,\,\,\,5.5$\pm$1.2 &  -      &    -   & 4382 & -  & -0.8 & F \\ 
 95 & 245.52211-50.10055 &  16:22:5.310 & -50:06:1.960 & 1.28$\times$0.99 & 117 & \,\,\,\,\,\,\,\,\,1.6$\pm$0.7 & \,\,\,\,\,\,4.5$\pm$0.9 & 1.80$\times$1.46 & \,\,\,95 & \,\,\,\,\,\,1.5$\pm$0.8\,\,\, & \,\,\,\,\,\,3.3$\pm$0.4 &  -      &    -   & 4518 & -  & 0.4 & F \\
\end{tabular}}
\tablefoot
{
The coordinates are given at the J2000 Epoch. 
(1) Estimated from the measures at 1.3\,mm. 
(2) Indicates if the source found by \textsl{getsf} was also found by \textsl{GExt2D} (True or False). 
The $^*$ next to the uncertainty indicates that the true value is $<$0.05. 
}
\end{table}
\end{landscape}

%% file: W51-IRS2-catpap.tex
\begin{landscape}
\begin{table}
\caption{Compact sources extracted by \textsl{getsf} in W51-IRS2, after smoothing.}
\label{t:core-extraction-smoothed-W51-IRS2}
\centerline{ 
\addtolength{\tabcolsep}{-2pt}
\begin{tabular}{cccc|cccc|cccc|ccccccc}
\hline\hline 
\noalign{\smallskip}
  &           &             &              &      \multicolumn{4}{c}{1.3\,mm}                                                               &  \multicolumn{4}{c}{3\,mm}                            \\ 
n & Core name & RA          & DEC          & $a  \times b $                                  & PA    &$S^{\rm Peak}$                   &$S^{\rm int}$                 & $a \times b$               & PA    &$S^{\rm Peak}$                 &$S^{\rm int}$            & T    & Mass$^{\,(1)}$& $Size^{\,(1)}$ & $\alpha_{\rm BE}^{\,(1)}$ & $\alpha$ & $\textsl{GExt2D}^{\,(2)}$\\ 
& & [IRCS]                & [ICRS]                 & [$\arcsec\times\arcsec$]            & [deg]              & [mJy.Beam$^{-1}$]           & [mJy]      & [$\arcsec\times\arcsec$]            & [deg]              & [mJy.Beam$^{-1}$]           & [mJy]          & [K]            & [$M_{\odot}$] & [au]   &                   &                      &          &                          \\ 
\noalign{\smallskip}
\hline 
\noalign{\smallskip} 
  1 & 290.91687+14.518190 &  19:23:40.05 &  14:31:5.480 & 0.72$\times$0.66 & \,\,\,89 & \,\,\,886.9$\pm$16.1 & 1816.0$\pm$21.5 & 0.64$\times$0.55 & \,\,\,74 & \,\,\,82.0$\pm$8.4\,\,\, & 151.4$\pm$11.8 & 300$\pm$100      &    \,\,\,63.0$\pm$0.6\,\,\, & 3664 & 0.07 & 3.1 & T \\ 
  2 & 290.91647+14.518160 &  19:23:39.95 &  14:31:5.370 & 0.63$\times$0.59 & 106 & \,\,\,346.2$\pm$15.1 & 489.5$\pm$12.8 & 0.58$\times$0.52 & \,\,\,66 & \,\,\,40.6$\pm$8.6\,\,\, & \,\,\,56.5$\pm$8.6 & 100$\pm$50      &    \,\,\,56.8$\pm$1.1\,\,\, & 3248 & 0.02 & 2.7 & T \\ 
  4 & 290.91072+14.511610 &  19:23:38.57 &  14:30:41.78 & 0.74$\times$0.61 & 150 & \,\,\,\,\,\,70.6$\pm$1.2 & 193.6$\pm$2.7 & 0.68$\times$0.60 & 118 & \,\,\,\,\,\,4.7$\pm$0.2\,\,\, & \,\,\,\,\,\,9.6$\pm$0.3 & 100$\pm$50      &    \,\,\,17.2$\pm$0.2\,\,\, & 3502 & 0.08 & 3.7 & T \\
\end{tabular}}
\end{table}
\end{landscape}

\begin{landscape}
\begin{table}
\caption{Continuation of Table~\ref{t:core-extraction-smoothed-W51-IRS2}}
\label{t:core-extraction-smoothed-w51irs2-suite2}
\centerline{
\addtolength{\tabcolsep}{-2pt}
\begin{tabular}{cccc|cccc|cccc|ccccccc}
  &           &             &              &      \multicolumn{4}{c}{1.3\,mm}                                                               &  \multicolumn{4}{c}{3\,mm}                            \\ 
n & Core name & RA          & DEC          & $a  \times b $                                  & PA    &$S^{\rm Peak}$                   &$S^{\rm int}$                 & $a \times b$               & PA    &$S^{\rm Peak}$                 &$S^{\rm int}$            & T    & Mass$^{\,(1)}$& $Size^{\,(1)}$ & $\alpha_{\rm BE}^{\,(1)}$ & $\alpha$ & $\textsl{GExt2D}^{\,(2)}$\\ 
\noalign{\smallskip}
\hline 
\noalign{\smallskip} 
 49 & 290.91144+14.512710 &  19:23:38.75 &  14:30:45.75 & 0.84$\times$0.76 & 102 & \,\,\,\,\,\,\,\,\,5.3$\pm$0.7 & \,\,\,13.7$\pm$0.9 &        -         &      -      &        \,\,\,-      &       \,\,\,\,\,\,-        & \,\,\,31$\pm$08      &    \,\,\,\,\,\,4.3$\pm$0.3\,\,\, & 4246 & 0.12 &  -  & T \\ 
 51 & 290.91169+14.510730 &  19:23:38.81 &  14:30:38.64 & 0.63$\times$0.50 & 124 & \,\,\,\,\,\,\,\,\,5.6$\pm$0.8 & \,\,\,\,\,\,6.8$\pm$0.6 &        -         &      -      &        \,\,\,-      &       \,\,\,\,\,\,-        & \,\,\,28$\pm$07      &    \,\,\,\,\,\,2.4$\pm$0.2\,\,\, & 2900 & 0.14 &  -  & T \\ 
 52 & 290.91681+14.514650 &  19:23:40.03 &  14:30:52.75 & 0.66$\times$0.57 & \,\,\,57 & \,\,\,\,\,\,\,\,\,3.7$\pm$0.4 & \,\,\,\,\,\,5.1$\pm$0.4 &        -         &      -      &        \,\,\,-      &       \,\,\,\,\,\,-        & \,\,\,28$\pm$07      &    \,\,\,\,\,\,1.7$\pm$0.1\,\,\, & 3222 & 0.21 &  -  & T \\ 
 53 & 290.91961+14.506820 &  19:23:40.71 &  14:30:24.55 & 0.55$\times$0.50 & 156 & \,\,\,\,\,\,\,\,\,3.7$\pm$0.4 & \,\,\,\,\,\,3.9$\pm$0.3 &        -         &      -      &        \,\,\,-      &       \,\,\,\,\,\,-        & \,\,\,28$\pm$07      &    \,\,\,\,\,\,1.4$\pm$0.1\,\,\, & 2772 & 0.23 &  -  & T \\
\end{tabular}}
\end{table}
\end{landscape}

\begin{landscape}
\begin{table}
\caption{Continuation of Table~\ref{t:core-extraction-smoothed-W51-IRS2}}
\label{t:core-extraction-smoothed-w51irs2-suite2}
\centerline{
\addtolength{\tabcolsep}{-2pt}
\begin{tabular}{cccc|cccc|cccc|ccccccc}
  &           &             &              &      \multicolumn{4}{c}{1.3\,mm}                                                               &  \multicolumn{4}{c}{3\,mm}                            \\ 
n & Core name & RA          & DEC          & $a  \times b $                                  & PA    &$S^{\rm Peak}$                   &$S^{\rm int}$                 & $a \times b$               & PA    &$S^{\rm Peak}$                 &$S^{\rm int}$            & T    & Mass$^{\,(1)}$& $Size^{\,(1)}$ & $\alpha_{\rm BE}^{\,(1)}$ & $\alpha$ & $\textsl{GExt2D}^{\,(2)}$\\ 
\noalign{\smallskip}
\hline 
\noalign{\smallskip} 
 99 & 290.92043+14.513530 &  19:23:40.90 &  14:30:48.70 & 0.66$\times$0.63 & \,\,\,65 & \,\,\,\,\,\,\,\,\,1.9$\pm$0.6 & \,\,\,\,\,\,2.8$\pm$0.5 &        -         &      -      &        \,\,\,-      &       \,\,\,\,\,\,-        & \,\,\,29$\pm$07      &    \,\,\,\,\,\,0.9$\pm$0.2\,\,\, & 3454 & 0.43 &  -  & T \\ 
100 & 290.90944+14.509660 &  19:23:38.26 &  14:30:34.78 & 0.64$\times$0.59 & \,\,\,53 & \,\,\,\,\,\,\,\,\,2.7$\pm$1.1 & \,\,\,\,\,\,3.7$\pm$0.8 &        -         &      -      &        \,\,\,-      &       \,\,\,\,\,\,-        & \,\,\,29$\pm$07      &    \,\,\,\,\,\,1.2$\pm$0.3\,\,\, & 3292 & 0.32 &  -  & T \\ 
101 & 290.91386+14.517350 &  19:23:39.33 &  14:31:2.440 & 0.66$\times$0.65 & 103 & \,\,\,\,\,\,\,\,\,3.1$\pm$1.1 & \,\,\,\,\,\,4.5$\pm$0.9 &        -         &      -      &        \,\,\,-      &       \,\,\,\,\,\,-        & \,\,\,34$\pm$09      &    \,\,\,\,\,\,1.2$\pm$0.2\,\,\, & 3512 & 0.39 &  -  & T \\ 
102 & 290.91539+14.521770 &  19:23:39.69 &  14:31:18.36 & 0.72$\times$0.58 & \,\,\,86 & \,\,\,\,\,\,\,\,\,1.9$\pm$0.5 & \,\,\,\,\,\,2.7$\pm$0.4 &        -         &      -      &        \,\,\,-      &       \,\,\,\,\,\,-        & \,\,\,30$\pm$08      &    \,\,\,\,\,\,0.9$\pm$0.1\,\,\, & 3364 & 0.47 &  -  & T \\
\end{tabular}}
\tablefoot
{
The coordinates are given at the J2000 Epoch. 
(1) Estimated from the measures at 1.3\,mm. 
(2) Indicates if the source found by \textsl{getsf} was also found by \textsl{GExt2D} (True or False). 
The $^*$ next to the uncertainty indicates that the true value is $<$0.05. 
}
\end{table}
\end{landscape}